\documentclass[reprint,superscriptaddress,float,11pt,longbibliography,onecolumn]{revtex4-2}
\usepackage{amsmath,amssymb}
\usepackage{rotating}
\usepackage{graphicx}
\usepackage{dcolumn}
\usepackage{color,braket}
\usepackage{calligra}
\usepackage{soul}
\usepackage{bm}
\usepackage{subfigure}
\usepackage{tikz}
\usepackage{overpic}
\usepackage{placeins}
\usepackage{lipsum}
\usepackage{blindtext}
\usepackage{enumitem}
\usepackage{amsfonts}
\usepackage{ragged2e}
\usepackage{xr}
\usepackage{hyperref}
\usepackage{cleveref}
\makeatletter
\newcommand*{\addFileDependency}[1]{
  \typeout{(#1)}
  \@addtofilelist{#1}
  \IfFileExists{#1}{}{\typeout{No file #1.}}
}
\makeatother

\usepackage[]{bigints}
\hypersetup{
 colorlinks=true,
 linkcolor=blue,
 filecolor=blue,
 urlcolor=cyan,
}

\makeatletter
\renewcommand{\section}{\@startsection{section}{1}{0mm}
  {-\baselineskip}{0.5\baselineskip}{\bf\leftline}}
  
\makeatother

 \makeatletter
\renewcommand{\subsection}{\@startsection{subsection}{1}{0mm}
  {-\baselineskip}{0.5\baselineskip}{\bf\leftline}}
 
\makeatother
\begin{document}
    \title{Experimental identification of the second-order non-Hermitian skin effect with physics-graph-informed machine learning}

\author{Ce Shang}
\thanks{These authors contributed equally}
\affiliation{King Abdullah University of Science and Technology (KAUST), Physical Science and Engineering Division (PSE), Thuwal 23955-6900, Saudi Arabia.}
\author{Shuo Liu}
\thanks{These authors contributed equally}
\affiliation{State Key Laboratory of Millimeter Waves, Southeast University, Nanjing 210096, China.}
\author{Ruiwen Shao}
\thanks{These authors contributed equally}
\affiliation{State Key Laboratory of Millimeter Waves, Southeast University, Nanjing 210096, China.}
\author{Peng Han}
\affiliation{King Abdullah University of Science and Technology (KAUST), Computer, Electrical, and Mathematical Sciences and Engineering Division (CEMSE), Thuwal 23955-6900, Saudi Arabia.}

\author{Xiaoning Zang}
\affiliation{King Abdullah University of Science and Technology (KAUST), Physical Science and Engineering Division (PSE), Thuwal 23955-6900, Saudi Arabia.}

\author{Xiangliang Zhang}
\affiliation{Department of Computer Science and Engineering, University of Notre Dame, Notre Dame, IN 46556, USA}
\affiliation{King Abdullah University of Science and Technology (KAUST), Computer, Electrical, and Mathematical Sciences and Engineering Division (CEMSE), Thuwal 23955-6900, Saudi Arabia.}

\author{Khaled Nabil Salama}
\affiliation{King Abdullah University of Science and Technology (KAUST), Computer, Electrical, and Mathematical Sciences and Engineering Division (CEMSE), Thuwal 23955-6900, Saudi Arabia.}

\author{Wenlong Gao}
\affiliation{Paderborn University, Department of Physics, Warburger Str. 100, 33098
Paderborn, Germany}

\author{Ching Hua Lee}
\affiliation{Department of Physics, National University of Singapore, Singapore 117551, Republic of Singapore}

\author{Ronny Thomale}
\affiliation{Institut f\"{u}r Theoretische Physik und Astrophysik, Universität Würzburg, W\"{u}rzburg, Germany.}

\author{Aur\'elien Manchon}
\email{manchon@cinam.univ-mrs.fr}
\affiliation{CINaM, Aix-Marseille University, CNRS, Marseille, France.}

\author{Shuang Zhang}
\email{shuzhang@hku.hk}
\affiliation{Department of Physics, The University of Hong Kong, Hong Kong, China}

\author{Tie Jun Cui}
\email{tjcui@seu.edu.cn}
\affiliation{State Key Laboratory of Millimeter Waves, Southeast University, Nanjing 210096, China.}

\author{Udo Schwingenschl\"{o}gl}
\email{udo.schwingenschlogl@kaust.edu.sa}
\affiliation{King Abdullah University of Science and Technology (KAUST), Physical Science and Engineering Division (PSE), Thuwal 23955-6900, Saudi Arabia.}


\begin{abstract}
Topological phases of matter are conventionally characterized by the bulk-boundary correspondence in  Hermitian systems: The topological invariant of the bulk in $d$ dimensions corresponds to the number of $(d-1)$-dimensional boundary states. By extension, higher-order topological insulators reveal a bulk-edge-corner correspondence, such that $n$-th order topological phases feature $(d-n)$-dimensional boundary states. The advent of non-Hermitian topological systems sheds new light on the emergence of the non-Hermitian skin effect (NHSE) with an extensive number of boundary modes under open boundary conditions.  Still,  the higher-order NHSE
remains largely unexplored, particularly in the experiment. We introduce an unsupervised approach -- physics-graph-informed machine learning (PGIML) --  to enhance the data mining ability of machine learning with limited domain knowledge.  Through PGIML, we experimentally demonstrate the second-order NHSE in a two-dimensional non-Hermitian topolectrical circuit. The admittance spectra of the circuit exhibit an extensive number of corner skin modes and extreme sensitivity of the spectral flow to the boundary conditions. The violation of the conventional bulk-boundary correspondence in the second-order NHSE implies that modification of the topological band theory is inevitable in higher dimensional non-Hermitian systems.
\end{abstract}
\maketitle
\section*{Introduction}
Conceptual theories about topological phases of matter are at the forefront of contemporary research. In Hermitian systems, the guiding principle of topological insulators (TIs) is the bulk-boundary correspondence, stating that the topological invariants of the bulk determine the number of gapless boundary modes \cite{RevModPhys.82.3045,RevModPhys.83.1057,RevModPhys.88.021004}. With  progress in research, higher-order TIs have revealed a novel bulk-edge-corner correspondence, where $n$-th order topological phases in $d$ dimensions feature $(d-n)$-dimensional boundary modes \cite{doi:10.1126/science.aah6233,Benalcazar61,PhysRevB.96.245115,PhysRevB.95.235143,PhysRevLett.119.246402,PhysRevLett.119.246401,Schindler2018sc,PhysRevB.97.155305,PhysRevLett.123.256402,PhysRevLett.123.216803,PhysRevLett.124.036803,PhysRevLett.124.216601,PhysRevLett.124.166804}. Building up on the categories of Hermitian systems, non-conservative systems without Hermiticity reveal a plethora of unconventional physical principles, phenomena, and applications. Among many others, this includes parity-time symmetry \cite{ruter2010observation,regensburger2012parity,peng2014parity}, exceptional points  \cite{zhang2018phonon}, exceptional Fermi arcs \cite{zhou2018observation}, sensing \cite{hodaei2017enhanced,chen2017exceptional}, and lasing \cite{hodaei2014parity,brandstetter2014reversing}. Recently, the concept of non-Hermiticity has been intertwined with topological phases of matter \cite{weimann2017topologically,bahari2017nonreciprocal,bandres2018topological,harari2018topological} to yield the non-Hermitian skin effect (NHSE) with an extensive number of boundary modes and the necessity to assess non-Hermitian topological properties beyond Bloch band theory \cite{PhysRevLett.121.026808,PhysRevX.9.041015,PhysRevLett.123.066404}.


Despite a fast-growing number of theoretical predictions for non-Hermitian topological systems \cite{PhysRevLett.121.086803,PhysRevLett.123.170401,PhysRevLett.123.073601,PhysRevLett.123.016805,PhysRevLett.125.126402,PhysRevLett.125.226402,PhysRevB.102.205118,PhysRevB.102.241202,PhysRevLett.124.086801,Li2020,2020chinghua,PhysRevB.103.045420,2021Linhu}, experimental explorations are still at an early stage \cite{Helbig2020,PhysRevResearch.2.023265,Liu2021,2021jianhua,2021lucas}. To date, the first-order NHSE has been realized in photonic \cite{Weidemann2020} and in circuitry \cite{Helbig2020,PhysRevResearch.2.023265,Liu2021} environments, whereas the experimental realization of the higher-order NHSE remains open. Although skin corner modes have been observed in very recent research \cite{2021jianhua,2021lucas},  the unique features of the higher-order NHSE need to be fully demonstrated, both the extensive number of boundary modes under open boundary conditions and the extreme sensitivity of the spectral flow to the boundary conditions. To analyze the spectral flow in higher dimensions, traditional methodologies are challenged by the large-scale data generated. The data size will grow exponentially with the dimension, and additional boundary conditions make it more difficult to analyze the outcome. Machine learning (ML) is a promising way to process large amounts of data \cite{RevModPhys.91.045002,2019Pankaj,2019mark}. The existing approaches, however, are unable to efficiently extract the crucial observables, in particular with a largely unexplored state of matter at hand. There is a pressing need for integrating fundamental physical laws and domain knowledge by teaching ML models the governing physical rules, which can, in turn, provide informative priors, i.e., theoretical constraints and inductive understanding of the observable features. To this end, physics-informed ML, using informative priors for the phenomenological description of the world, can be leveraged to improve the performance of the learning algorithm \cite{Karniadakis2021}.

In this article, we report two significant advances: (i) The methodology of physics-graph-informed machine learning (PGIML) is introduced to enforce identification of an unrevealed physical phenomenon by integrating physical principles, graph visualization of features, and ML. The informative priors provided by PGIML enable an analysis that remains robust even in the presence of imperfect data (such as missing values, outliers, and noise) to make accurate and physically consistent predictions of phenomenological parameters. (ii) The second-order NHSE, characterized by skin corner modes and the violation of the conventional bulk-boundary correspondence, is realized in a two-dimensional (2D) non-Hermitian topoelectrical circuit. We achieve the first experimental demonstration of the extreme sensitivity of the spectral flow to (fully controlled) boundary conditions (PBC$x$-PBC$y$, PBC$x$-OBC$y$, OBC$x$-PBC$y$, and OBC$x$-OBC$y$, where PBC (OBC) represents a periodic (open) boundary condition and $x$ $(y)$ represents direction), and observe corner skin modes under OBC$x$-OBC$y$ as well as edge skin modes under PBC$x$-OBC$y$. Prospectively, the powerful tool of PGIML can be applied more widely to solve digital twin problems \cite{kritzinger_digital_2018,tao_make_2019,singh_digital_2021}, thus bridging the physical and digital worlds by linking the flow of data/information between them \cite{gelernter1993mirror,grieves_product_2005}.
\section*{Results}
\subsection*{Physics-graph-informed machine learning}
The PGIML framework is implemented in the context of a circuitry environment. In an electrical circuit, the scattering matrix ($S$-matrix) relates the voltage of the waves incident to ports to those of the waves reflected from ports (see Supplementary Material Sec.\ \textcolor{blue}{S1}), providing a complete description of the circuit \cite{pozar2011microwave}.
 According to graph theory (network topology), a $N$-port electrical circuit can be converted into a  matrix ${\textit{\textbf{G}}}=({\textit{\textbf{P}}},{\textit{\textbf{S}}})$ of complex-weighted directed bipartite graphs $G_{ab}=(P_{ab},S_{ab})$ with the matrix ${\textit{\textbf{P}}}$  of positions $P_{ab}=({a},{b})$ and  the $S$-matrix ${\textit{\textbf{S}}}$ of scattering-parameters ($S$-parameters) $S_{ab}$, where $a,b \in \{1,2,\dots, N\} $ denotes the ports \cite{west2001introduction}. We define the set of graphs as $\mathcal{G}=(\mathcal{P},\mathcal{S})=\{G_{ab}| a,b \in \{1,2,\dots, N\}\}$ with the set of positions $\mathcal{P}$ and the set of $S$-parameters $\mathcal{S}$. To identify the characteristic features of the circuit, especially of a large circuit, cluster analysis can be used to detect graphs with similar properties. Here, a $K$-means clustering algorithm \cite{likas_global_2003,wu_advances_2012} is employed to partition $\mathcal{G}$ into $K$ clusters ${\mathcal{G}}_\kappa$ based on the value of the $S$-parameter, where $\mathcal{G}=\bigcup\limits_{\kappa = 1}^K { {{\mathcal{G}}_\kappa} }$. The axiom of choice \cite{moore2012zermelo}
 states that for every indexed ${\mathcal{G}}_{\kappa}$ we can find a representative graph $\hat{G}_{\kappa}$ such that ${ \hat{G}_{\kappa}\in {\mathcal{G}}_{\kappa}}$. In a digital twin scenario of simulation and experiment, the set of simulated graphs $\mathcal{G}_{\rm{sim.}}$ is generated to describe the numerical outcome that imitates the set of experimental graphs $\mathcal{G}_{\rm{exp.}}$.  As $\mathcal{G}_{\rm{sim.}}$ and $\mathcal{G}_{\rm{exp.}}$ are isomorphic, the subsets $\mathcal{G}_{\rm{sim.},\kappa}$ and $\mathcal{G}_{\rm{exp.},\kappa}$ are isomorphic \cite{miller1979graph}. Therefore,  PGIML can  be understood in the  teacher-student  scenario in the sense that the  teacher ($\mathcal{G}_{\rm{sim.}}$)  imparts informative priors ($\hat{G}_{\rm{sim.}}$)  to  the  student ($\mathcal{G}_{\rm{exp.}}$).

  \begin{figure}[htbp]
\centering
\includegraphics[width=1\linewidth]{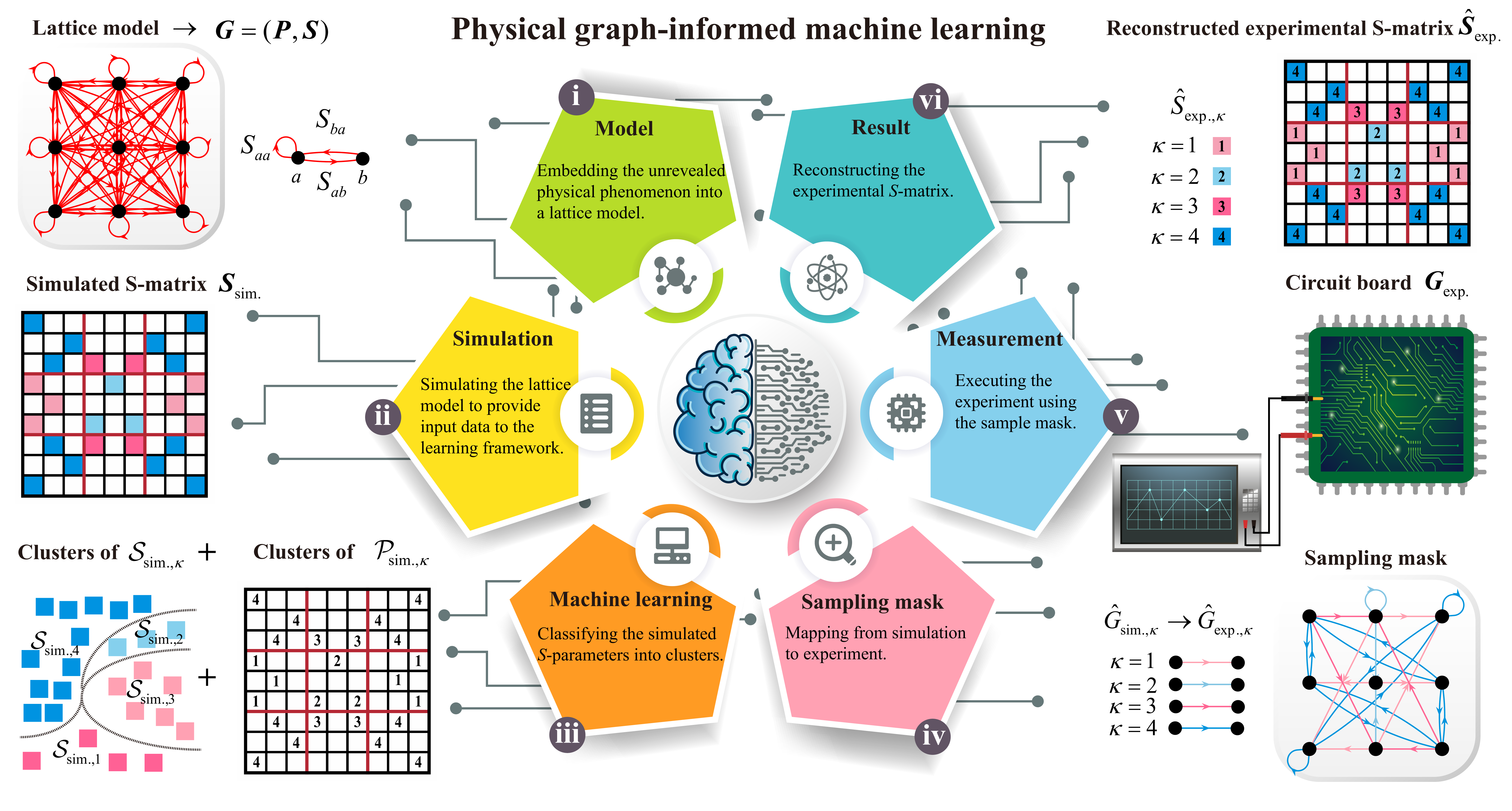}
\caption{{\bf{PGIML framework.}} (i) A lattice model embedding the unrevealed physical phenomenon is generated. The directional red circles and lines correspond to $S_{aa}$ and $S_{ab}$, respectively. (ii) The simulated $S$-matrix ${\textit{\textbf{S}}}_{\rm{sim.}}$ of the $L \times L$ lattice model is constructed and $N\times N$ ($N=L^2$) elements of a learning set $\mathcal{G}_{\rm{sim.}}=(\mathcal{P}_{\rm{sim.}},\mathcal{S}_{\rm{sim.}})$  are accumulated. (iii) $\mathcal{P}_{\rm{sim.}}$ and $\mathcal{S}_{\rm{sim.}}$ are classified into  clusters  $\mathcal{P}_{\rm{sim.},\kappa}$ and  $\mathcal{S}_{\rm{sim.},\kappa}$ using the $K$-means method. (iv) A sampling mask corresponding to the graph-to-graph mapping $\hat{G}_{\rm{sim.},\kappa}\to \hat{G}_{\rm{exp.},\kappa}$ is generated. (v) The representative experimental $S$-parameters $\hat{S}_{\rm{exp.},\kappa}$ are measured in the circuit. (vi)  The reconstructed experimental $S$-matrix $\hat{\textit{\textbf{S}}}_{\rm{exp.}}$ is retrieved.}
\label{fig:1}
\end{figure}

We depict the PGIML framework in Fig.\ \ref{fig:1}: (i) A lattice model that embeds the unrevealed physical phenomenon is generated and converted into a matrix of graphs {\textit{\textbf{G}}}. (ii) The simulated $S$-matrix ${\textit{\textbf{S}}}_{\rm{{sim.}}}$ of the circuit is constructed and  a learning set  ${\cal G}_{\rm{sim.}}=({\cal P}_{\rm{sim.}},{\cal S}_{\rm{sim.}})$ is accumulated. (iii) The set of simulated positions $ {{\cal P}_{{\rm{sim.}}} }$ and the set of simulated $S$-parameters $ { {{\cal S}_{{\rm{sim.}}} }}$ are classified into clusters  $ { {{\cal P}_{{\rm{sim.},\kappa}} }}$ and $ { {{\cal S}_{{\rm{sim.},\kappa}} }}$ using the $K$-means method (see Supplementary Material Sec.\ \textcolor{blue}{S3}). (iv) The graph-to-graph mapping ${{\hat G}_{{\rm{sim}}.,\kappa }} \to  {{\hat G}_{\exp .,\kappa }} $ is translated into a  sampling mask that mirrors the clustering information. (v) The representative experimental $S$-parameters ${\hat S}_{{\rm{exp}}.,\kappa}$ are measured in the circuit. (vi) The $S$-matrix is encoded with the measured features $\bigcup\limits_{\kappa  = 1}^K {\{ {{\hat S}_{\exp.,\kappa }}\} }$ and  the reconstructed experimental $S$-matrix ${\hat{{\textit{\textbf{S}}}}_{\rm{exp.}}}$ is retrieved.  The experimental $S$-matrix ${\textit{\textbf{S}}}_{\rm{{exp.}}}$
is then given by
 \begin{equation}
\begin{array}{l}
{{\textit{\textbf{S}}}_{\rm{exp.}}} \sim {\hat{{\textit{\textbf{S}}}}_{\rm{exp.}}}  =\sum\limits_{\kappa = 1}^K {{\hat{S}}_{{\rm{exp.}},\kappa}} \sum\limits_{(a,b) \in {{\mathcal{P}}_{{\rm{sim.}},\kappa}}} {{{\textit{\textbf{E}}}_{ ab}}},
\end{array}
 \end{equation}
 where ${\textit{\textbf{E}}}_{ab}$ is a single-entry matrix (element $ab$ is one and the other elements are zero) \cite{petersen2008matrix}. Compared to conventional measurements of $N^2$ elements, the PGIML method is $N^2/K$ times faster, as it filters out redundancies, especially efficient for circuits that are too complex for a human to process.

\subsection*{Second-order non-Hermitian skin effect}
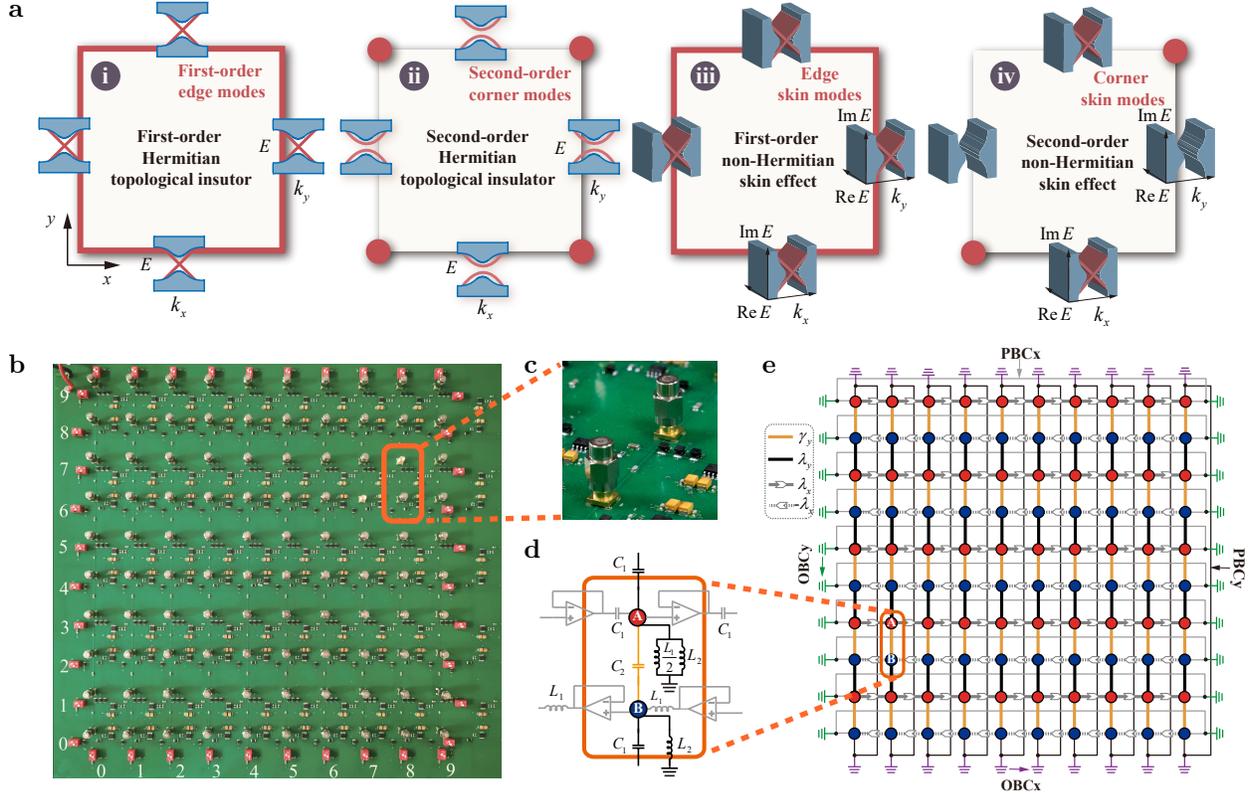
\begin{figure*}[htbp!]
\begin{minipage}{1\linewidth}
\subfigure{\includegraphics[height=4.5cm]{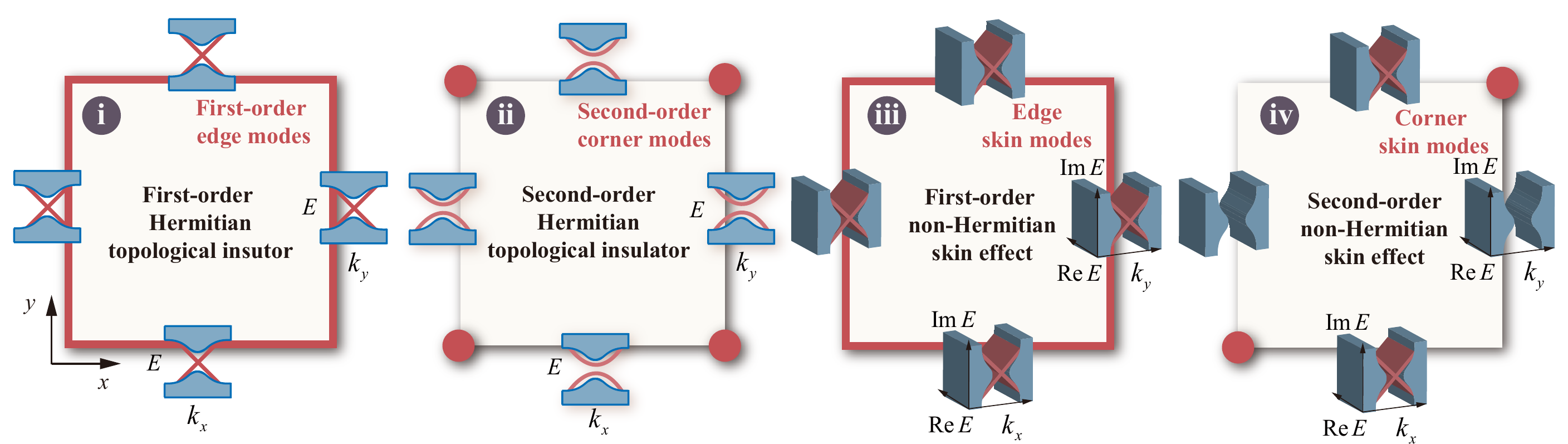}\label{fig:2a}}
\end{minipage}
\begin{minipage}{0.38\linewidth}
\subfigure{\includegraphics[height=5.5cm]{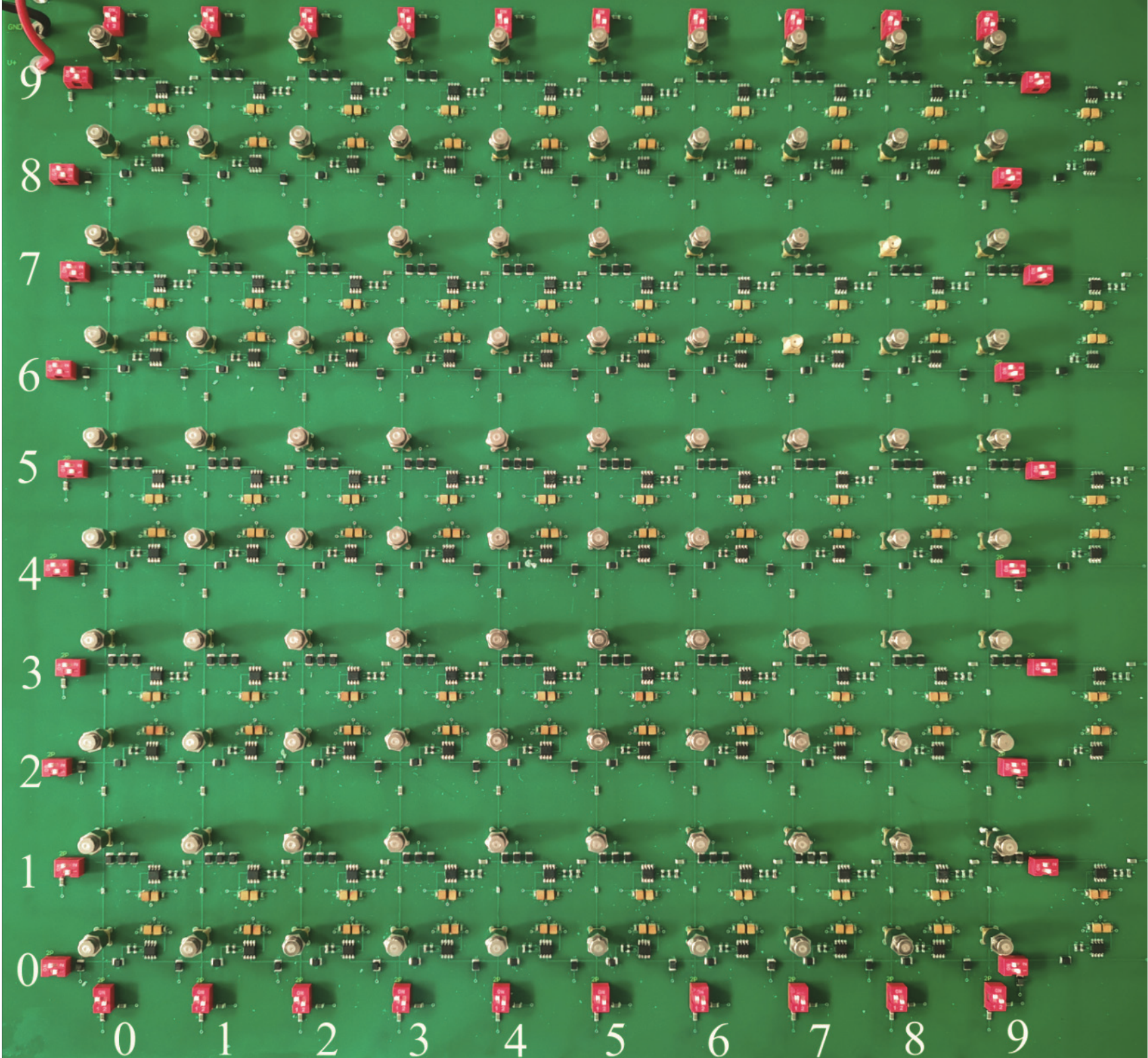}\label{fig:2b}}
\end{minipage}
\hspace{-10mm}
\begin{minipage}{0.3\linewidth}
\subfigure{\includegraphics[height=2.1cm]{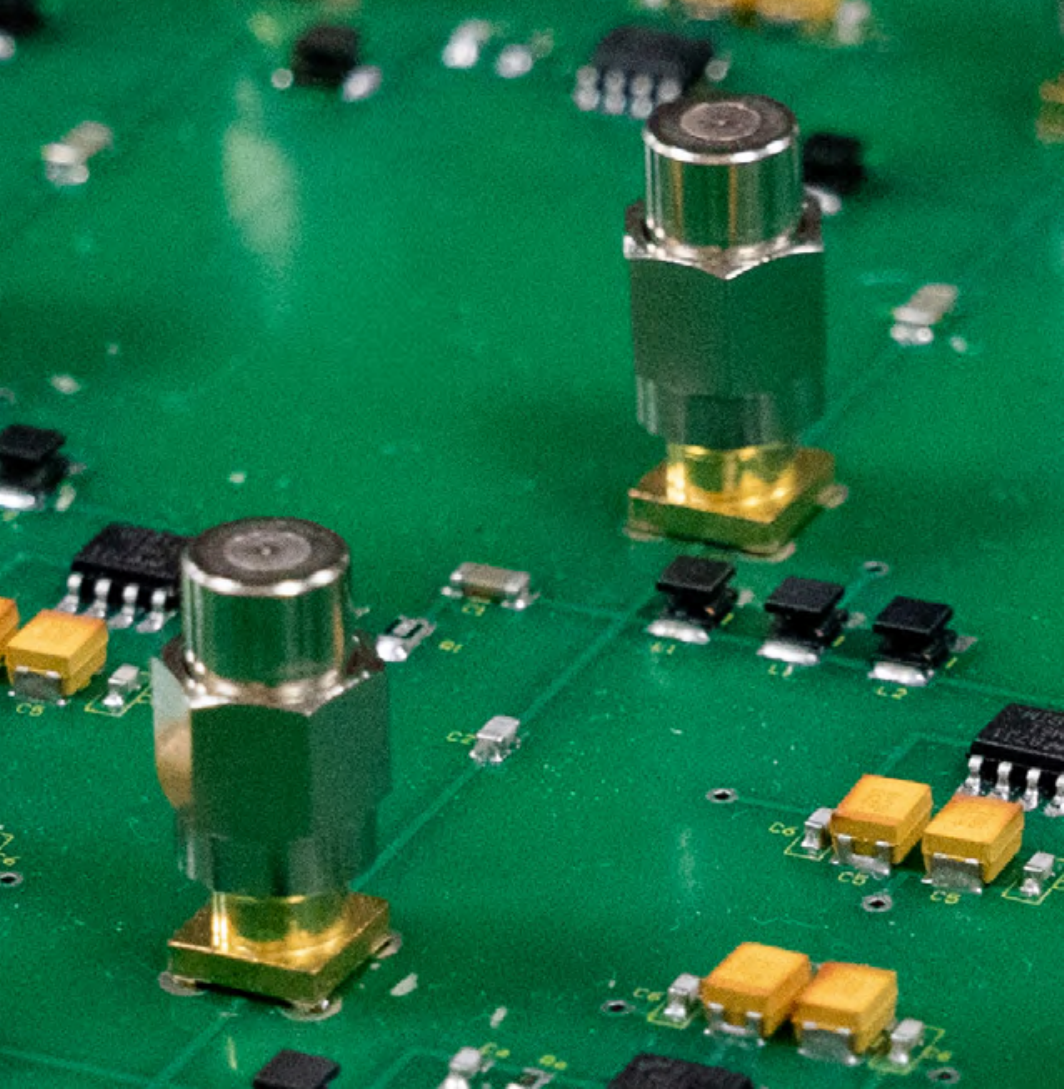}\label{fig:2c}}
\subfigure{\includegraphics[height=3.1cm]{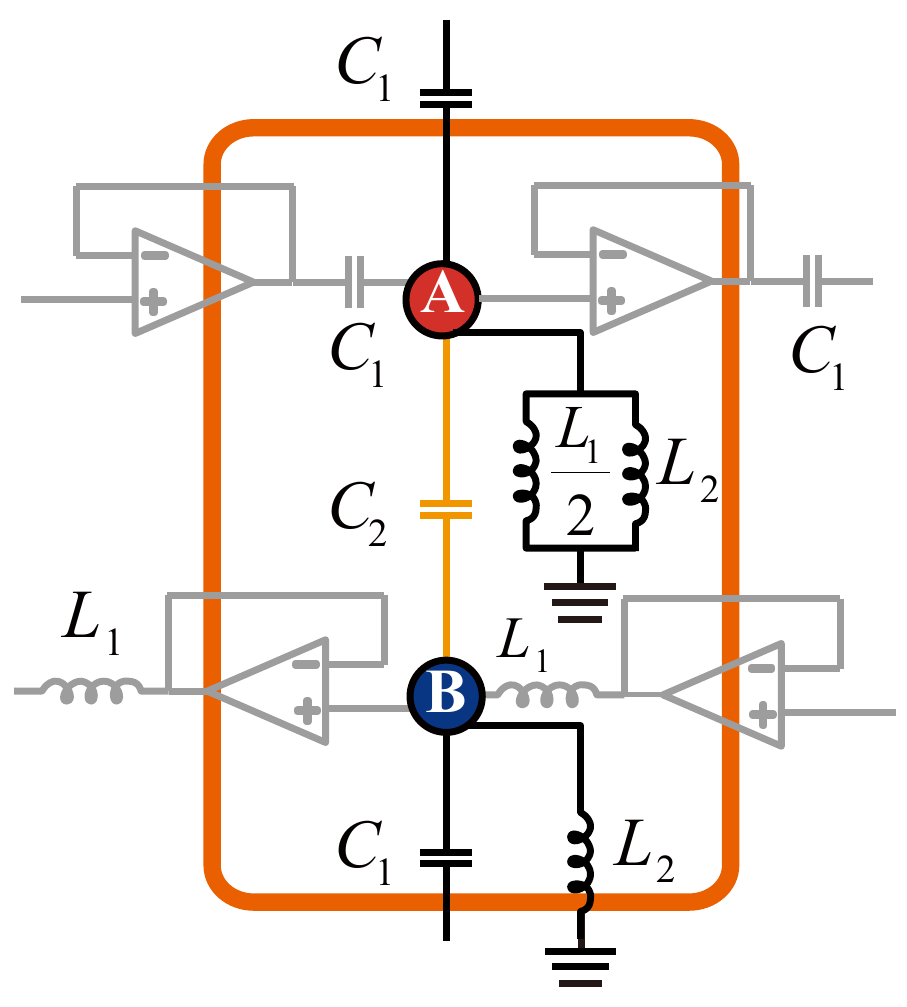}\label{fig:2d}}
\end{minipage}%
\hspace{-10mm}
\begin{minipage}{0.4\linewidth}
\subfigure{\includegraphics[height=6.1cm]{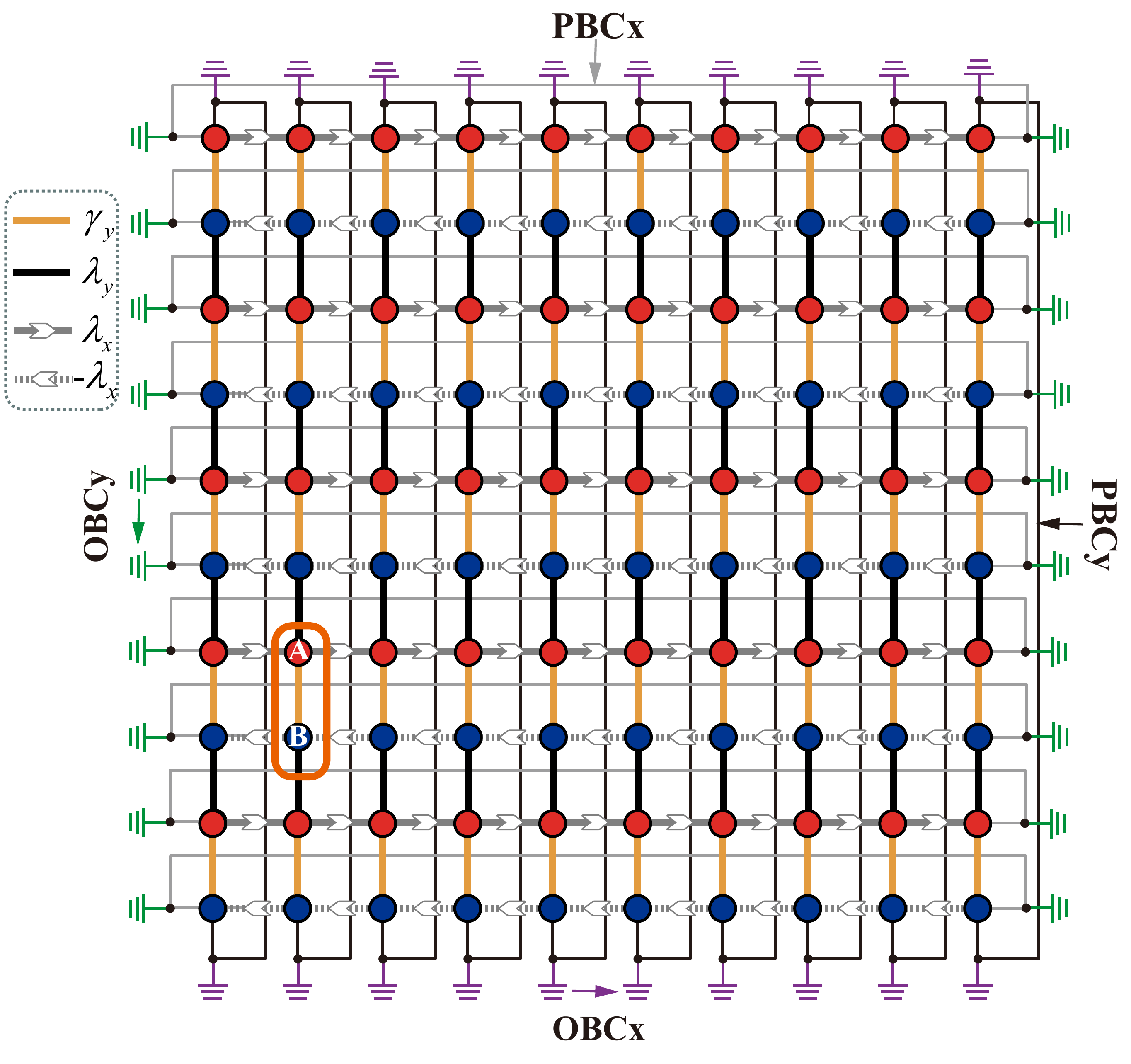}\label{fig:2e}}
\end{minipage}
\hspace{-8mm}
\begin{picture}(0,0)
\put(-455,215){\bf{a}}\put(-455,80){\bf{b}} \put(-260,80){\bf{c}} \put(-260,10){\bf{d}} \put(-170,80){\bf{e}} 

\put(-300,51){\begin{tikzpicture}
\definecolor{dkor}{rgb} {1,0.4,0.16}
\draw [dashed,color=dkor,line width=2pt](0.1,1) -- (1.95,2.1);
\end{tikzpicture} }
\put(-304,23){\begin{tikzpicture}
\definecolor{dkor}{rgb} {1,0.4,0.16}
\draw [dashed,color=dkor,line width=2pt](0,0) -- (2,0.05);
\end{tikzpicture} }

\put(-314,23){\begin{tikzpicture}
\definecolor{dkor}{rgb} {1,0.4,0.16}
\draw[draw=dkor, rounded corners=3,line width=2pt] (0,0)
-- (0.5,0)
-- (0.5,1)
-- (0,1)
 -- cycle;
\end{tikzpicture} }
\put(-190,-12){\begin{tikzpicture}
\definecolor{dkor}{rgb} {1,0.4,0.16}
\draw [dashed,color=dkor,line width=2pt](0,0) -- (2.5,-0.4);
\end{tikzpicture} }
\put(-190,-64){\begin{tikzpicture}
\definecolor{dkor}{rgb} {1,0.4,0.16}
\draw [dashed,color=dkor,line width=2pt](0,0) -- (2.5,1);
\end{tikzpicture} }
\end{picture}
\caption{{\bf{Second-order NHSE in a non-Hermitian circuit.}} $ \bf{a}$, Schematic diagrams of topological band theory for the first-order TI, second-order TI,  first-order NHSE, and second-order NHSE. $\bf{b}$,  Photograph of the circuit. $\bf{c}$, Photograph of the unit cell. $\bf{d}$, Scheme of the unit cell. $\bf{e}$,  Tight-binding analog of the circuit. The circuit components are represented by the intracell  coupling $\gamma_y$ $(C_2)$,  intercell coupling $\lambda_y$ $(C_1)$, and intercell non-reciprocal couplings $\lambda_x$ ($C_1$ connected to a voltage follower) and $-\lambda_x$ ($L_1$ reversely connected to  a voltage follower). The grounding components are $L_1L_2/(L_1+2L_2)$ and $L_2$ for sublattices A and B, respectively. PBC$x$ (grey), PBC$y$ (black), OBC$x$ (green), and OBC$y$ (purple) are controlled by the switches connecting the boundaries  (see Supplementary Material Sec.\ \textcolor{blue}{S4}).}
\label{fig:2}
\end{figure*}

We are now set up to explore the second-order NHSE, which gives rise to new types of boundary modes as a result of higher-order non-Hermitian topology. 
In a $L \times L$ lattice model, a first-order TI  has  ${\cal{O} }(L)$  edge modes with a gapless edge spectrum  in the $x$ and $y$-directions. A second-order TI has ${\cal{O} }(1)$ corner modes with a gapped edge spectrum in the $x$ and $y$-directions. The first-order NHSE features extensive ${\cal{O} }(L^2)$  edge skin modes with a gapless complex-valued edge spectrum in the $x$ and $y$-directions. Distinct from the Hermitian limit and the first-order NHSE, the second-order NHSE features ${\cal{O} }(L)$ corner skin modes with a  gapless complex-valued edge spectrum in one direction and no edge spectrum in the other direction (see Supplementary Material Sec.\ \textcolor{blue}{S2}). Schematic diagrams of these four situations are shown in Fig.\ \ref{fig:2a}. The explicit violation of the conventional bulk-boundary correspondence clearly demonstrates that modification of the topological band theory is inevitable in higher-dimensional non-Hermitian systems.

To realize the second-order NHSE experimentally, we design a topoelectrical circuit that represents a 2D non-Hermitian two-band model.
The  $10 \times 10$ circuitry lattice is shown in Fig.\ \ref{fig:2b} and the unit cell is shown in Fig.\ \ref{fig:2c} as photograph and in Fig.\ \ref{fig:2d} as scheme.  The tight-binding analog of the circuit is shown in Fig.\ \ref{fig:2e} with intracell couplings $\gamma_y$, intercell couplings $\lambda_y$ in the $y$-direction, and intercell non-reciprocal couplings $\pm \lambda_x$ in the $x$-direction.

 \begin{figure*}[b!]
\begin{minipage}{0.19\textwidth}
\subfigure{\includegraphics[width=.99\textwidth]{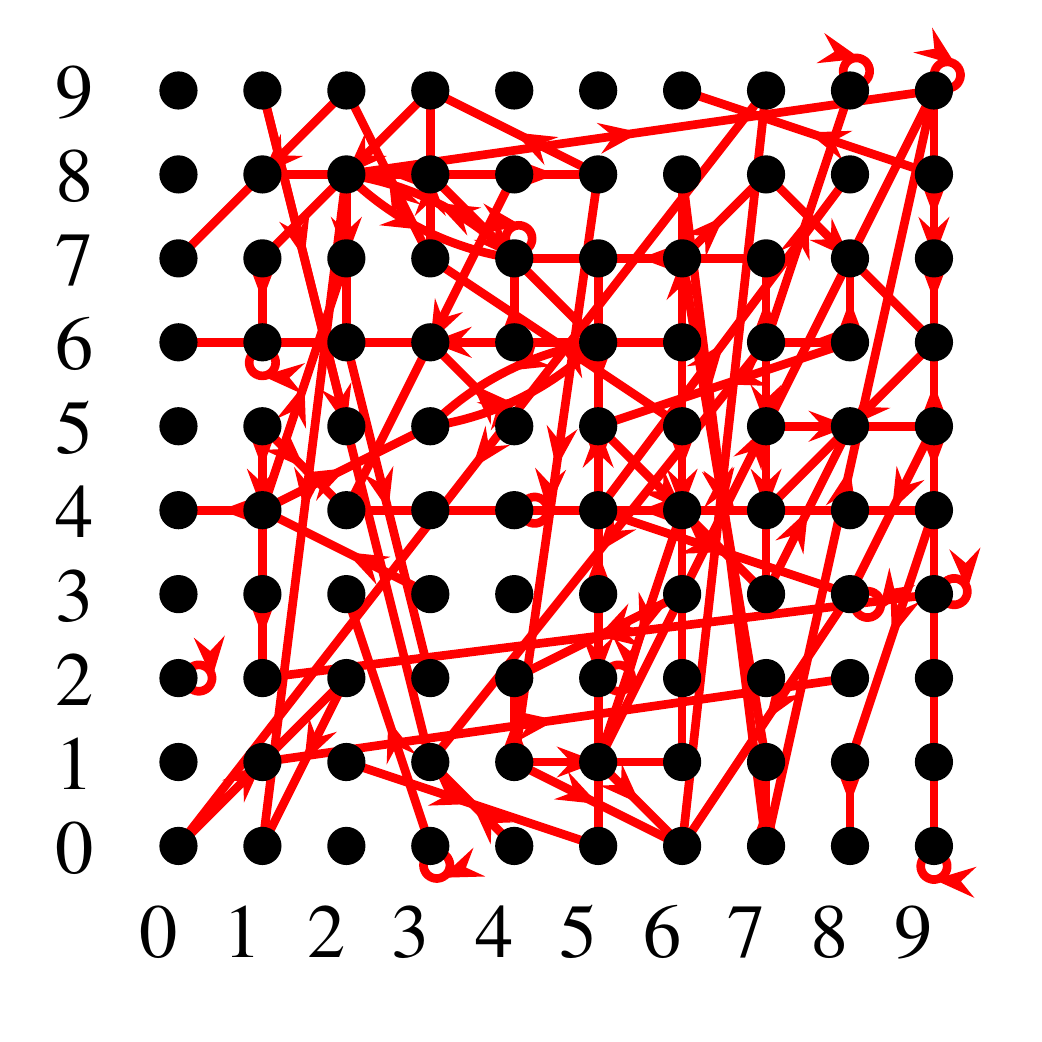}\label{fig:3a}}
\end{minipage}
\begin{minipage}{0.19\textwidth}
\subfigure{\includegraphics[width=.99\textwidth]{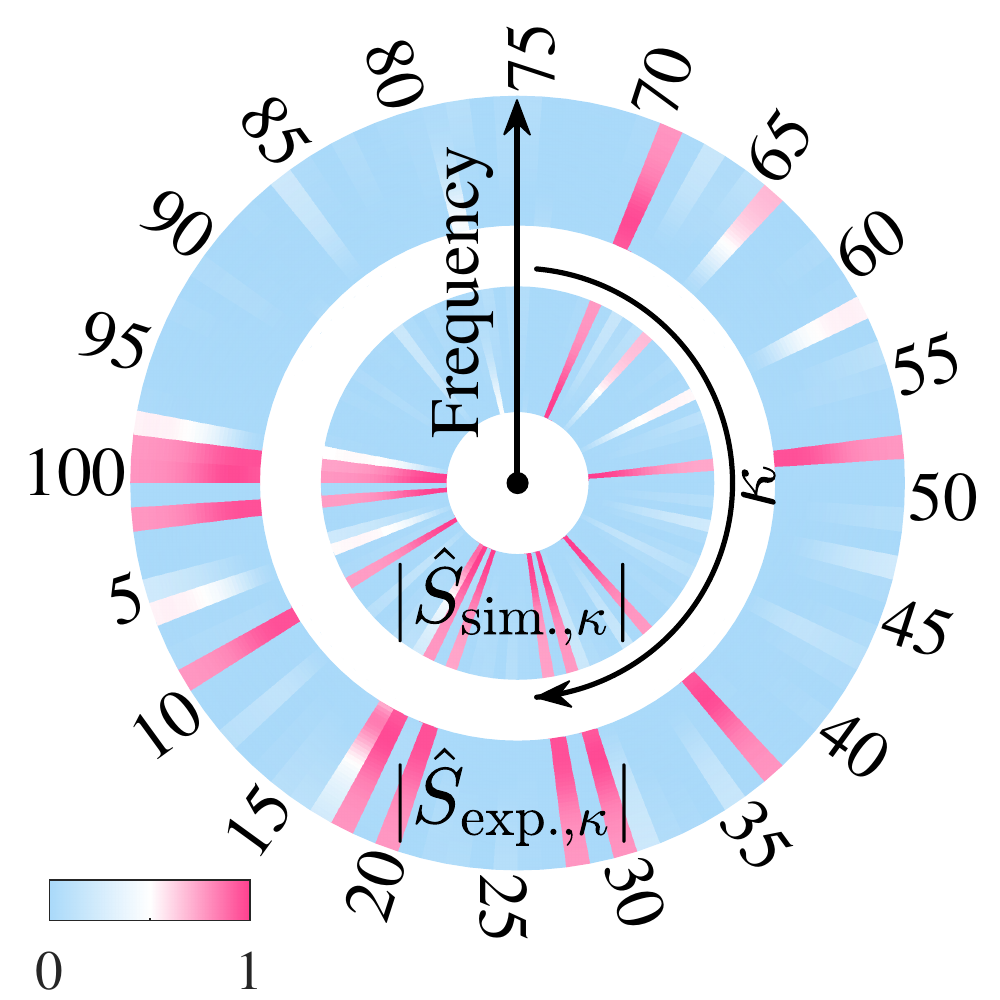}\label{fig:3b}}
\end{minipage}
\begin{minipage}{0.19\textwidth}
\subfigure{\includegraphics[width=.99\textwidth]{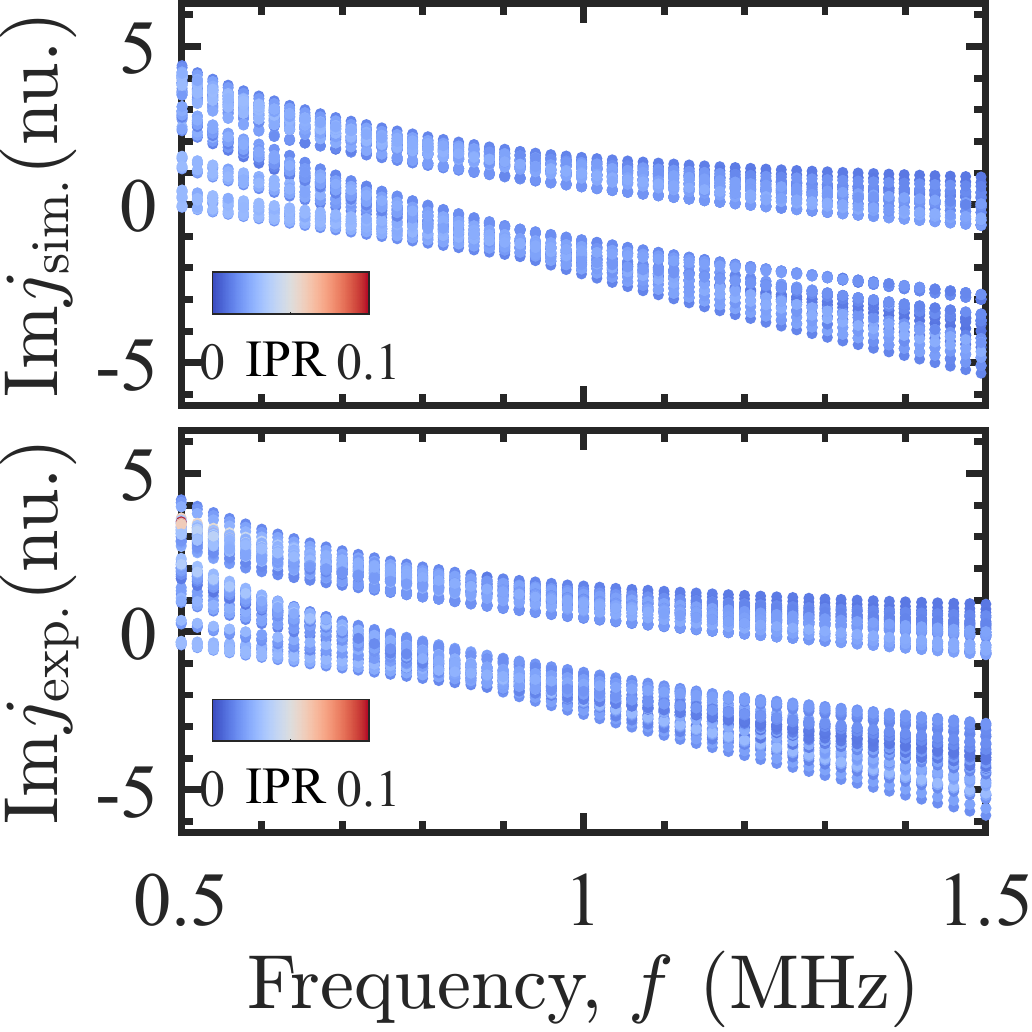}\label{fig:3c}}
\end{minipage}
\begin{minipage}{0.19\textwidth}
\subfigure{\includegraphics[width=.99\textwidth]{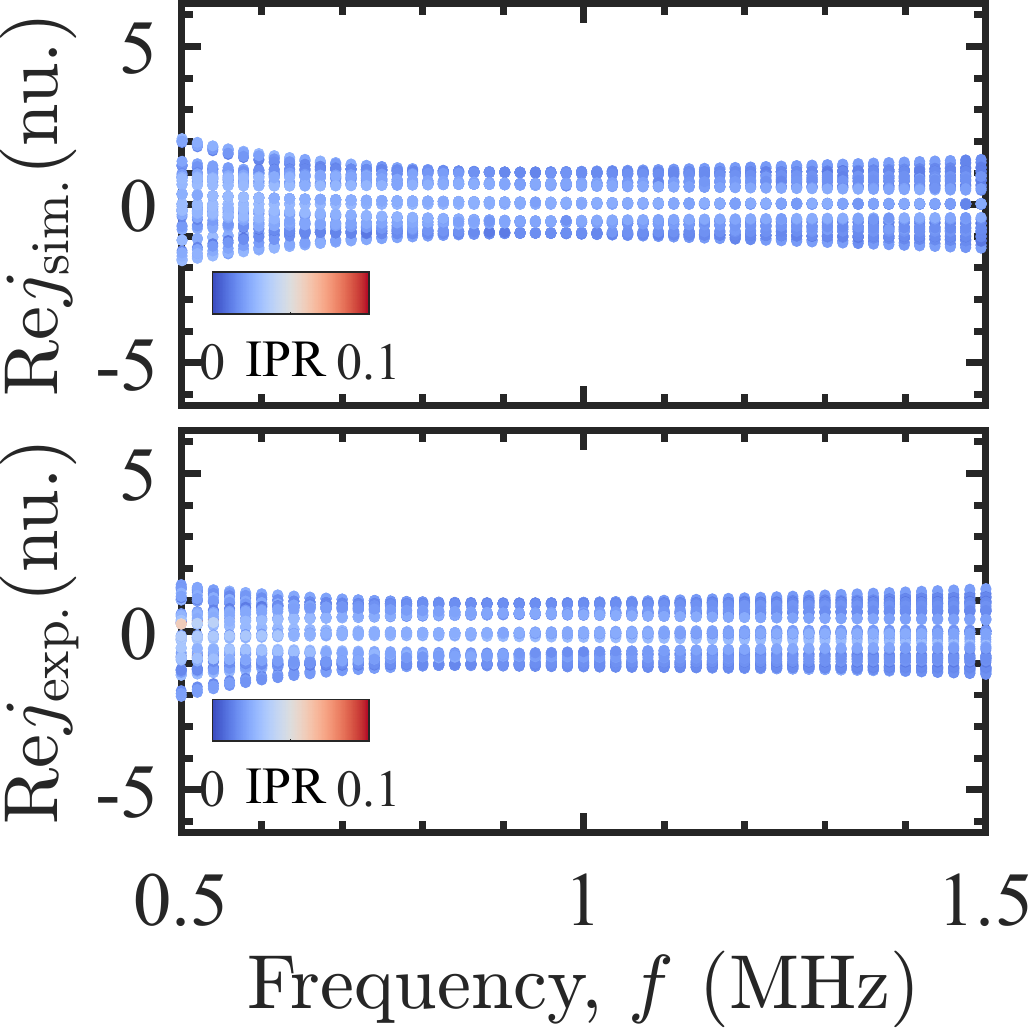}\label{fig:3d}}
\end{minipage}
\begin{minipage}{0.19\textwidth}
\subfigure{\includegraphics[width=.99\textwidth]{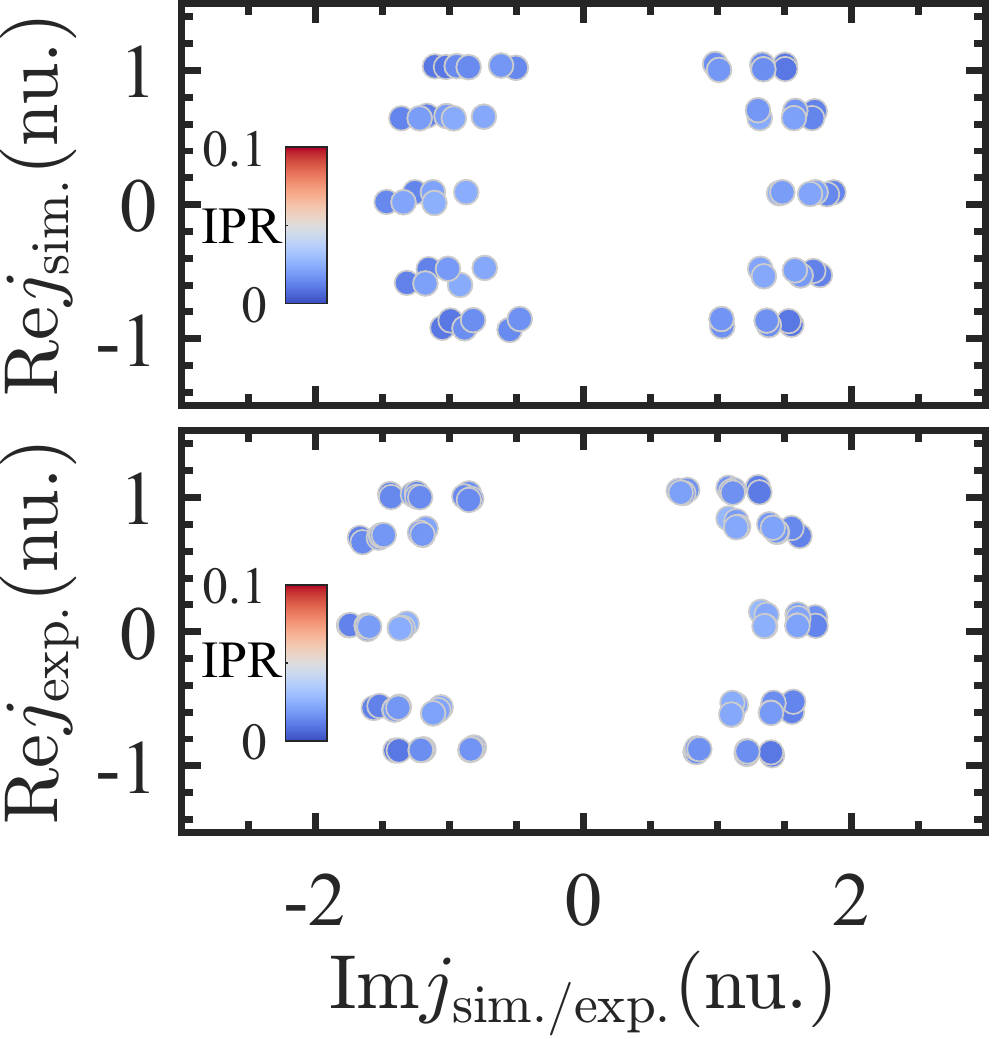}\label{fig:3e}}
\end{minipage}
\begin{minipage}{0.19\textwidth}
\subfigure{\includegraphics[width=.99\textwidth]{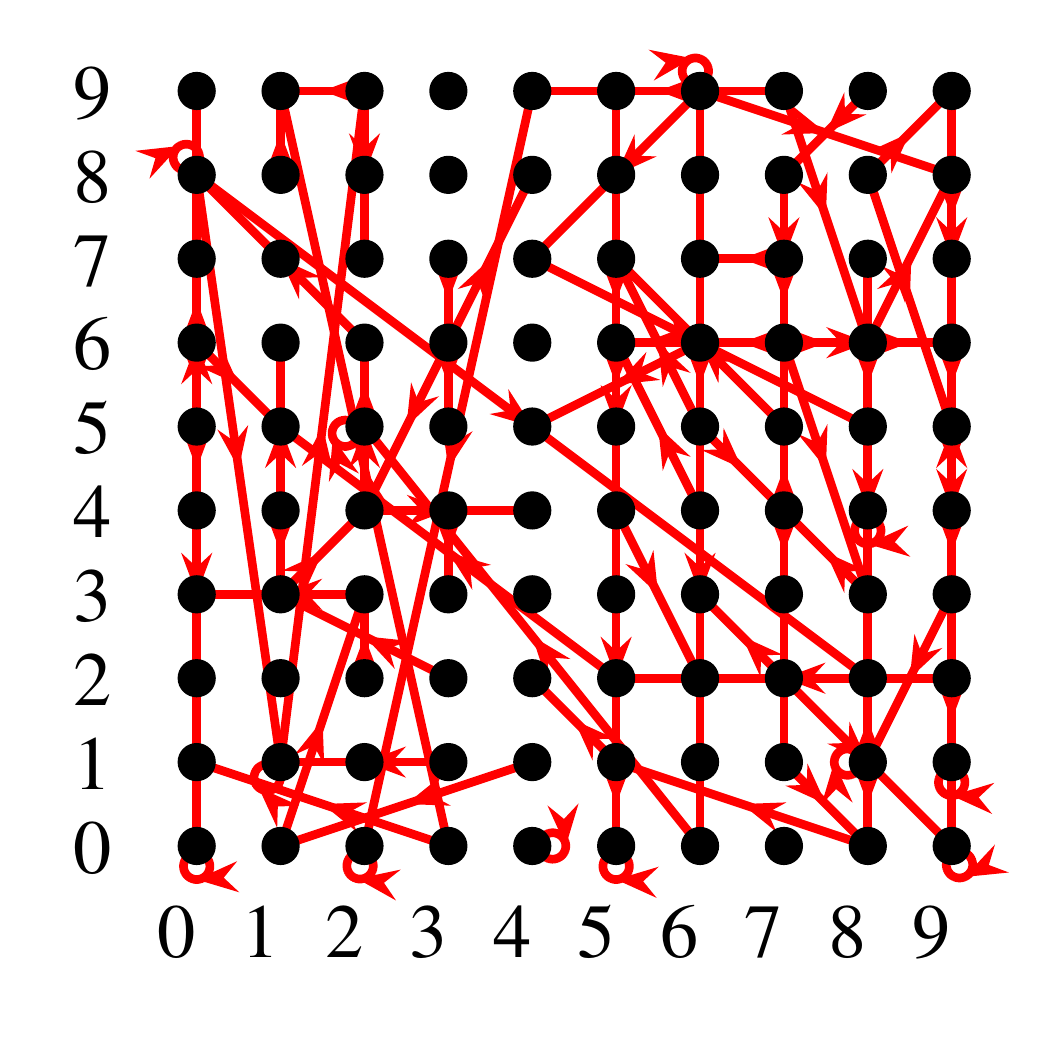}\label{fig:3f}}
\end{minipage}
\begin{minipage}{0.19\textwidth}
\subfigure{\includegraphics[width=.99\textwidth]{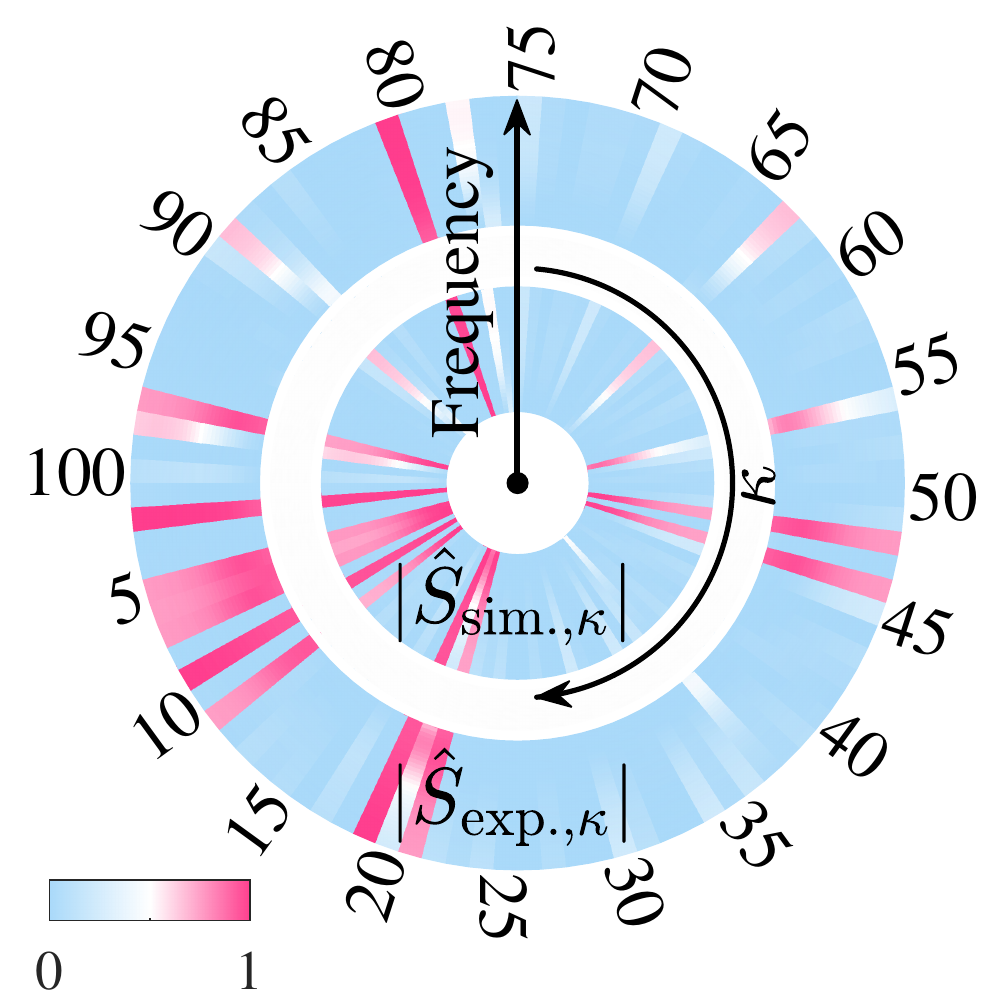}\label{fig:3g}}
\end{minipage}
\begin{minipage}{0.19\textwidth}
\subfigure{\includegraphics[width=.99\textwidth]{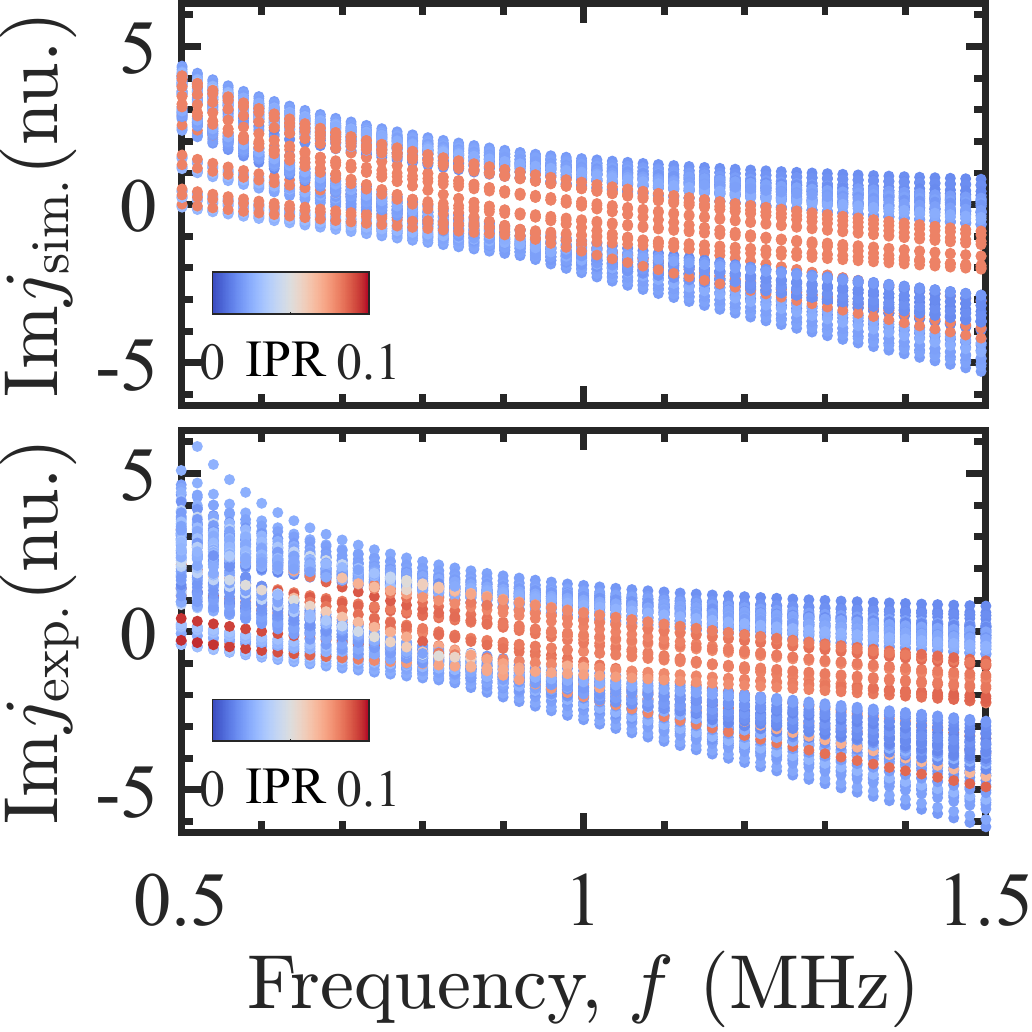}\label{fig:3h}}
\end{minipage}
\begin{minipage}{0.19\textwidth}
\subfigure{\includegraphics[width=.99\textwidth]{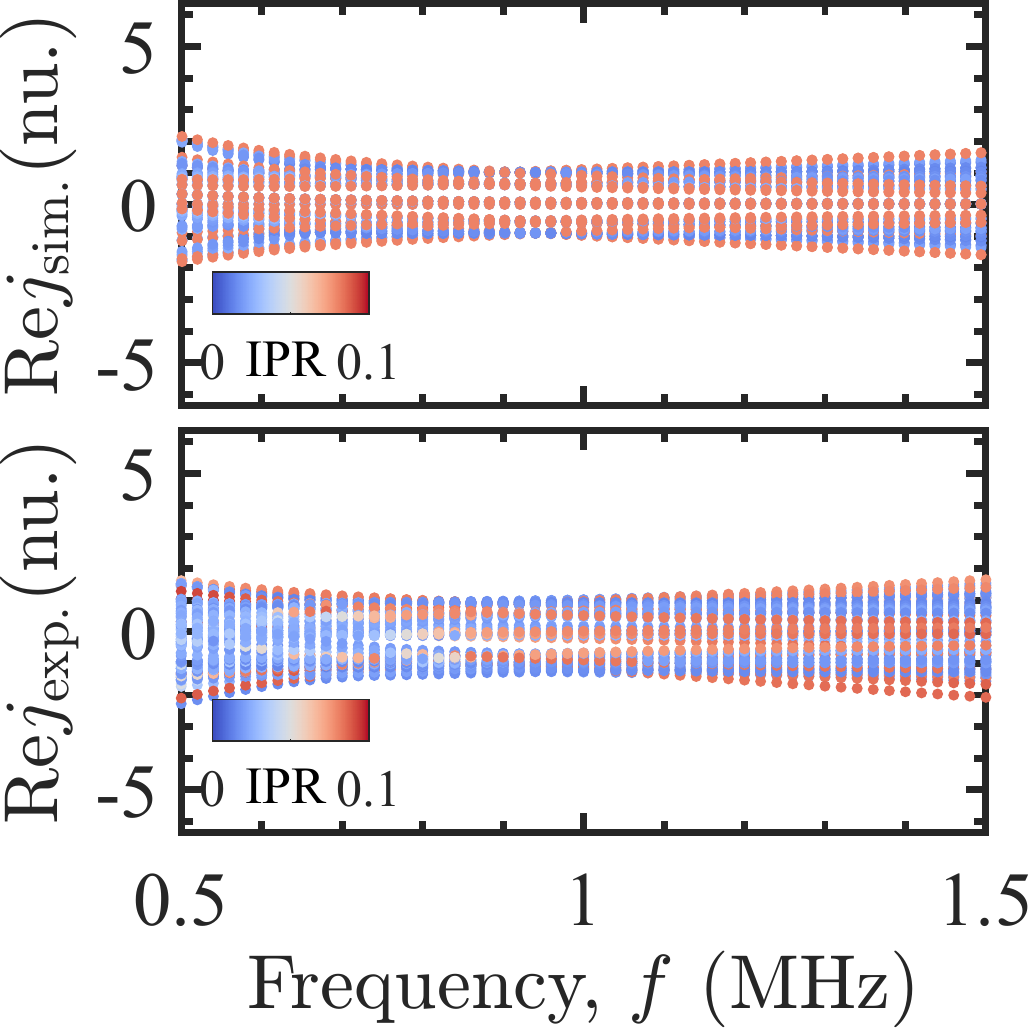}\label{fig:3i}}
\end{minipage}
\begin{minipage}{0.19\textwidth}
\subfigure{\includegraphics[width=.99\textwidth]{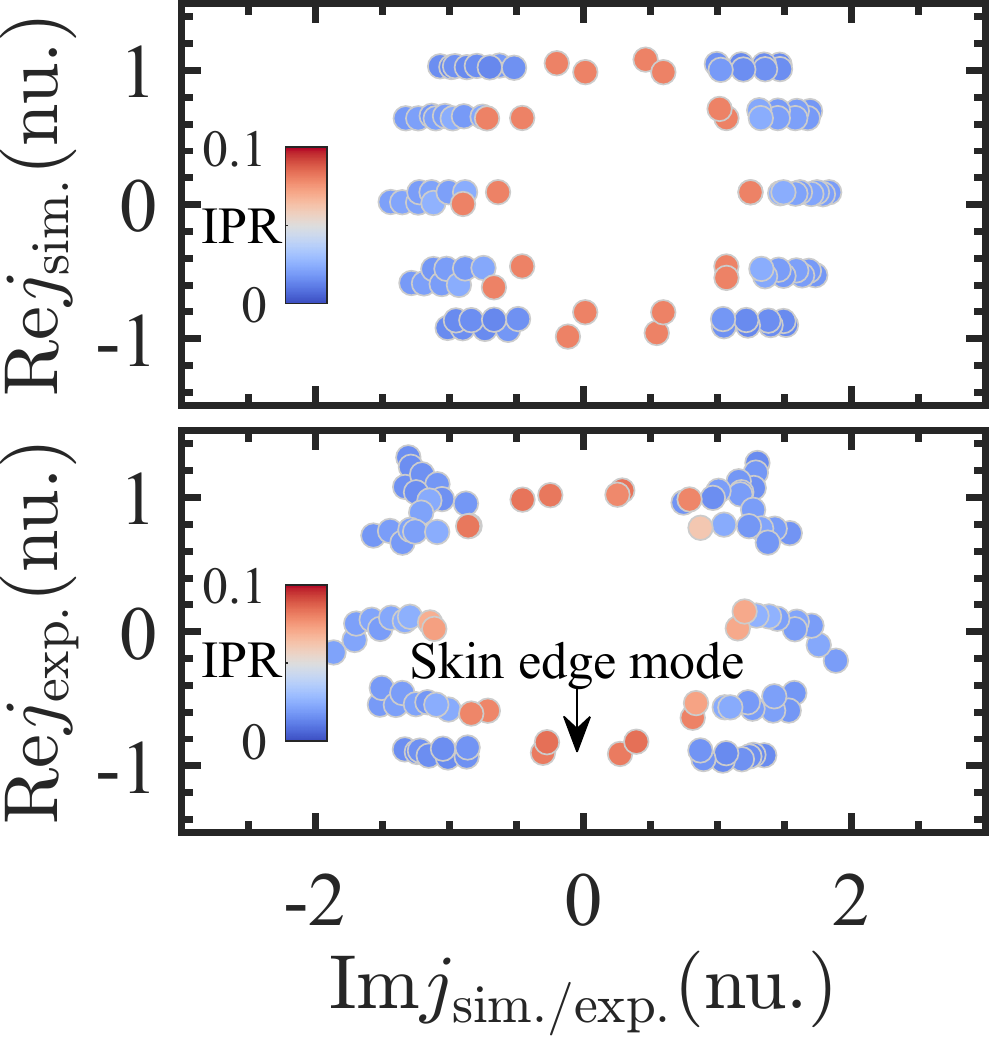}\label{fig:3j}}
\end{minipage}
\begin{minipage}{0.19\textwidth}
\subfigure{\includegraphics[width=.99\textwidth]{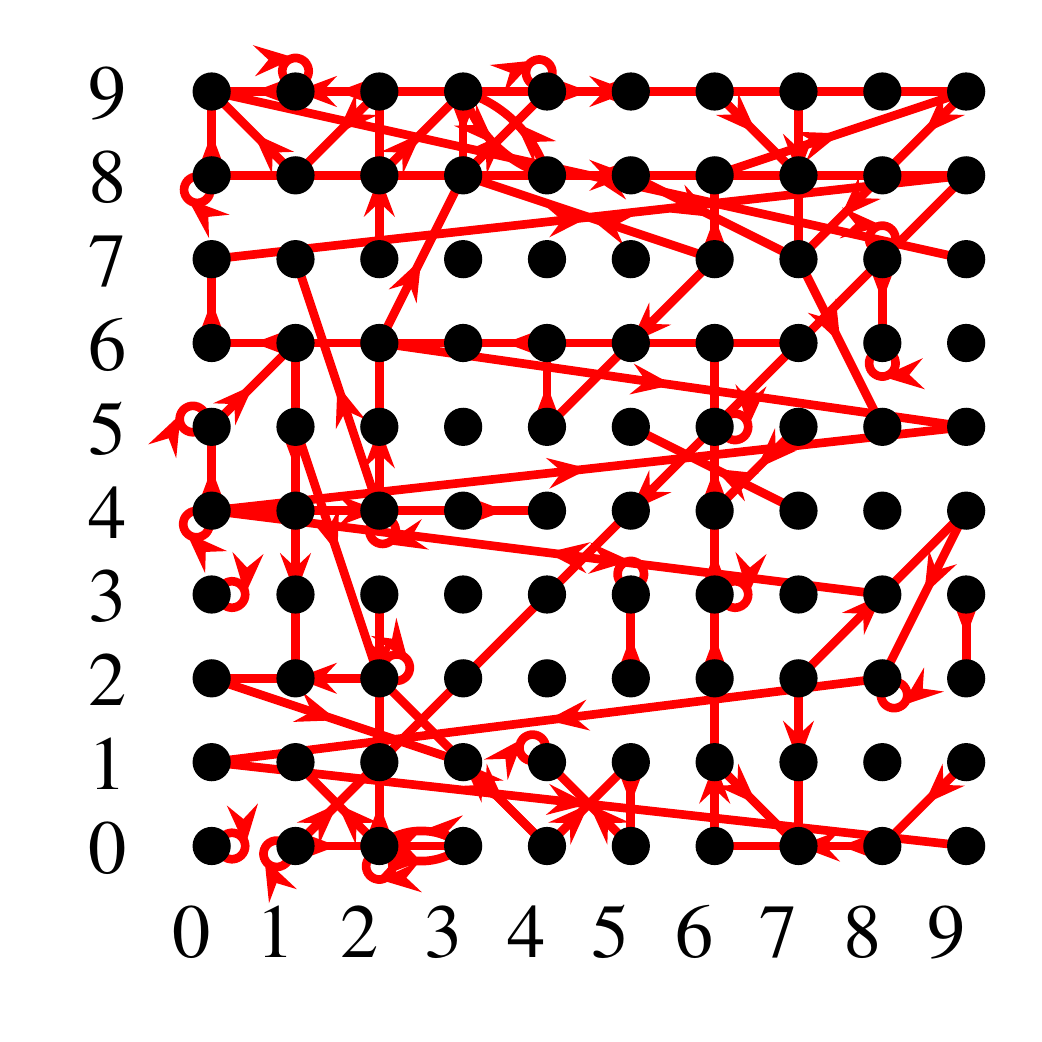}\label{fig:3k}}
\end{minipage}
\begin{minipage}{0.19\textwidth}
\subfigure{\includegraphics[width=.99\textwidth]{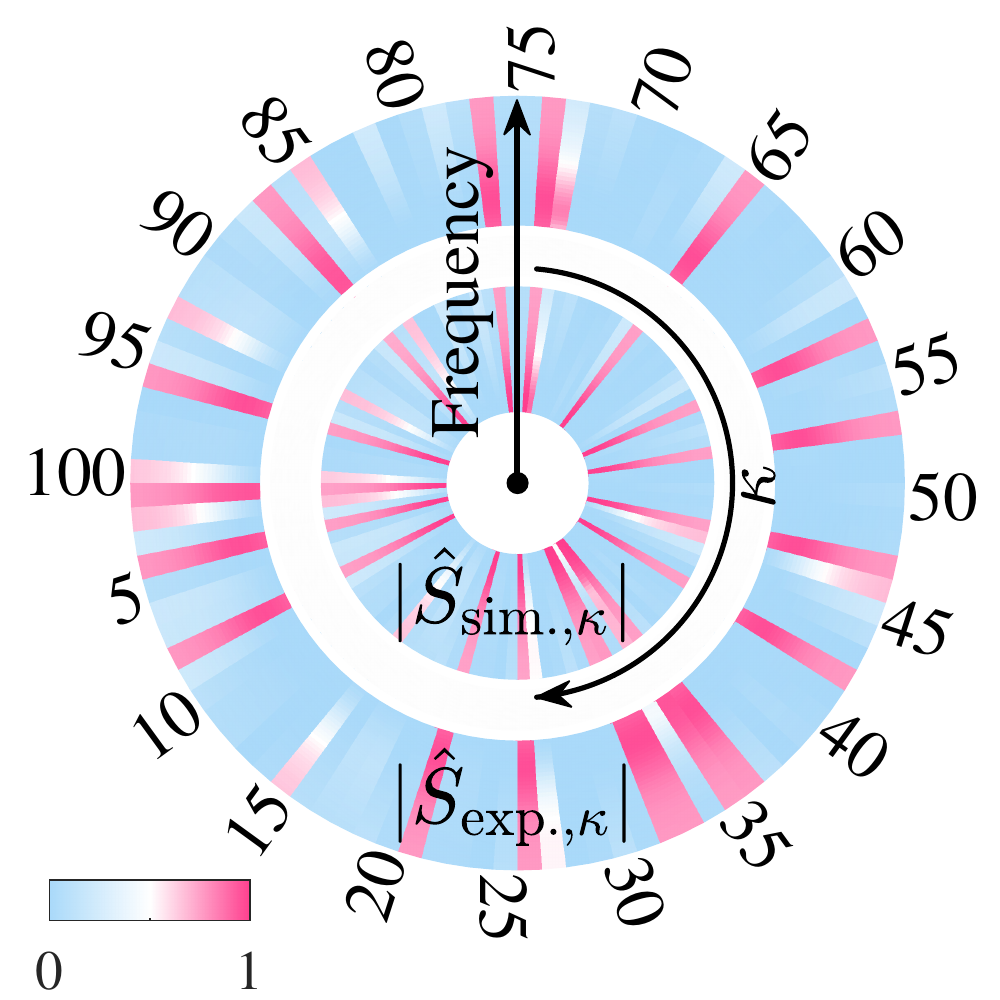}\label{fig:3l}}
\end{minipage}
\begin{minipage}{0.19\textwidth}
\subfigure{\includegraphics[width=.99\textwidth]{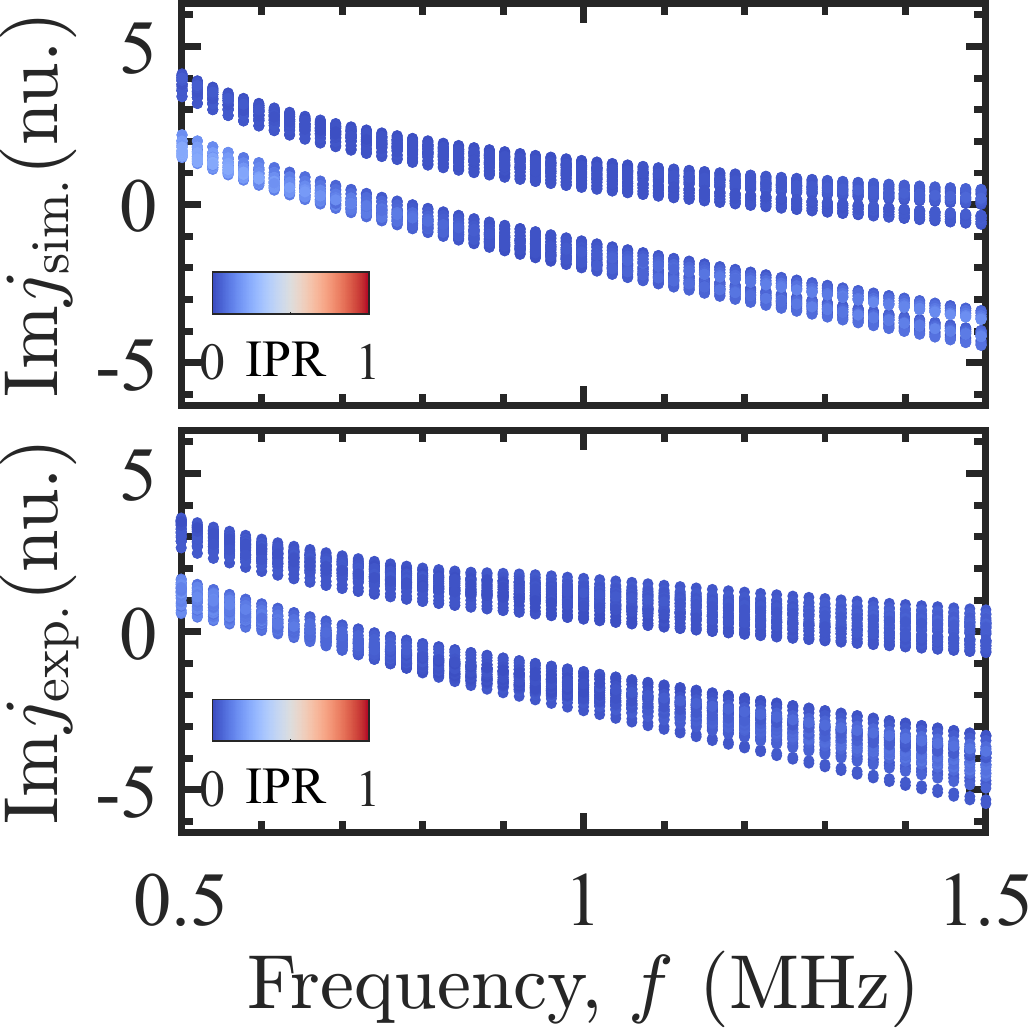}\label{fig:3m}}
\end{minipage}
\begin{minipage}{0.19\textwidth}
\subfigure{\includegraphics[width=.99\textwidth]{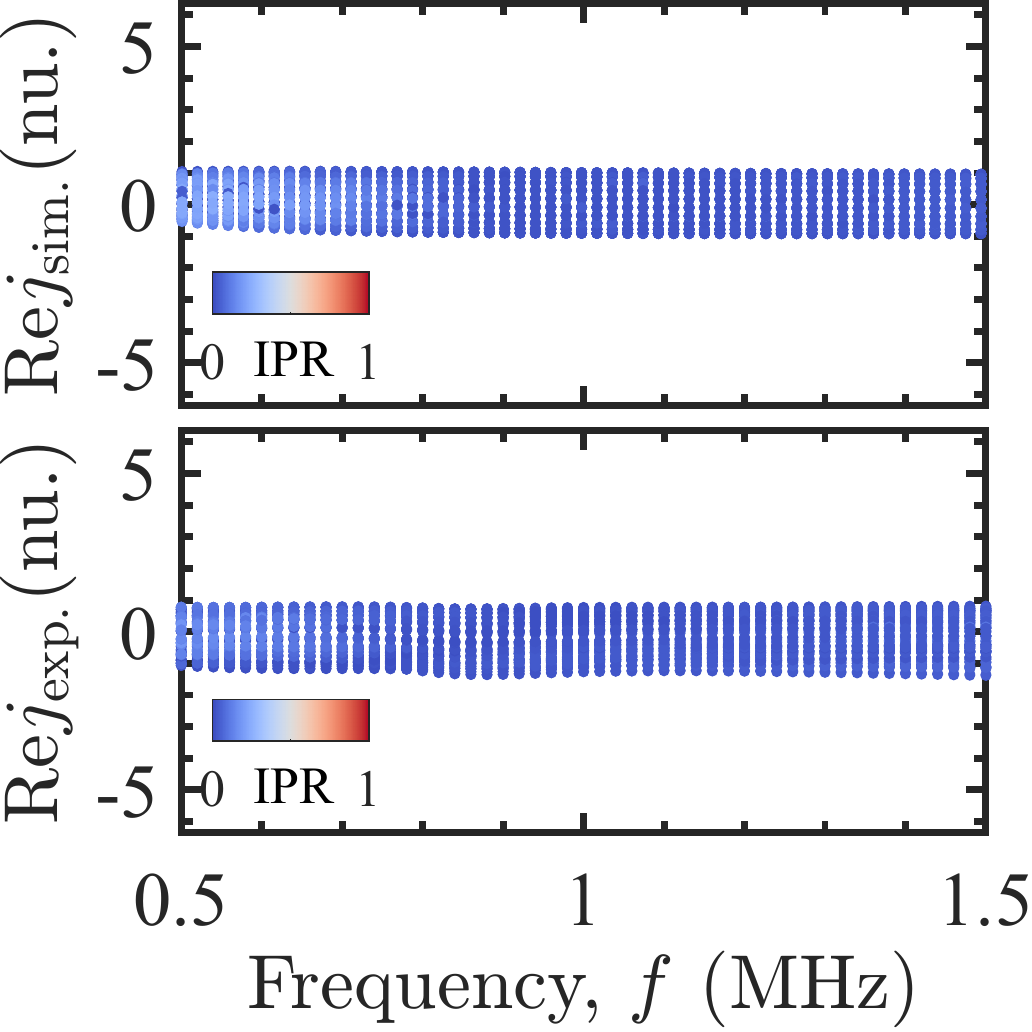}\label{fig:3n}}
\end{minipage}
\begin{minipage}{0.19\textwidth}
\subfigure{\includegraphics[width=.99\textwidth]{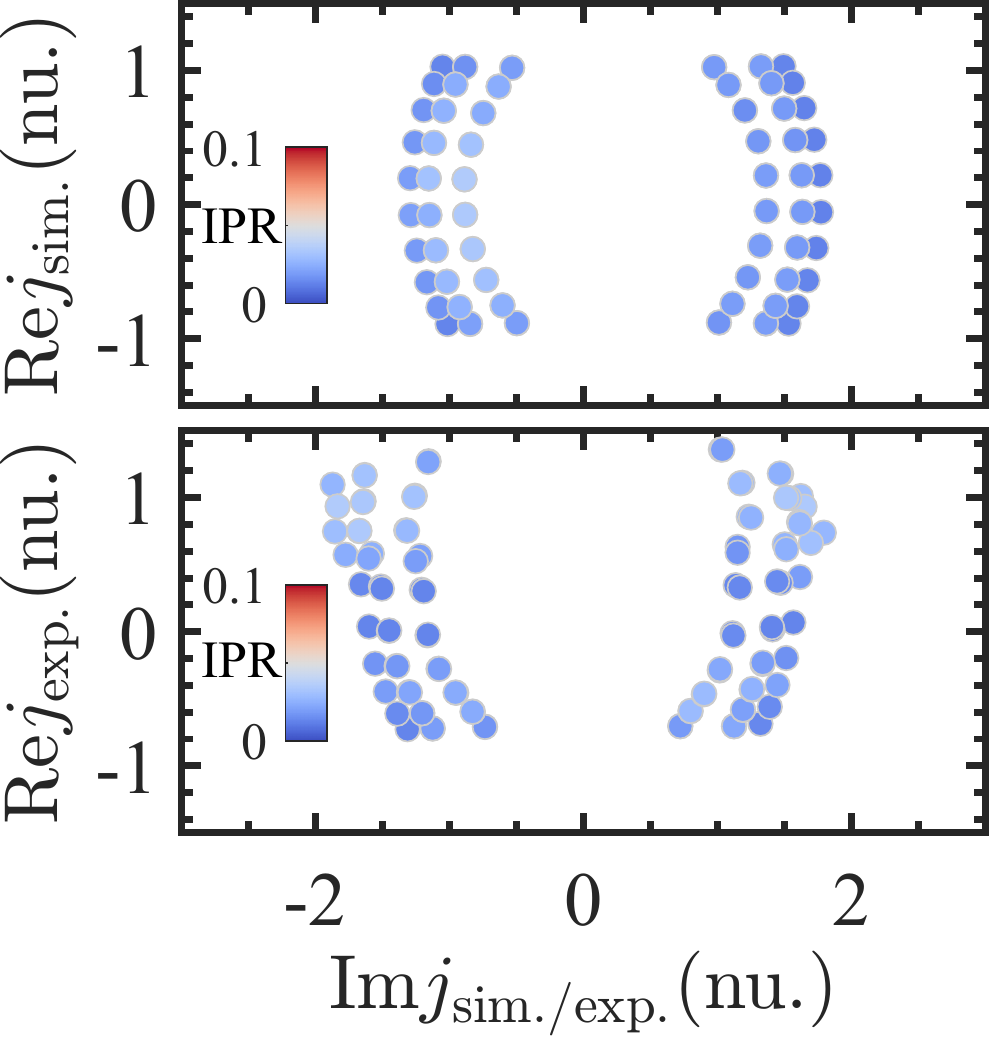}\label{fig:3o}}
\end{minipage}
\begin{minipage}{0.19\textwidth}
\subfigure{\includegraphics[width=.99\textwidth]{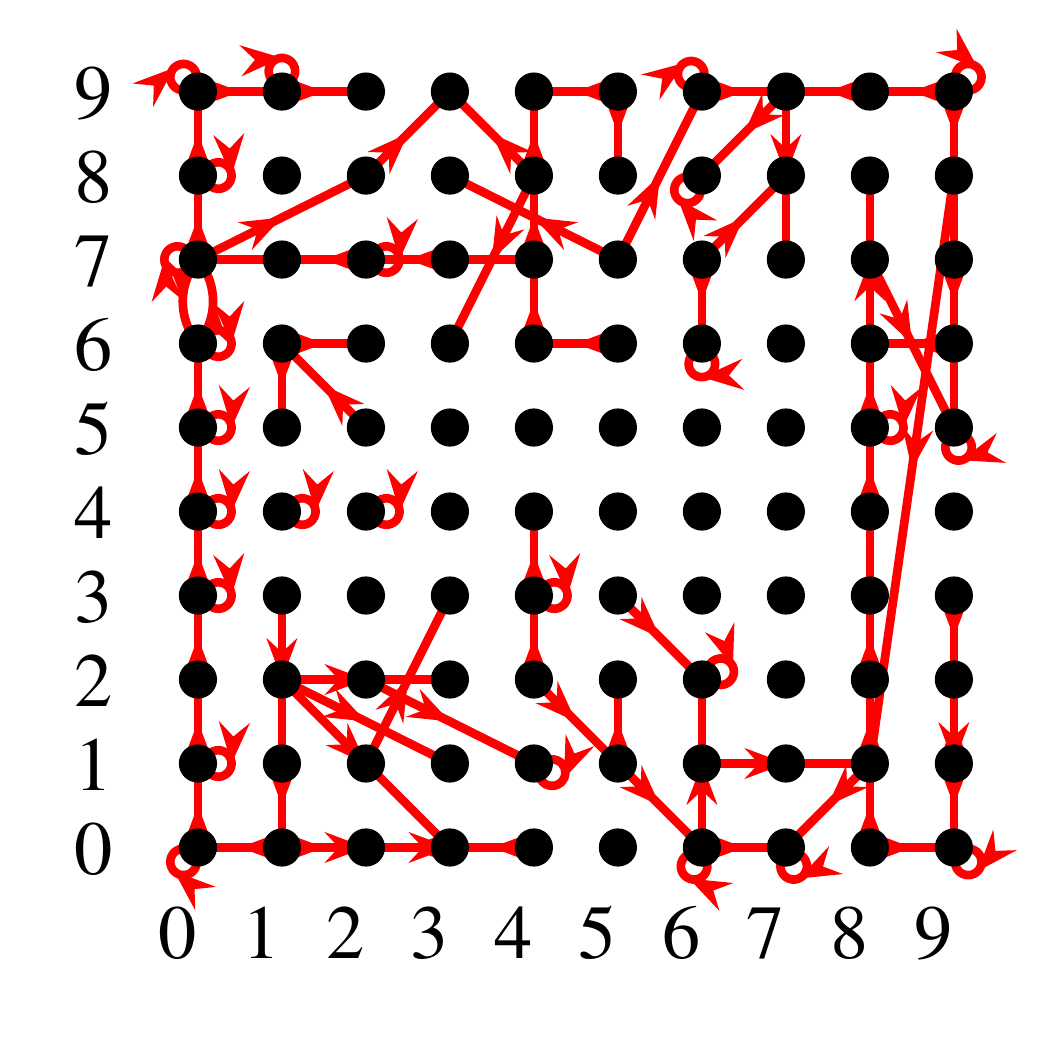}\label{fig:3p}}
\end{minipage}
\begin{minipage}{0.19\textwidth}
\subfigure{\includegraphics[width=.99\textwidth]{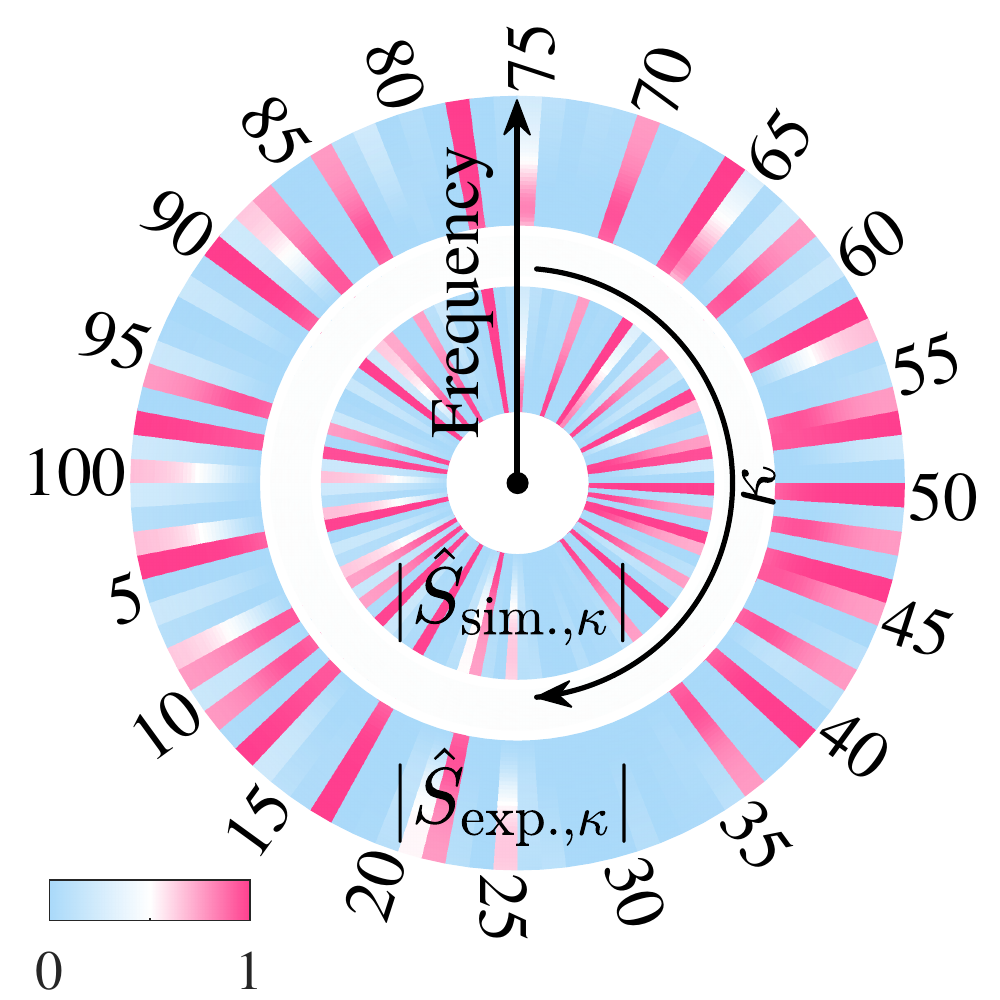}\label{fig:3q}}
\end{minipage}
\begin{minipage}{0.19\textwidth}
\subfigure{\includegraphics[width=.99\textwidth]{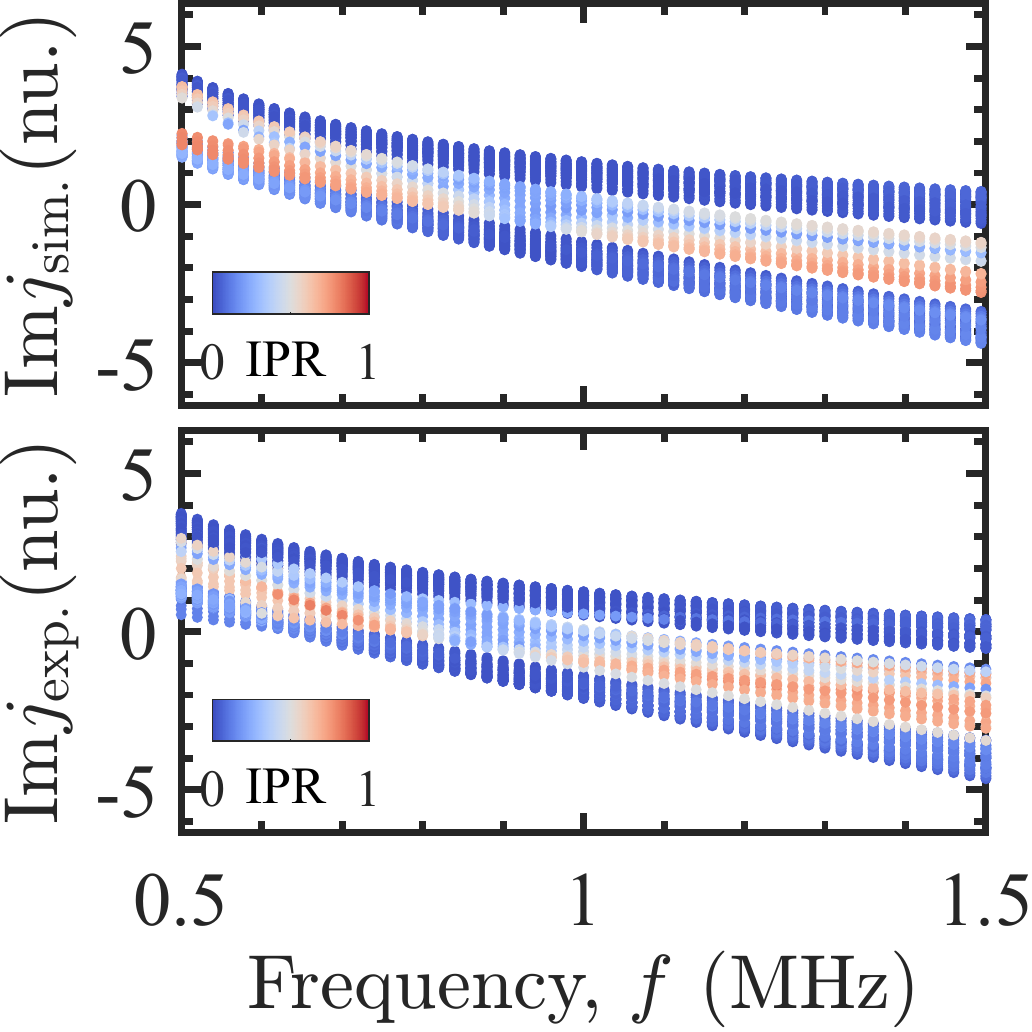}\label{fig:3r}}
\end{minipage}
\begin{minipage}{0.19\textwidth}
\subfigure{\includegraphics[width=.99\textwidth]{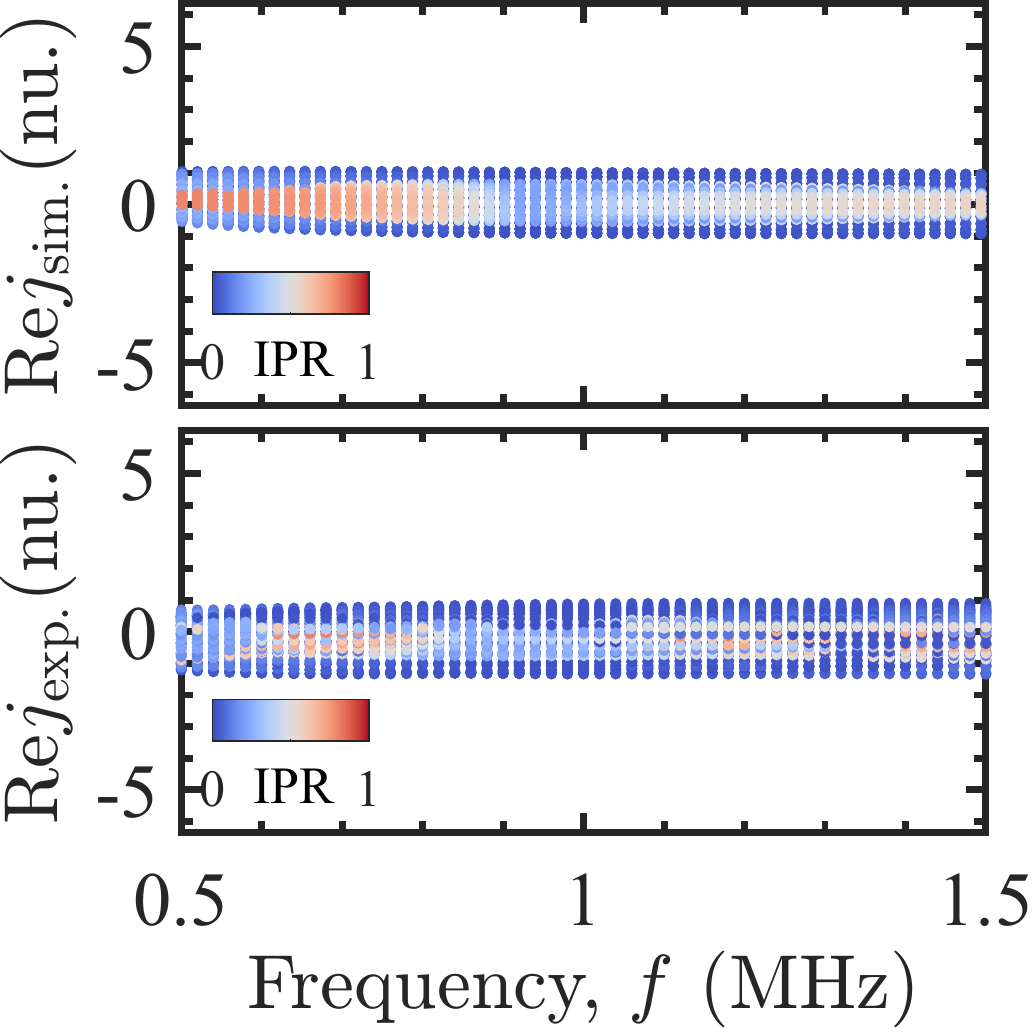}\label{fig:3s}}
\end{minipage}
\begin{minipage}{0.19\textwidth}
\subfigure{\includegraphics[width=.99\textwidth]{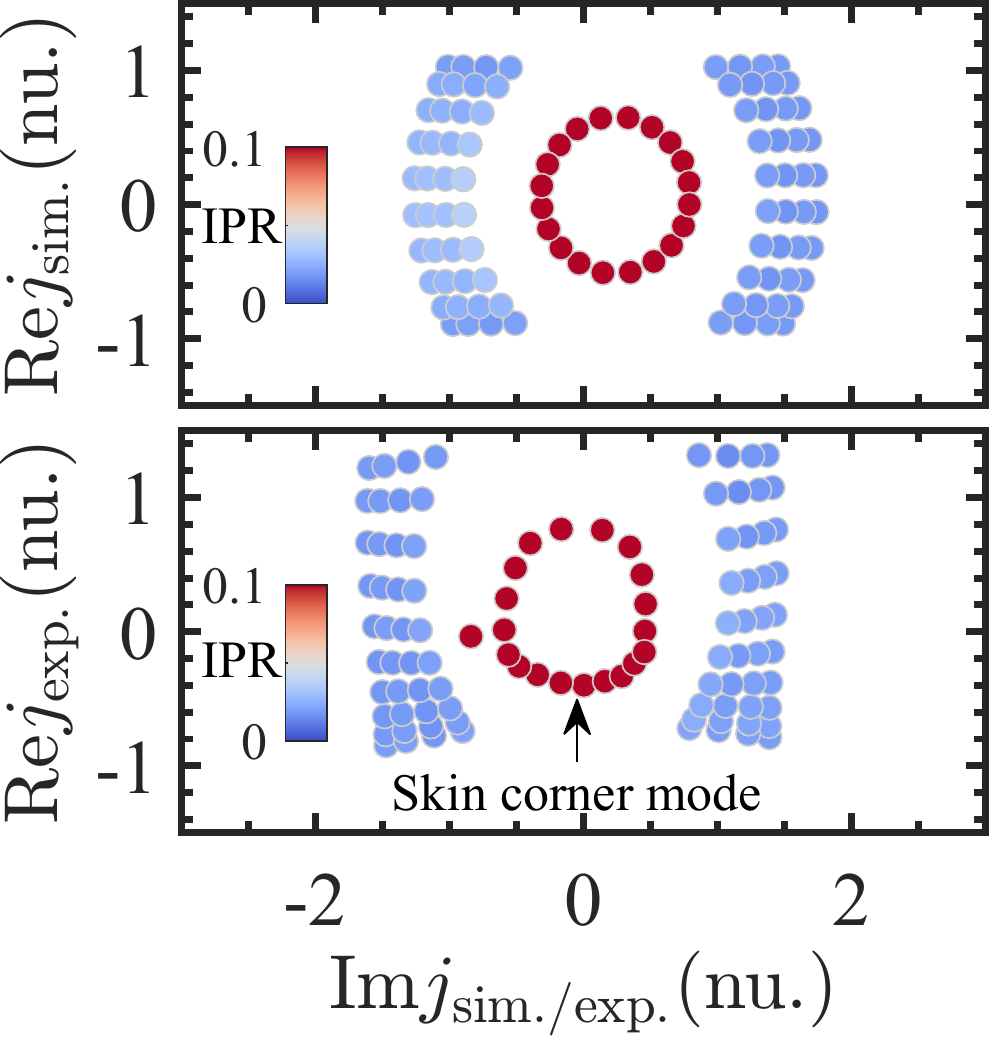}\label{fig:3t}}
\end{minipage}\\
\begin{picture}(0,0)
\put(-230,388){\bf{a}}\put(-140,388){\bf{b}} \put(-50,388){\bf{c}}\put(43,388){\bf{d}}\put(135,388){\bf{e}}
\put(-230,290){\bf{f}}\put(-140,290){\bf{g}} \put(-50,290){\bf{h}}\put(43,290){\bf{i}}\put(135,290){\bf{j}}
\put(-230,194){\bf{k}}\put(-140,194){\bf{l}} \put(-50,194){\bf{m}}\put(43,194){\bf{n}}\put(135,194){\bf{o}}
\put(-230,96){\bf{p}}\put(-140,96){\bf{q}}
\put(-50,96){\bf{r}}\put(43,96){\bf{s}}\put(135,96){\bf{t}}
 \put(-237,326){\rotatebox{90}{\colorbox{lightgray}{\tiny{PBC$x$-PBC$y$}}}}
 \put(-237,228){\rotatebox{90}{\colorbox{lightgray}{\tiny{PBC$x$-OBC$y$}}}}
  \put(-237,130){\rotatebox{90}{\colorbox{lightgray}{\tiny{OBC$x$-PBC$y$}}}}
 \put(-237,32){\rotatebox{90}{\colorbox{lightgray}{\tiny{OBC$x$-OBC$y$}}}}
\end{picture}
\caption{{\bf{Comparison of experimental and simulated results for different boundary conditions.}}  $\bf{a, f , k, p}$, Representative $S$-parameters are measured between ports connected by red directional circles and lines for $\kappa=1,2,\cdots,100$. $\bf{b, g , l, q}$, Frequency response of $|{{\hat{S}}_{\rm{sim.},\kappa}|}$ (inner circle) and $|{{\hat{S}}_{\rm{exp.},\kappa}|}$ (outer circle) for each $\kappa$, showing excellent agreement. $\bf{c, d, h, i, m, n, r, s}$,
Imaginary and real parts of the admittance spectra  $j_{\rm{sim.}}$ (top panel) and $j_{\rm{exp.}}$ (bottom panel) as  functions of the driving frequency $f$, weighted by the IPR.  $\bf{e, j , o, t,}$ Complex admittance spectra  for the resonance frequency $f_0\sim  0.876 \mathop{\rm{MHz}}$. }
\label{fig:3}
\end{figure*}

According to Kirchhoff’s laws, any circuit can be described by the block diagonal admittance matrix (circuit Laplacian) ${\textit{\textbf{J}}}( \omega  ) =  {{\rm{i}}\omega {\textit{\textbf{C}}} + \frac{1}{{{\rm{i}}\omega }}{{\textit{\textbf{W}}}}}$,
where ${\textit{\textbf{C}}}$ and ${\textit{\textbf{W}}}$ are the Laplacian matrices of the capacitance and inverse inductance, respectively. For a given input current of frequency $\omega= 2\pi f$, we obtain the non-reciprocal two-band admittance matrix (see Supplementary Material Sec.\ \textcolor{blue}{S1})
  \begin{equation} \label{eq2}
{\textit{\textbf{J}}}({\textit{\textbf{k}}},\omega ) = {\rm{i}}\omega \left[ {\begin{array}{*{20}{c}}
{\frac{{{L_1}{L_2}}}{{\left( {{L_1} + 2{L_2}} \right){\omega ^2}}} - 2{C_1} - {C_2} + {C_1}{e^{ - {\rm{i}}{k_x}}}}&{{C_2}+{C_1}{e^{ - {\rm{i}}{k_y}}}}\\
{{C_2}+{C_1}{e^{{\rm{i}}{k_y}}}}&{\frac{{{L_1}{L_2}}}{{\left( {{L_1} + {L_2}} \right){\omega ^2}}} - {C_1} - {C_2} - \frac{{{L_1}}}{{{\omega ^2}}}{e^{{\rm{i}}{k_x}}}}
\end{array}} \right],
 \end{equation}
where two pairs of capacitors and inductors, $(C_1, L_1)$ and $(C_2, L_2)$, with the same resonance frequency ${\omega _0} = {1 \mathord{\left/
 {\vphantom {1 {\sqrt {{L_1}{C_1}} }}} \right.
 \kern-\nulldelimiterspace} {\sqrt {{L_1}{C_1}} }} = {1 \mathord{\left/
 {\vphantom {1 {\sqrt {{L_2}{C_2}} }}} \right.
 \kern-\nulldelimiterspace} {\sqrt {{L_2}{C_2}} }}$ are used to couple the nodes. This implies
 \begin{equation}
{\textit{\textbf{J}}}({{\textit{\textbf{k}}}},{\omega _0}) = {\rm{i}}\sqrt {{{{C_1}} \mathord{\left/
 {\vphantom {{{C_1}} {{L_1}}}} \right.
 \kern-\nulldelimiterspace} {{L_1}}}} \left[ - {\rm{i}}{\lambda _x}\sin {k_x}{\sigma _0}+{\lambda _x}\cos {k_x}{\sigma _z} + {\lambda _y}\sin {k_y}{\sigma _y} + \left( {{\gamma _y} + {\lambda _y}\cos {k_y}} \right){\sigma _x}\right].
 \end{equation}
For ${C_1}=1000\mathop{\rm{pH}}$, ${C_2} = 330\mathop{\rm{pH}}$, ${L_1} = 33\mathop\mu{\rm{F}}$, and ${L_2} = 100\mathop\mu{\rm{F}}$, we arrive at $\lambda _x=1$, $\lambda _y=1$, and $\gamma_y=0.33$. The eigenvalues of ${\textit{\textbf{J}}}({{\textit{\textbf{k}}}},{\omega _0})$ are given by
 \begin{equation}
   j({\textit{\textbf{k}}},{\omega _0}) = {\rm{i}}\sqrt {{C_1}/{L_1}} ( \pm \sqrt {{\lambda _x}^2{{\cos }^2}{k_x} + 2{\lambda _y}{\gamma _y}\cos {k_y} + {\lambda _y}^2 + \gamma _y^2} {\rm{ - i}}{\lambda _x}\sin {k_x}).
 \end{equation}

\begin{figure*}[htbp!]
\begin{minipage}{0.49\linewidth}
\subfigure{\includegraphics[height=4cm]{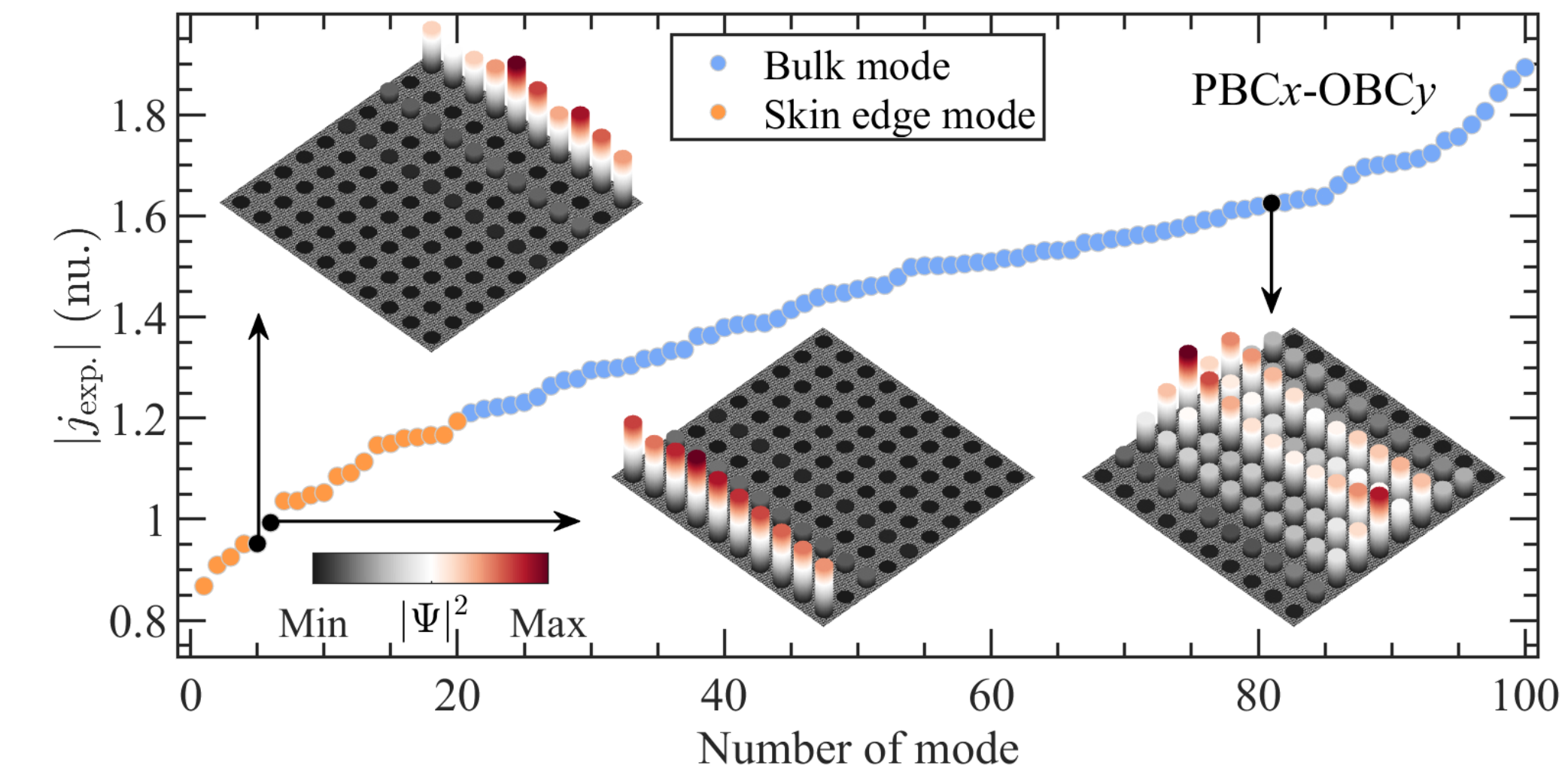}\label{fig:4a}}
\end{minipage}
\begin{minipage}{0.49\linewidth}
\subfigure{\includegraphics[height=4cm]{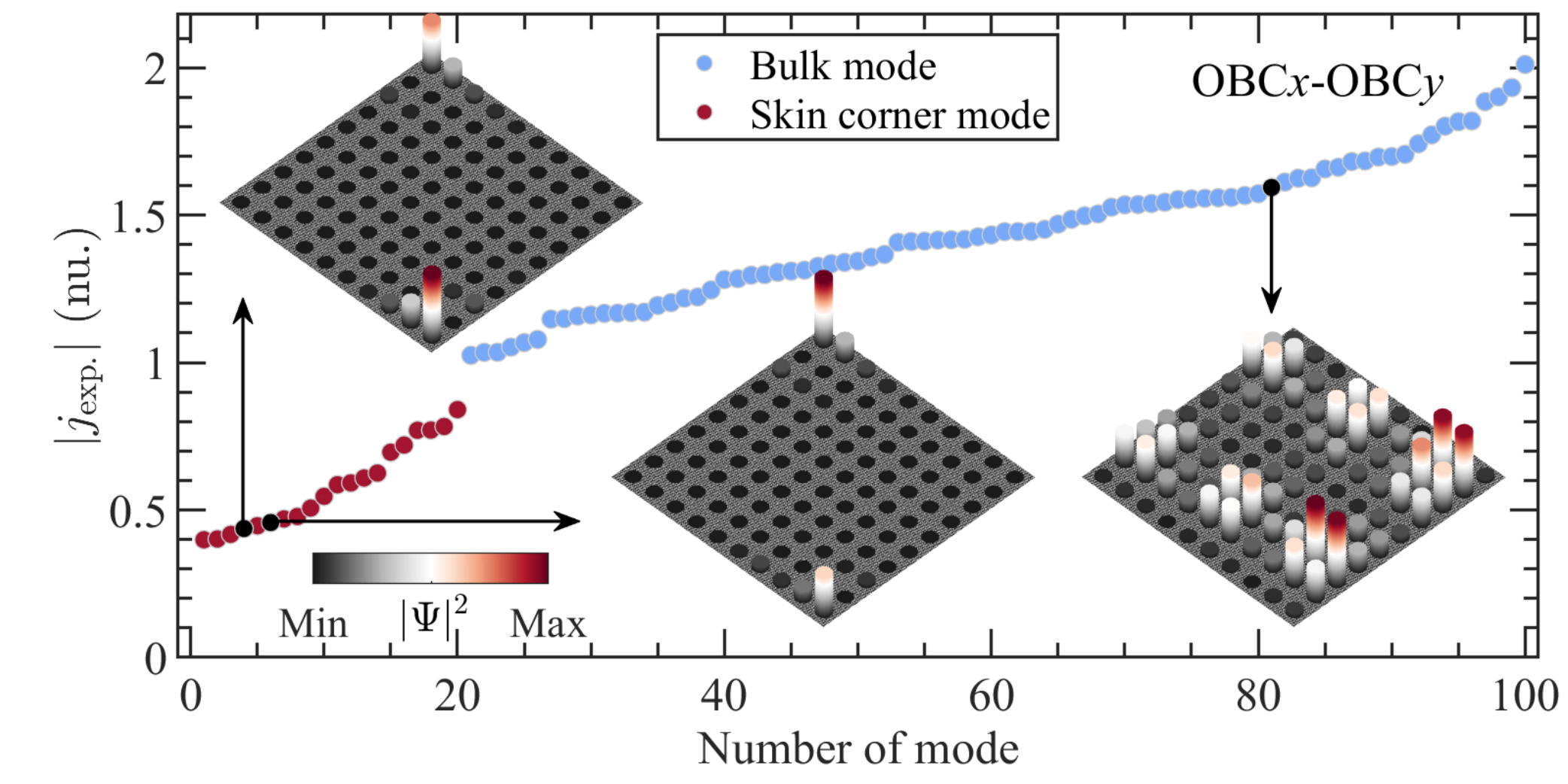}\label{fig:4b}}
\end{minipage}
\begin{picture}(0,0)
 \put(-460,60){\bf{a}}  \put(-230,60){\bf{b}}
\end{picture}
\caption{{\bf{Admittance spectra and mode distributions.}} $\bf{a}$,  Absolute values of $j_{\rm{exp.}}$ for  PBC$x$-OBC$y$. $\bf{b}$, Absolute values of $j_{\rm{exp.}}$ for  OBC$x$-OBC$y$. The insets show the bulk and skin edge/corner modes.}
\label{fig:4}
\end{figure*}
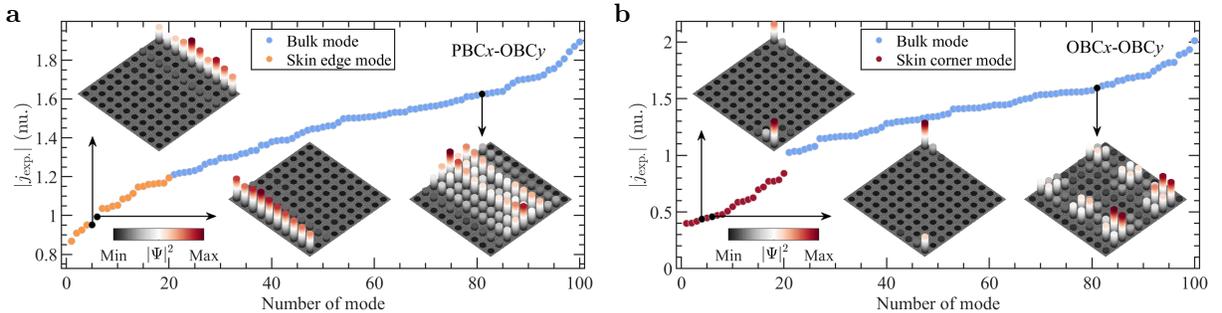

As the boundary connections can be customized, we can observe phase transitions through differences in the spectral flow, enabling the study of the topological modes at any choice of boundary conditions.
The admittance eigenvalues and eigenstates are accessible by an $S$-parameter measurement using the PGIML framework. We address the circuit for PBC$x$-PBC$y$ in  Figs.\ \ref{fig:3a}-\textcolor{blue}{e}, for PBC$x$-OBC$y$ in  Figs.\ \ref{fig:3f}-\textcolor{blue}{j}, for OBC$x$-PBC$y$ in  Figs.\ \ref{fig:3k}-\textcolor{blue}{o}, and for PBC$x$-PBC$y$ in  Figs.\ \ref{fig:3p}-\textcolor{blue}{t}. Figures\ \ref{fig:3a},\textcolor{blue}{f},\textcolor{blue}{k},\textcolor{blue}{p} show the  ports selected for measuring the representative $S$-parameters (see Supplementary Material Sec.\ \textcolor{blue}{S4}). According to Figs. \ref{fig:3b},\textcolor{blue}{g},\textcolor{blue}{l},\textcolor{blue}{q}, the frequency response of  $|{{\hat{S}}_{\rm{exp.},\kappa}|}$ (outer circle) agrees well with that of $|{{\hat{S}}_{\rm{sim.},\kappa}|}$ (inner circle). Figures\ \ref{fig:3c},\textcolor{blue}{h},\textcolor{blue}{m},\textcolor{blue}{r} and \ref{fig:3d},\textcolor{blue}{i},\textcolor{blue}{n},\textcolor{blue}{s} show the imaginary and real parts, respectively, of the simulated (top panel) and  experimental (bottom panel) admittance spectra as functions of the driving frequency $f$, weighted by the inverse participation ratio $\rm{IPR}=\sum\limits_n {{{| {{\Psi_n}} |}^4}} /{( {\sum\limits_n {{{| {{\Psi_n}} |}^2}} } )^2}$, where  $\Psi_n$ is the $n$-th eigenmode. A larger IPR corresponds to a more localized mode. For simplicity, the results are given in normalized units (nu.) as multiplies of $\sqrt {{L_1}/{C_1}} \mathop \Omega^{-1}$. Figures.\ \ref{fig:3e},\textcolor{blue}{j},\textcolor{blue}{o},\textcolor{blue}{t} show the simulated and experimental admittance spectra in the complex plane at the resonance frequency $f_0=\left(2 \pi \sqrt{L_{1} C_{1}}\right)^{-1} \sim 0.876 \mathop {\mathrm{MHz}}$. The system has trivial topology without NHSE for PBC$x$-PBC$y$ and OBC$x$-PBC$y$, and non-trivial topology with NHSE for PBC$x$-OBC$y$ and OBC$x$-OBC$y$. In particular, figure\ \ref{fig:3j} shows skin edge modes for PBC$x$-OBC$y$. We observe in Fig.\ \ref{fig:4a} a localized mode distribution at the left/right boundary, in contrast to the delocalized bulk modes. Remarkably,  the skin corner modes in Fig.\ \ref{fig:3t} form a circle in the complex-energy plane for OBC$x$-OBC$y$, analytically given by $j_{\rm{sim.}} = 0.33\sqrt {{C_1}/{L_1}} {e^{{\rm{i}}\theta }},\theta  \in \left[ {0,2\pi } \right]$ (see Supplementary Material Sec.\ \textcolor{blue}{S2}). They are localized at the corners while the bulk modes are delocalized, as can be seen in Fig.\ \ref{fig:4b}. A non-Bloch 2D winding number $v_{2D}=1$  characterizes the higher-order NHSE (Methods). For all the $L^2$ eigenmodes, the number of corner skin modes is $2L$ while the number of delocalized bulk modes is $L^2-2L$.


\section*{Conclusion}
In times of digital research and measurement, many scientific disciplines produce large amounts of data that by far surpass conventional computational abilities for processing and analyzing.  Hence, we develop the PGIML method by integrating physical principles,  graph visualization of features, and ML to enforce the identification of an unrevealed physical phenomenon. At the example of a topoelectrical circuit, we embed the physical principles of the second-order NHSE into the circuit, observe the skin corner modes, demonstrate the violation of the conventional bulk-boundary correspondence, and reveal an intriguing interplay between higher-order topology and non-Hermiticity. Our results suggest that the PGIML method provides a paradigm shift in processing and analyzing data, opening new avenues to understanding complex systems in higher dimensions.
\section*{Methods}\label{Method}
\subsection*{Topological invariant}
According to point-gap topology \cite{PhysRevX.8.031079,PhysRevX.9.041015,PhysRevLett.124.086801}, we derive a topological characterization
of the NHSE.  A non-Hermitian Hamiltonian $H$ has a point gap at a reference point $E\in  {\mathbb{C}}$ if and only if its complex spectrum does not cross $E$, i.e., $\det(H- E)  \ne 0$. The topological invariant is given by the winding number
\begin{equation}
w(E) = \int_0^{2\pi } {\frac{{dk}}{{2\pi {\rm{i}}}}\frac{d}{{dk}}\log \det [H(k) - E]},
 \end{equation}
where $H(k)$ is the non-Hermitian Bloch Hamiltonian. The second-order NHSE occurs when $w(E) \ne 0$. The non-Hermitian topology of $H(k)$ can also be understood in terms of the extended Hermitian Hamiltonian
\begin{equation}
\tilde H(k,E) = \left( {\begin{array}{*{20}{c}}
0&{H(k) - E}\\
{{H^\dag }(k) - {E^ * }}&0
\end{array}} \right),
 \end{equation}
 which is topologically nontrivial with a finite energy gap if and only if  $H(k)$ is topologically nontrivial with a point gap at $E$.

 To clarify the topological property of the second-order NHSE \cite{Hayashi2018,Hayashi2019,PhysRevB.100.235302}, we define the extended Hermitian admittance Hamiltonian
 \begin{equation} \label{eq7}
\tilde J({{\textit{\textbf{k}}}},{\omega _0}) = \left( {\begin{array}{*{20}{c}}
0&{J({{\textit{\textbf{k}}}},{\omega _0}) - j}\\
{{J^\dag }({{\textit{\textbf{k}}}},{\omega _0}) - {j^*}}&0
\end{array}} \right),
 \end{equation}
and perform the unitary transformation $\tilde{\cal{J}}({{\textit{\textbf{k}}}},{\omega _0}) = \textit{\textbf{U}}\tilde J({{\textit{\textbf{k}}}},{\omega _0}){\textit{\textbf{U}}^\dag }$ using
    \begin{equation}
\textit{\textbf{U}} = \left( {\begin{array}{*{20}{c}}
0&0&0&{ - 1}\\
1&0&0&0\\
0&{ - 1}&0&0\\
0&0&1&0
\end{array}} \right).
 \end{equation}
 We obtain
 \begin{equation}
\tilde{\cal{J}}({{\textit{\textbf{k}}}},{\omega _0})= {\tilde{\cal{J}}_x}({k_x},{\omega _0}) \otimes {\tau _z} + {\sigma _0} \otimes {\tilde{\cal{J}}_y}({k_y},{\omega _0}),
 \end{equation}
 with
   \begin{equation}
\begin{array}{*{20}{c}}
{{\tilde{\cal{J}}_x}({k_x},{\omega _0}) =  - {\rm{i}}\sqrt {{{{C_1}} \mathord{\left/
 {\vphantom {{{C_1}} {{L_1}}}} \right.
 \kern-\nulldelimiterspace} {{L_1}}}} \left[ {{\lambda _x}\cos {k_x}{\sigma _x} + ({\lambda _x}\sin {k_x} - E){\sigma _y}} \right]},\\
{{\tilde{\cal{J}}_y}({k_y},{\omega _0}) =  - {\rm{i}}\sqrt {{{{C_1}} \mathord{\left/
 {\vphantom {{{C_1}} {{L_1}}}} \right.
 \kern-\nulldelimiterspace} {{L_1}}}} \left[ {({\lambda _y}\cos {k_y} + {\gamma _y}){\tau _x} - {\lambda _y}\sin {k_y}{\tau _y}} \right]}.
\end{array}
 \end{equation}
 Both ${\tilde{\cal{J}}_x}({k_x},{\omega _0})$ and ${\tilde{\cal{J}}_y}({k_x},{\omega _0})$ have chiral symmetry corresponding to $\sigma_z$ and $\tau_z$, respectively.  Since chirality and inversion symmetry here commute, the non-Hermitian topology of $\tilde{\cal{J}}({{\textit{\textbf{k}}}},{\omega _0})$ is  characterized by the chiral symmetry ${\cal{C}} = {\sigma _z} \otimes {\tau _z}$.  Thus, the second-order NHSE is characterized by the $\mathbb{Z}$ topological invariant  \cite{PhysRevB.100.235302,PhysRevB.102.241202}
\begin{equation}
{v_{2D}} = {w_x}{w_y},
 \end{equation}
with the winding numbers
\begin{equation}
{w_{\alpha}}(j) = \int_0^{2\pi } {\frac{{d{k_{\alpha}}}}{{2\pi {\rm{i}}}}\frac{d}{{d{k_{\alpha}}}}\log \det [\tilde{\cal{J}}_{\alpha}({k_{\alpha}},{\omega _0}) - j]},
\end{equation}
where $\alpha=x,y$. Thus, $w_x =1$ as $E\in(-\lambda_x,\lambda_x)$ and $w_y =1$ as $\lambda_y/{\gamma_y} >1$. Hence, we obtain a nonzero topological invariant ${v_{2D}} =1$ if and only if
 $E\in(-\lambda_x,\lambda_x)$ and $\lambda_y/{\gamma_y} >1$. ${v_{2D}}$ changes when the edge and bulk modes close the gap, establishing the second-order non-Hermitian topology.

\subsection*{Experiment}
Nonreciprocal couplings are realized by voltage feedback operational amplifiers (Texas Instruments, LM6171), which block the input current while maintaining the output current. To ensure small linewidths of the circuit Laplacian spectra, we use high-Q inductors (Murata, Q-factor $> 40$with $5{\%} $ component variation). Additional elements are added to the  circuit to increase the stability of the voltage feedback operational amplifiers, including a $5\mathop{\Omega} $ resistor connected in series at the output and a $2000\mathop{\Omega}$ resistor in shunt with a  $100 \mathop{\rm{pF}}$ capacitor connecting across the inverting input and output of the voltage feedback operational amplifier. The circuit Laplacian spectra are  obtained by measuring the $S$-parameters of the circuit at $10\mathop {\rm{kHz}}$ frequency resolution. We employ a vector network analyzer (Tektronix TTr500) and transform the $S$-matrix into the circuit Laplacian using 
 the impedance matrix, i.e., the inverse of the circuit Laplacian ${\textit{\textbf{J}}}^{-1} = {Z_0}({\textit{\textbf{S}}}+ {\mathbb{I}}){({\mathbb{I}} - {\textit{\textbf{S}}})^{ - 1}}$, where $\mathbb{I}$ is the identity matrix and $Z_0$ is the characteristic impedance. In an $S$-parameter measurement between two ports, the other ports are connected with  $50\mathop{\Omega}$ load terminators to ensure zero reflection. Note that the impedance matrix obtained by our method is equivalent to that obtained by current probes \cite{PhysRevX.5.021031, Lee2020nodal}, while the measurement is simplified dramatically and  the experimental stability is improved.

\section*{Acknowledgements}
C.S., P.H., X.Z., X.Z., K.N.S., and U.S. acknowledge funding from King Abdullah University of Science and Technology (KAUST). S.L., R.S., and T.J.C. acknowledge the National Key Research and Development Program of China under Grant Nos. 2017YFA0700201, 2017YFA0700202, and 2017YFA0700203. C.H.L. acknowledges the Singapore MOE Tier I grant WBS: R-144-000-435-133. R.T. acknowledges funding from the Deutsche Forschungsgemeinschaft (DFG, German Research Foundation) through Project-ID 258499086-SFB 1170 and the Würzburg-Dresden Cluster of Excellence on Complexity and Topology in Quantum Matter (ct.qmat Project-ID 390858490-EXC 2147). S.Z. acknowledges the Research Grants Council of Hong Kong (AoE/P-701/20 and 17309021).

\section*{Author contributions}
C.S. and S.L. conceived the idea. C.S. performed the theoretical analyses. C.S., S.L., and R.S. designed the circuits and performed the experiments. C.S. and P.H. developed the machine learning method. C.H.L. and R.T. evaluated the experimental and numerical results. A.M., S.Z., T.J.C., and U.S. guided the research. All the authors contributed to the discussions of the results and the preparation of the manuscript.
\section*{Data availability statement}
The datasets generated and analyzed in the current study are available from the corresponding author on reasonable request.
\section*{Competing interests}
The authors declare no competing interests.

\bibliography{apssamp}

\end{document}


\title{Supplementary Material for \\
Experimental identification of the second-order non-Hermitian skin effect with physics-graph-informed machine learning}

\author{Ce Shang}
\thanks{These authors contributed equally}
\affiliation{King Abdullah University of Science and Technology (KAUST), Physical Science and Engineering Division (PSE), Thuwal 23955-6900, Saudi Arabia.}
\author{Shuo Liu}
\thanks{These authors contributed equally}
\affiliation{State Key Laboratory of Millimeter Waves, Southeast University, Nanjing 210096, China.}
\author{Ruiwen Shao}
\thanks{These authors contributed equally}
\affiliation{State Key Laboratory of Millimeter Waves, Southeast University, Nanjing 210096, China.}
\author{Peng Han}
\affiliation{King Abdullah University of Science and Technology (KAUST), Computer, Electrical, and Mathematical Sciences and Engineering Division (CEMSE), Thuwal 23955-6900, Saudi Arabia.}

\author{Xiaoning Zang}
\affiliation{King Abdullah University of Science and Technology (KAUST), Physical Science and Engineering Division (PSE), Thuwal 23955-6900, Saudi Arabia.}

\author{Xiangliang Zhang}
\affiliation{Department of Computer Science and Engineering, University of Notre Dame, Notre Dame, IN 46556, USA}
\affiliation{King Abdullah University of Science and Technology (KAUST), Computer, Electrical, and Mathematical Sciences and Engineering Division (CEMSE), Thuwal 23955-6900, Saudi Arabia.}
\author{Khaled Nabil Salama}
\affiliation{King Abdullah University of Science and Technology (KAUST), Computer, Electrical, and Mathematical Sciences and Engineering Division (CEMSE), Thuwal 23955-6900, Saudi Arabia.}
\author{Wenlong Gao}
\affiliation{Paderborn University, Department of Physics, Warburger Str. 100, 33098
Paderborn, Germany}
\author{Ching Hua Lee}
\affiliation{Department of Physics, National University of Singapore, Singapore 117551, Republic of Singapore}
\author{Ronny Thomale}
\affiliation{Institut f\"{u}r Theoretische Physik und Astrophysik, Universität Würzburg, W\"{u}rzburg, Germany.}

\author{Aur\'elien Manchon}
\email{manchon@cinam.univ-mrs.fr}
\affiliation{CINaM, Aix-Marseille University, CNRS, Marseille, France.}

\author{Shuang Zhang}
\email{shuzhang@hku.hk}
\affiliation{Department of Physics, The University of Hong Kong, Hong Kong, China}

\author{Tie Jun Cui}
\email{tjcui@seu.edu.cn}
\affiliation{State Key Laboratory of Millimeter Waves, Southeast University, Nanjing 210096, China.}

\author{Udo Schwingenschl\"{o}gl}
\email{udo.schwingenschlogl@kaust.edu.sa}
\affiliation{King Abdullah University of Science and Technology (KAUST), Physical Science and Engineering Division (PSE), Thuwal 23955-6900, Saudi Arabia.}




\maketitle
\begin{textblock}{10}[0,0.5](0,2)
{\noindent This document includes:\\
Supplementary Sections S1 to S5\\
Supplementary Figures S1 to S9\\
Supplementary Tables S1 and S2\\
Supplementary Algorithms S1 and S2\\}
\end{textblock}

\newpage

\tableofcontents
\newpage
\section{Non-Hermitian circuit} \label{sec:1}
To realize the non-Hermitian skin effect (NHSE) in topoelectrical circuits, we start with the existing Hermitian formulation for grounded circuit Laplacians \cite{Lee2018} and obtain a general description for non-Hermitian circuits.

\subsection{Circuit Laplacian formulation} \label{CircuitLaplacianformulation}
According to Kirchhoff’s laws, the response of an electrical circuit can be described by the equation of motion 
\begin{equation} \label{s1}
\frac{d}{{dt}}{I_a} = {C_{ab}}\frac{{{d^2}}}{{d{t^2}}}{V_b} + {W_{ab}}{V_b},
\end{equation}
where $I_a$ is the current flowing out of node $a$ and $V_b$ is the electrical potential at node $b$. $C_{ab}$ and $W_{ab}$ are the capacitance and conductance between nodes $a$ and $b$, respectively. When we apply an alternating voltage $V(t)= V(0){e^{{\rm{i}}\omega t}}$ to the circuit, Eq.\ (\ref{s1}) yields
\begin{equation} \label{s2}
{I_a} = \sum\limits_b {({\rm{i}}\omega {C_{ab}} + \frac{1}{{{\rm{i}}\omega }}{W_{ab}})} {V_b} = \sum\limits_b {{J_{ab}}(\omega ){V_b}}.
\end{equation}
In matrix form, we have
\begin{equation} \label{s3}
{\textit{\textbf{J}}}( \omega  ) =  {{\rm{i}}\omega {\textit{\textbf{C}}} + \frac{1}{{{\rm{i}}\omega }}{{\textit{\textbf{W}}}}}, 
\end{equation}
where ${\textit{\textbf{J}}}( \omega  )$ is the circuit Laplacian and $\omega$ is the frequency. ${\textit{\textbf{C}}}$ and ${\textit{\textbf{W}}}$ are the Laplacian matrices of the capacitance and inverse inductance, respectively. The diagonal and off-diagonal elements represent the self-admittance via a certain node and mutual admittance between two nodes, respectively. We use $\omega$ for steady-state analysis of the circuit and obtain an adiabatic continuum of  spectra ${j}(\omega)$ corresponding to ${\textit{\textbf{J}}}(\omega)$.  As the capacitance and inductance explicitly depend on it, the frequency $\omega$ of the driving voltage  is a central tuning parameter of topolectrical circuits.

Due to the translational symmetry and Bloch's theorem,  a circuit can be represented by periodically repeated cells labelled by the coordinate vector ${\textit{\textbf{r}}}$ with the additional sublattice degree of freedom $\alpha \in {1,2,\dots,M}$. Discrete translational invariance leads to a Laplacian  ${J_{( {{\textit{\textbf{r}}},\alpha } ),( {{\textit{\textbf{r}}}',\alpha' } )}} = {J_{\alpha ,\alpha' }}( {{\textit{\textbf{r}}} - {\textit{\textbf{r}}}'} )$ that only depends on the  distance between the cells. Thus, the Bloch Hamiltonian of the unit cell is diagonalized by the spatial Fourier transformation 
\begin{equation}
\textit{\textbf{J}}({\textit{{\textit{\textbf{k}}}}},\omega)=\sum_{{\textit{\textbf{r}}}} e^{-{\rm{i}} {\textit{{\textit{\textbf{k}}}}} \cdot  {\textit{\textbf{r}}}} \textit{\textbf{J}}({\textit{\textbf{r}}},\omega),
\end{equation}
resulting in 
\begin{equation}
   \textit{\textbf{I}}({\textit{\textbf{k}}},\omega ) = \textit{\textbf{J}}({\textit{\textbf{k}}},\omega )\textit{\textbf{V}}({\textit{\textbf{k}}},\omega ).
\end{equation}
 By diagonalizing the $N\times N$ matrix  ${\textit{\textbf{J}}}({\textit{\textbf{k}}},\omega)$, we obtain $N$ eigenvalues, which constitute the band structure $j({\textit{\textbf{k}}},\omega)$ of the complex-valued admittance, parametrically depending on ${\textit{\textbf{k}}}$ and $\omega$.
\newline
\newline
\subsection{Impedance, admittance, and scattering matrices}\label{spa}

A circuit is most commonly studied through measurement of the two-point impedance    \cite{Imhof2018}
\begin{equation} \label{sn}
{Z_{ab}}= \frac{{{V_a} - {V_b}}}{{{I_{ab}}}}, 
\end{equation}
where $I_{ab}$ is the current between nodes $a$ and $b$. We can use the impedance or admittance matrices to relate the nodes to each other and arrive at a matrix description of the circuit \cite{montgomery1987principles,pozar2011microwave}. We begin by considering an arbitrary $N$-node circuit with the admittance matrix ${\textit{\textbf{J}}}$ defined by
\begin{equation}
\left[ {\begin{array}{*{20}{c}}
{{I_1}}\\
{{I_2}}\\
 \vdots \\
{{I_N}}
\end{array}} \right] = \left[ {\begin{array}{*{20}{c}}
{{J_{11}}}&{{J_{12}}}& \dots &{{J_{1N}}}\\
{{J_{21}}}&{{J_{22}}}& \dots &{{J_{2N}}}\\
 \vdots & \vdots & \ddots & \vdots \\
{{J_{N1}}}&{{J_{N2}}}& \dots &{{J_{NN}}}
\end{array}} \right]\left[ {\begin{array}{*{20}{c}}
{{V_1}}\\
{{V_2}}\\
 \vdots \\
{{V_N}}
\end{array}} \right],
\end{equation}
or in matrix form 
\begin{equation}
 {\textit{\textbf{I}}} = {\textit{\textbf{J}}}{\textit{\textbf{V}}}.  
\end{equation}
The impedance matrix ${\textit{\textbf{Z}}}$ is defined by
\begin{equation}
\left[ {\begin{array}{*{20}{c}}
{{V_1}}\\
{{V_2}}\\
 \vdots \\
{{V_N}}
\end{array}} \right] = \left[ {\begin{array}{*{20}{c}}
{{Z_{11}}}&{{Z_{12}}}& \dots &{{Z_{1N}}}\\
{{Z_{21}}}&{{Z_{22}}}& \dots &{{Z_{2N}}}\\
 \vdots & \vdots & \ddots & \vdots \\
{{Z_{N1}}}&{{Z_{N2}}}& \dots &{{Z_{NN}}}
\end{array}} \right]\left[ {\begin{array}{*{20}{c}}
{{I_1}}\\
{{I_2}}\\
 \vdots \\
{{I_N}}
\end{array}} \right],
\end{equation}
or in matrix form
\begin{equation}\label{eq10}
 {{\textit{\textbf{V}}}={\textit{\textbf{Z}}}{\textit{\textbf{I}}}}. 
\end{equation}
At the $a$-th node, the voltages $(V_a^+,V_a^-)$ and currents $(I_a^+,I_a^-)$ of the incident $(+)$ and reflected $(-)$ waves are given by
\begin{subequations}\label{c68}
    \begin{align}
{V_a} &= V_a^ +  + V_a^ -, \label{c68a}\\
{I_a} & = I_a^ +  - I_a^ -  = \frac{{V_a^ +  - V_a^ - }}{{{{Z_0}_a}}},\label{c68b}
    \end{align}
\end{subequations}
where ${{Z_0}_a}$ is the characteristic impedance of node $a$.  

The scattering matrix ${\textit{\textbf{S}}}$ ($S$-matrix) is defined by
\begin{equation}\label{c69}
\left[ {\begin{array}{*{20}{c}}
{V_1^ - }\\
{V_2^ - }\\
 \vdots \\
{V_N^ - }
\end{array}} \right] = \left[ {\begin{array}{*{20}{c}}
{{S_{11}}}&{{S_{12}}}& \dots &{{S_{1N}}}\\
{{S_{21}}}&{{S_{22}}}& \dots &{{S_{2N}}}\\
 \vdots & \vdots & \ddots & \vdots \\
{{S_{N1}}}&{{S_{N2}}}& \dots &{{S_{NN}}}
\end{array}} \right]\left[ {\begin{array}{*{20}{c}}
{V_1^ + }\\
{V_2^ + }\\
 \vdots \\
{V_N^ + }
\end{array}} \right],
\end{equation}
or in matrix form
\begin{equation}\label{equation13}
    {{{{\textit{\textbf{V}}}^ {-} }}  =  {\textit{\textbf{S}}}  {{{\textit{\textbf{V}}}^ + }} }.
\end{equation}
The elements of the $S$-matrix ($S$-parameters),
\begin{equation}
S_{a b}=\left.\frac{V_{a}^{-}}{V_{b}^{+}}\right|_{V_{c}^+ =  0 {\text { for }} c \neq b},
\end{equation}
are found by applying an incident wave of voltage $V^+_b$ to node $b$ and measuring the voltage of the reflected wave $V^-_a$ at node $a$. The incident waves are set to zero at all the other nodes, which thus must be terminated with matched loads to avoid reflections. Consequently, $S_{aa}$ is the reflection coefficient of node $a$ and $S_{a b}$ is the transmission coefficient from node $a$ to node $b$.

We assume that the characteristic impedances of
all the nodes are identical,
\begin{equation}
{{{Z_0}_1} = {{Z_0}_2}= \dots = {{Z_0}_N} = {Z_0}}.
\end{equation}
 Equations (\ref{eq10}), (\ref{c68a}), and (\ref{c68b}) imply
\begin{equation}
{\textit{\textbf{V}}}= {{\textit{\textbf{V}}}^ + } + {{\textit{\textbf{V}}}^ - } = {\textit{\textbf{Z}}}{\textit{\textbf{I}}} = {\textit{\textbf{Z}}}( {{{\textit{\textbf{I}}}^ + } - {{\textit{\textbf{I}}}^ - }} ) = {\textit{\textbf{Z}}}\frac{{{{\textit{\textbf{V}}}^ + } - {{\textit{\textbf{V}}}^ - }}}{{{Z_0}}},
\end{equation}
which can be rewritten as
\begin{equation}
(\mathbb{I} + \frac{1}{{{Z_0}}}{\textit{\textbf{Z}}} ){{\textit{\textbf{V}}}^ - } = (\frac{1}{{{Z_0}}}{\textit{\textbf{Z}}} - {\mathbb{I}}){{\textit{\textbf{V}}}^ + },  
\end{equation}
where $\mathbb{I}$ is the identity matrix. Using Eq.\ (\ref{equation13}), we obtain
\begin{equation}
{\textit{\textbf{S}}}={( {{\textit{\textbf{Z}}} + {Z_0}{\mathbb{I}}} )^{ - 1}}( {{\textit{\textbf{Z}}}- {Z_0}{\mathbb{I}}} )
\end{equation}
and
\begin{equation}
{\textit{\textbf{Z}}} = {Z_0}({\textit{\textbf{S}}}+ {\mathbb{I}}){({\mathbb{I}} - {\textit{\textbf{S}}})^{ - 1}}.
\end{equation}
 The impedance matrix, admittance matrix, and $S$-matrix each provide a complete description of the circuit. While the impedance and admittance matrices relate the total voltages to the total currents at the nodes, the $S$-matrix relates the voltages of the waves incident to the nodes to those of the waves reflected from the nodes. The $S$-parameters can be measured directly by a vector network analyzer. Once they are known, conversion to other matrix parameters can be performed.

\subsection{Non-Hermitian circuit}\label{sec:1D}

We design a circuit without Hermiticity and reciprocity to study the second-order NHSE. The unit cell with nodes A and B is schematically depicted in Fig.\ \textcolor{blue}{2d} (main text) and the corresponding circuit board is shown in Fig.\ \textcolor{blue}{2b} (main text). The circuit contains intracell connections of capacitance $C_2$ in the $y$-direction, intercell connections of capacitance $C_1$ in the $y$-direction,   intercell non-reciprocal connections of capacitance $C_1$ connected to  a voltage follower in the $x$-direction, and intercell non-reciprocal connections of inductance $L_1$ reversely connected to a voltage follower in the $x$-direction. The currents out of the nodes relate to the voltages at the nodes as
\begin{equation}
 \begin{split}
&{I_{\rm{A}}}( {\textit{\textbf{r}}} ) = {\rm{i}}\omega \left\{ {{C_1}\left[ {{V_{\rm{B}}}( {\textit{\textbf{r}}} ) - {V_{\rm{A}}}({\textit{\textbf{r}}} )} \right] + {C_2}\left[ {{V_{\rm{B}}}( {{\textit{\textbf{r}}} + {\textit{\textbf{y}}}} ) - {V_{\rm{A}}}( {\textit{\textbf{r}}} )} \right] + {C_1}\left[ {{V_{\rm{A}}}( {{\textit{\textbf{r}}} + {\textit{\textbf{x}}}} ) - {V_{\rm{A}}}( {\textit{\textbf{r}}} )} \right] - \frac{{  {V_{\rm{A}}}( {\textit{\textbf{r}}} )}}{{{\omega ^2}{G_{\rm{A}}}}}} \right\},\\
&{I_{\rm{B}}}( {\textit{\textbf{r}}} )= {\rm{i}}\omega \left\{ {{C_1}\left[ {{V_{\rm{A}}}( {\textit{\textbf{r}}} ) - {V_{\rm{B}}}( {\textit{\textbf{r}}} )} \right] + {C_2}\left[ {{V_{\rm{A}}}( {{\textit{\textbf{r}}} - {\textit{\textbf{y}}}} ) - {V_{\rm{B}}}( {\textit{\textbf{r}}} )} \right] - \frac{{{V_{\rm{B}}}( {{\textit{\textbf{r}}} - {\textit{\textbf{x}}}} ) - {V_{\rm{B}}}( {\textit{\textbf{r}}})}}{{{\omega ^2}{L_1}}}  - \frac{{ {V_{\rm{B}}}( {\textit{\textbf{r}}} )}}{{{\omega ^2}{G_{\rm{B}}}}}} \right\},
\end{split}
\end{equation}
with the position vector ${\textit{\textbf{r}}}=(m,n)$, lattice vectors ${\textit{\textbf{x}}}=(1,0)$ and ${\textit{\textbf{y}}}=(0,1)$, and grounding components $G_{\rm{A}}={{2{L_1}{L_2}} \mathord{\left/
 {\vphantom {{2{L_1}{L_2}} {(2{L_1} + {L_2})}}} \right.
 \kern-\nulldelimiterspace} {(2{L_1} + {L_2})}}$ and $G_{\rm{B}}=L_2$.  We obtain
\begin{equation}
    \left( {\begin{array}{*{20}{c}}
{{I_{\rm{A}}}}\\
{{I_{\rm{B}}}}
\end{array}} \right) = {\textit{\textbf{J}}}( {\textit{\textbf{r}}},\omega ) \left( {\begin{array}{*{20}{c}}
{{V_{\rm{A}}}}\\
{{V_{\rm{B}}}}
\end{array}} \right)
\end{equation}
with
\begin{equation} \label{eqf27}
{\textit{\textbf{J}}}( {\textit{\textbf{r}}},\omega ) =  {\rm{i}}\omega \left[ \begin{aligned}
{\left( - 2{C_1} - {C_2} + \frac{2}{{{\omega ^2}{L_1}}} + \frac{1}{{{\omega ^2}{L_2}}}\right){\delta _{{\textit{\textbf{r}}},(0,0)}} + {C_1}{\delta _{{\textit{\textbf{r}}},(1,0)}}}  \qquad \qquad  {{C_2}{\delta _{{\textit{\textbf{r}}},(0,0)}} + {C_1}{\delta _{{\textit{\textbf{r}}},(0,1)}}}\\
{{C_2}{\delta _{{\textit{\textbf{r}}},(0,0)}} + {C_1}{\delta _{{\textit{\textbf{r}}},(0, - 1)}}}  {\qquad \qquad } \left( - {C_1} - {C_2} + \frac{1}{{{\omega ^2}{L_1}}} + \frac{1}{{{\omega ^2}{L_2}}}\right){\delta _{{\textit{\textbf{r}}},(0,0)}} - \frac{{{\delta _{{\textit{\textbf{r}}},( - 1,0)}}}}{{{\omega ^2}{L_1}}}
\end{aligned} \right],
\end{equation}
where $\delta_{{\textit{\textbf{r}}},{\textit{\textbf{r}}}'}$ is the two-dimensional (2D) discrete delta function. We arrive at the non-reciprocal two-band admittance matrix
  \begin{equation} \label{eq2}
{\textit{\textbf{J}}}({\textit{\textbf{k}}},\omega ) = {\rm{i}}\omega \left[ {\begin{aligned}
{\frac{{{L_1}{L_2}}}{{( {{L_1} + 2{L_2}} ){\omega ^2}}} - 2{C_1} - {C_2} + {C_1}{e^{ - {\rm{i}}{k_x}}}}&{\qquad \qquad \qquad {{C_2}+{C_1}{e^{ - {\rm{i}}{k_y}}}}}\\
{{{C_2}+{C_1}{e^{{\rm{i}}{k_y}}}}\qquad \qquad \qquad}&{\frac{{{L_1}{L_2}}}{{( {{L_1} + {L_2}} ){\omega ^2}}} - {C_1} - {C_2} - \frac{{{L_1}}}{{{\omega ^2}}}{e^{{\rm{i}}{k_x}}}}
\end{aligned}} \right],
 \end{equation}
where two pairs of capacitors and inductors, $(C_1, L_1)$ and $(C_2, L_2)$, with the same resonance frequency ${\omega _0} = {1 \mathord{\left/
 {\vphantom {1 {\sqrt {{L_1}{C_1}} }}} \right.
 \kern-\nulldelimiterspace} {\sqrt {{L_1}{C_1}} }} = {1 \mathord{\left/
 {\vphantom {1 {\sqrt {{L_2}{C_2}} }}} \right.
 \kern-\nulldelimiterspace} {\sqrt {{L_2}{C_2}} }}$ are used to couple the nodes. Equation (\ref{eq2}) implies
 \begin{equation}\label{s10}
 \begin{split}
{\textit{\textbf{J}}}({{\textit{\textbf{k}}}},{\omega _0}) &= {\rm{i}}\sqrt {{{{C_1}} \mathord{\left/
 {\vphantom {{{C_1}} {{L_1}}}} \right.
 \kern-\nulldelimiterspace} {{L_1}}}} \left[ - {\rm{i}}{\lambda _x}\sin {k_x}{\sigma _0}+{\lambda _x}\cos {k_x}{\sigma _z} + {\lambda _y}\sin {k_y}{\sigma _y} + ( {{\gamma _y} + {\lambda _y}\cos {k_y}} ){\sigma _x}\right]\\
 &=: {\rm{i}}\sqrt {{{{C_1}} \mathord{\left/ {\vphantom {{{C_1}} {{L_1}}}} \right.
 \kern-\nulldelimiterspace} {{L_1}}}}H({\bf{k}}).
 \end{split}
 \end{equation}
For ${C_1}=1000\mathop{\rm{pH}}$, ${C_2} = 330\mathop{\rm{pH}}$, ${L_1} = 33\mathop{\mu \rm{F}}$, and ${L_2} = 100\mathop{\mu \rm{F}}$,  we  arrive at $\lambda _x=1$, $\lambda _y=1$, and $\gamma _y=0.33$.

The circuit Laplacian can be represented by the real-space tight-binding Hamiltonian
\begin{equation}\label{s12}
H = {\lambda _x}\sum\limits_{\textit{\textbf{r}}}{(c_{{\textit{\textbf{r}}} + {\textit{\textbf{x}}},{\rm{A}}}^\dag {c_{{\textit{\textbf{r}}},{\rm{A}}}} - c_{{\textit{\textbf{r}}},{\rm{B}}}^\dag {c_{{\textit{\textbf{r}}} - {\textit{\textbf{x}}},{\rm{B}}}})}  + \sum\limits_{\textit{\textbf{r}}} {({\lambda _y}c_{{\textit{\textbf{r}}},{\rm{A}}}^\dag {c_{{\textit{\textbf{r}}}+ {\textit{\textbf{y}}},{\rm{B}}}} + {\gamma _y}c_{{\textit{\textbf{r}}},{\rm{A}}}^\dag {c_{{\textit{\textbf{r}}},{\rm{B}}}} + h.c.)} , 
\end{equation}
where $c_{{\textit{\textbf{x}}},\alpha}^\dag$ ($c_{{\textit{\textbf{x}}},\alpha}$) is the creation (annihilation) operator of sublattice $\alpha$ in the unit cell at position ${\textit{\textbf{r}}}$,  $\lambda_{x}$ and $\lambda_{y}$ are the intracell hopping amplitudes, and $\gamma_y$ is the intercell hopping amplitude.

\section{Boundary modes}\label{sec:2}

 We solve the non-Hermitian Hamiltonian of  Eq.\ (\ref{s12}) for different boundary conditions and obtain the edge and corner skin modes.

\subsection{Generalized Bloch band theory}
The bulk properties of an open boundary condition (OBC) Hamiltonian can be characterized by a Bloch periodic boundary condition (PBC) Hamiltonian \cite{PhysRevLett.117.076804}. As the OBC Hamiltonian may host the NHSE \cite{PhysRevLett.121.086803, PhysRevLett.123.170401}, the spectra of the OBC and PBC Hamiltonians can be different. In both Hermitian and non-Hermitian systems, the OBC Hamiltonian can be encoded using the generalized Brillouin zone  \cite{PhysRevLett.123.066404, PhysRevLett.125.126402, PhysRevLett.125.226402} (although the OBC breaks the translational symmetry) by providing a generalized Bloch Hamiltonian for which the boundary scattering can be regarded as a perturbation. We use the generalized Bloch band theory to determine the generalized Brillouin zone $C_\beta$ for $\beta :={e^{{\rm{i}}k}},k \in \mathbb{C}$. For a one-dimensional (1D) tight-binding OBC Hamiltonian with a unit cell of $q$ degrees of freedom and hopping of any range, we can write
\begin{equation}
H=\sum_{n} \sum_{i=-N}^{N} \sum_{\mu, \nu=1}^{q} t_{i, \mu \nu} c_{n+i, \mu}^{\dagger} c_{n, \nu},
\end{equation}
where  $c_{n, \mu}^{\dagger}$ ($c_{n, \mu}$) is the creation (annihilation) operator in the $n$-th unit cell and $t_{i, \mu \nu}$ is the hopping  to the $i$-th nearest unit cell.   The eigenvector $\left| \psi  \right\rangle ={( {{\psi _{1,1}}, {\psi _{1,2}},\dots, {\psi _{n,q}}} )^T}$  resulting from the real-space eigenvalue equation  $H\left| \psi  \right\rangle  = E\left| \psi  \right\rangle $  can be written as a linear combination of wave functions, 
\begin{equation}\label{b1}
\psi_{n, \mu}=\sum_{j} \phi_{n, \mu}^{(j)}.
\end{equation}
 Since $\phi_{n, \mu}^{(j)}=\beta_{j}^{n} \phi_{\mu}^{(j)}$, we have
 \begin{equation}\label{b2}
\psi_{n, \mu}= \beta_{j}^{n} \phi_{\mu}^{(j)}.
\end{equation}
 $H\left| \psi  \right\rangle  = E\left| \psi  \right\rangle$ can be rewritten as
\begin{equation}\label{b3}
\sum_{\nu=1}^{q}[\mathcal{H}(\beta)]_{\mu \nu} \phi_{\nu}=E \phi_{\mu}
\end{equation}
with the generalized Bloch Hamiltonian
\begin{equation}
[\mathcal{H}(\beta)]_{\mu \nu}=\sum_{i=-N}^{N} t_{i, \mu \nu} \beta^{i}.
\end{equation}
The eigenvalue equation is solved as
\begin{equation} \label{b6}
\det \left[ {\mathcal{H}( \beta  ) - E} \right] = 0,
\end{equation}
yielding $2M$ $(=2qN)$ solutions $\beta_j$  with
\begin{equation}
\left|\beta_{1}\right| \leq\left|\beta_{2}\right| \leq \dots \leq\left|\beta_{2 M-1}\right| \leq\left|\beta_{2 M}\right|.
\end{equation}
For a chain with $L$ unit cells, Eq.\ (\ref{b2}) implies
\begin{equation}
\psi_{n, \mu}=\sum_{j=1}^{2 M}\beta_{j}^{n} \phi_{\mu}^{(j)} \quad (n=1,2, \dots, L; \mu=1,2, \dots, q).
\end{equation}
As the ratio ${{\phi _\mu ^{(j)}} \mathord{\left/
 {\vphantom {{\phi _\mu ^{(j)}} {\phi _\mu ^{(j')}}}} \right.
 \kern-\nulldelimiterspace} {\phi _\mu ^{(j')}}}$ is equal to a constant  for all $j \ne j'$,  $\mu=1,2, \dots, q$ reduces to a single value $\mu=1$. As a consequence,  we obtain 
\begin{equation} \label{b8}
\sum_{j=1}^{2 M} f_{i}(\beta_{j}, E, \mathcal{D}) \phi_{1}^{(j)}=0 \quad(i=1,2, \dots, M)
\end{equation}
 at the left end of the chain ($n = 1$)  and  
\begin{equation} \label{b9}
\sum_{j=1}^{2 M} g_{i}(\beta_{j}, E, \mathcal{D})\beta_{j}^{L} \phi_{1}^{(j)}=0 \quad(i=1,2, \dots, M)
\end{equation}
at the right end of the chain ($n = L$), where $\mathcal{D}$ is the set of hopping parameters $t_{i, \mu \nu}$.
By combining Eqs.\ (\ref{b8}) and (\ref{b9}), we have to fulfill
\begin{equation} \label{b11}
\rm{det}\left[ {\begin{array}{*{20}{c}}
{{f_1}( {{\beta _1},E,{\cal D}} )}& \dots &{{f_1}( {{\beta _{2M}},E,{\cal D}} )}\\
 \vdots & \vdots & \vdots \\
{{f_M}( {{\beta _1},E,{\cal D}} )}& \dots &{{f_M}( {{\beta _{2M}},E,{\cal D}} )}\\
{{g_1}( {{\beta _1},E,{\cal D}} ){{{{\beta _1^L}}}}}& \dots &{{g_1}( {{\beta _{2M}},E,{\cal D}} ){{{{\beta _{2M}^L}}}}}\\
 \vdots & \vdots & \vdots \\
{{g_M}( {{\beta _1},E,{\cal D}} ){{{{\beta _1^L}}}}}& \dots &{{g_M}( {{\beta _{2M}},E,{\cal D}} ){{{{\beta _{2M}^L}}}}}
\end{array}} \right] =0
\end{equation}
for a nontrivial solution.  Equation\ (\ref{b11}) is an algebraic
equation for $\beta_j$  and, for $L$ sufficiently large,  can be written as
\begin{equation} \label{b12}
\sum_{P, Q} F(\beta_{i \in P}, \beta_{j \in Q}, E, \mathcal{D}) \prod_{k \in P}\beta_{k}^{L}=0,
\end{equation}
where $P$ and $Q$ are two disjoint subsets of $\{1,2,\dots,2M\}$. As $L$ grows, the energy levels become dense and asymptotically continuous. The asymptotic distribution of $\beta$ is the generalized Brillouin zone  $C_\beta$. In Hermitian systems, $\beta$ equals unity, implying that the eigenmodes extend over the bulk. In non-Hermitian systems, $\beta$ does not necessarily equal unity and the eigenmodes may be localized at either end of the chain. To distinguish them from the bulk bands, they are called the continuum bands. 

We next consider the asymptotic behavior of the solutions
of Eq.\ (\ref{b12}) for large $L$. When $\left| {{\beta _M}} \right| \ne \left| {{\beta _{M + 1}}} \right|$, we have a leading term of Eq.\ (\ref{b12}) proportional to ${( {{\beta _M}{\beta _{M+1}} \dots {\beta _{2M }}} )^L}$, and obtain
\begin{equation} \label{b13}
F(\beta_{i \in P_0}, \beta_{j \in Q_0}, E, \mathcal{D})=0
\end{equation}
with $P_0=\{M+1,M+2,\dots ,2M\}$ and $Q_0=\{1,2,\dots ,M\}$. 
Equation\ (\ref{b13}) does not depend on $L$ and does not include the continuum bands. When $\left| {{\beta _M}} \right| = \left| {{\beta _{M + 1}}} \right|$, we have leading terms of Eq.\ (\ref{b12}) proportional to ${( {{\beta _M}{\beta _{M + 2}} \dots {\beta _{2M}}} )^L}$ and  ${( {{\beta _{M+1}}{\beta _{M + 2}} \dots {\beta _{2M}}} )^L}$, and obtain
\begin{equation}
\frac{F(\beta_{i \in P_{1}}, \beta_{j \in Q_{1}}, E, \mathcal{D})}{F(\beta_{i \in P_{2}}, \beta_{j \in Q_{2}}, E, \mathcal{D})}=-\left(\frac{\beta_{M}}{\beta_{M+1}}\right)^{L}
\end{equation}
with $P_1=\{M+1,M+2,\dots ,2M\}$, $Q_1=\{1,2,\dots ,M\}$, $P_2=\{M, M+2,\dots ,2M\}$, and $Q_2=\{1,2,\dots ,M-1, M+1\}$. The phase between ${{\beta _M}}$ and ${{\beta _{M+1}}}$ changes almost continuously for large $L$, giving rise to the continuum bands.  The trajectories of ${{\beta _M}}$ and ${{\beta _{M+1}}}$ with $\left| {{\beta _M}} \right| = \left| {{\beta _{M + 1}}} \right|$ determine $C_{\beta}$. Irrespective of the boundary condition, the spectrum of a long chain asymptotically approaches the  continuum bands of $C_{\beta}$. When the numbering of the sites is reversed by setting
$ n'= L + 1- n$ for $n=1,2,\dots, L$ and replacing $\beta$ by $\beta'=1/\beta$, then $\left| {{\beta' _M}} \right| = \left| {{\beta' _{M + 1}}} \right|$.
 
The described method can be extended to higher dimensions. In 2D systems, for example, two parameters  $\beta_x := {e^{{\rm{i}}k_x}}$ with $k_x \in \mathbb{C}$ and  $\beta_y  := {e^{{\rm{i}}k_y}}$ with $k_y \in \mathbb{C}$ are introduced. Then the eigenvalue equation is given by
$\det \left[ {{\mathcal{H}}( {{\beta _x},{\beta _y}} ) - E} \right] = 0$, where ${\mathcal{H}}( {{\beta _x},{\beta _y}} )$ is a 2D generalized Bloch Hamiltonian. If we fix $\beta_y$ $(\beta_x)$, we have a 1D system and $| {{\beta _{x,{M_x}}}} | = | {{\beta _{x,{M_x} + 1}}} |$ $( {| {{\beta _{y,{M_y}}}} | = | {{\beta _{y,{M_y} + 1}}} |} )$, where $2M_x$ $(2M_y)$ is the dimension of the eigenvalue equation for $\beta_x$ $(\beta_y)$.

\subsection{OBC in the $y$-direction} \label{2a}

We consider a semi-infinite 2D lattice that has edges in the $y$-direction. Equation\ (\ref{s12}) can be written in terms of the momentum $k_x$ in the $x$-direction and the real space lattice index $n$ in the $y$-direction ($k_y$ is no longer a good quantum number, since the system is not translational invariant in the $y$-direction).  Fourier transformation with respect to $k_y$ yields a 1D tight-binding Hamiltonian that depends only on $k_x$ \cite{Knig2008},
\begin{equation}
{H_{{k_x}}} = \sum\limits_{{k_{x,}}n} {( {{\rm{\Psi }}_{{k_x},n}^\dag {M_{{k_x}}}{{\rm{\Psi }}_{{k_x},n}} + {\rm{\Psi }}_{{k_x},n + 1}^\dag {T_{y - }}{{\rm{\Psi }}_{{k_x},n}} + {\rm{\Psi }}_{{k_x},n}^\dag {T_{y+ }}{{\rm{\Psi }}_{{k_x},n - 1}}} )}
\end{equation}
with 
\begin{subequations}
    \begin{align}
     {M_{{k_x}}} &=  - {\rm{i}}{\lambda _x}\sin {k_x}{\sigma _0} + {\lambda _x}\cos {k_x}{\sigma _z} + {\gamma _y}{\sigma _x}, \\
      {T_{y \pm }}& = {\lambda _y}\frac{{{\sigma _x} \pm {\rm{i}}{\sigma _y}}}{2}. 
    \end{align}
\end{subequations}
We construct Harper’s equation for the 2D wave function as
\begin{equation}
{M_{{k_x}}}{\Psi _{{k_x},n}} + {T_{y - }}{\Psi _{{k_x},n + 1}} + {T_{y + }}{\Psi _{{k_x},n - 1}} = {E_{{k_x}}} {\Psi _{{k_x},n}}
\end{equation}
with the boundary conditions
\begin{subequations}\label{cab}
    \begin{align}
 {M_{{k_x}}}{{\rm{\Psi }}_{{k_x},1}} + {T_{y - }}{{\rm{\Psi }}_{{k_x},2}} &= {E_{{k_x}}}{{\rm{\Psi }}_{{k_x},1}}, \label{cab43a}\\
{M_{{k_x}}}{{\rm{\Psi }}_{{k_x},{L_y}}} + {T_{y + }}{{\rm{\Psi }}_{{k_x},{L_y} - 1}} &= {E_{{k_x}}}{{\rm{\Psi }}_{{k_x},{L_y}}}.\label{cab43b}
    \end{align}
\end{subequations}
Equations\ (\ref{cab43a}) and (\ref{cab43b}) can be simplified as 
\begin{equation} \label{cab20}
{T_{y + }}{{\rm{\Psi }}_{{k_x},0}} = {T_{y - }}{{\rm{\Psi }}_{{k_x},{L_y} + 1}} = 0.
\end{equation}
The generalized Bloch Hamiltonian is
\begin{equation} \label{cab21}
{\mathcal{H}}_{{k_x}}( {{\beta _y}} ) = {M_{{k_x}}} + {\beta _y}{T_{y - }} + \beta _y^{ - 1}{T_{y + }}
\end{equation}
and we obtain the eigenvalue equation
\begin{equation}\label{cab22}
E_{{k_x}}^2 + 2{\rm{i}}{\lambda _x}{E_{{k_x}}}\sin {k_x} - ( {{\beta _y} + \beta _y^{ - 1}} ){\lambda _y}{\gamma _y} - \lambda _x^2 - \lambda _y^2 - \gamma _y^2 = 0,
\end{equation}
which results in two eigenvalues ${\beta _{y \pm }}$ for given $k_x$ and $E_{{k_x}}$. The corresponding eigenvectors are
\begin{equation}\label{cab23}
{\Phi _{y \pm }} = \left( {\begin{split}
{\frac{{{E_{{k_x}}} + {\lambda _x}{e^{{\rm{i}}{k_x}}}}}{{\sqrt {{{( {{E_{{k_x}}} + {\lambda _x}{e^{{\rm{i}}{k_x}}}} )}^2} + {{( {{\gamma _y} + {\beta _{y \pm }}{\lambda _y}} )}^2}} }}}\\
{\frac{{{\gamma _y} + {\beta _{y \pm }}{\lambda _y}}}{{\sqrt {{{( {{E_{{k_x}}} + {\lambda _x}{e^{{\rm{i}}{k_x}}}} )}^2} + {{( {{\gamma _y} + {\beta _{y \pm }}{\lambda _y}} )}^2}} }}}
\end{split}} \right).
\end{equation}
To fulfill Eq.\ (\ref{cab20}), ${\Phi _{y + }}$ and ${\Phi _{y - }}$ must be linearly dependent. Therefore, we can reduce the condition for the existence of  edge modes as
\begin{equation}\label{cab24}
{\rm det}\left[ {{\Phi _{y + }}\quad{\Phi _{y - }}} \right] = 0.
\end{equation}
Combining Eqs.\ (\ref{cab22}) and (\ref{cab24}) implies
\begin{subequations}
    \begin{align}
{E_{{k_x}}} &=  - {\lambda _x}{e^{{\rm{i}}{k_x}}},\\
{\beta _y} &=  - \frac{{{\gamma _y}}}{{{\lambda _y}}}.
    \end{align}
\end{subequations}
The edge modes are localized for $|\beta_y|<1$ and no edge modes exist for $|\beta_y|\ge 1$.

\subsection{OBC in the $x$-direction}\label{sec:3B}

Following the procedure of Sec.\ \ref{2a}, Fourier transformation with respect to $k_x$ yields a 1D tight-binding Hamiltonian that depends only on  $k_y$,
\begin{equation}
{H_{{k_y}}} = \sum\limits_{{k_{y,}}m} {( {{\rm{\Psi }}_{{k_y},m}^\dag {M_{{k_y}}}{{\rm{\Psi }}_{{k_y},m}} + {\rm{\Psi }}_{{k_y},m + 1}^\dag {T_{x - }}{{\rm{\Psi }}_{{k_y},m}} + {\rm{\Psi }}_{{k_y},m}^\dag {T_{x + }}{{\rm{\Psi }}_{{k_y},m - 1}}} )}
\end{equation}
with 
\begin{subequations} 
    \begin{align}
{M_{{k_y}}} &= ( {{\gamma _y} + {\lambda _y}\cos {k_y}} ){\sigma _x} + {\lambda _y}\sin {k_y}{\sigma _y},\\
{T_{x \pm }} &= {\lambda _x}\frac{{{\sigma _0} \pm {\sigma _z}}}{2}.
    \end{align}
\end{subequations}
We construct Harper’s equation for the 2D wave function as
\begin{equation}
{M_{{k_y}}}{\Psi _{{k_y},m}} + {T_{x - }}{\Psi _{{k_y},m + 1}} + {T_{x + }}{\Psi _{{k_y},m - 1}} = E_{k_y}{\Psi _{{k_y},m}}
\end{equation}
with the boundary conditions
\begin{subequations}\label{cab29}
    \begin{align}
{M_{{k_y}}}{{\rm{\Psi }}_{{k_y},1}} + {T_{x - }}{{\rm{\Psi }}_{{k_x},2}} &= E_{k_y}{{\rm{\Psi }}_{{k_y},1}}, \label{cab53a}\\
{M_{{k_y}}}{{\rm{\Psi }}_{{k_y},{L_x}}} + {T_{x + }}{{\rm{\Psi }}_{{k_y},{L_x} - 1}} &= E_{k_y}{{\rm{\Psi }}_{{k_y},{L_x}}}.\label{cab53b}
    \end{align}
\end{subequations}
Equations (\ref{cab53a}) and (\ref{cab53b}) can be simplified as 
\begin{equation} \label{cab30}
{T_{x + }}{{\rm{\Psi }}_{{k_y},0}} = {T_{x - }}{{\rm{\Psi }}_{{k_y},{L_x} + 1}} = 0.
\end{equation}
The generalized Bloch Hamiltonian is
\begin{equation} \label{cab31}
{\mathcal{H}}_{{k_y}}( {{\beta _x}} ) = {M_{{k_y}}} + {\beta _x}{T_{x - }} + \beta _x^{ - 1}{T_{x + }}
\end{equation}
and we obtain the eigenvalue equation
\begin{equation}\label{cab32}
E_{{k_y}}^2 + ( {{\beta _x} - \beta _x^{ - 1}} ){\lambda _x}{E_{{k_y}}} - 2{\lambda _y}{\gamma _y}\cos {k_y} - \lambda _x^2 - \lambda _y^2 - \gamma _y^2 = 0,
\end{equation}
which results in two eigenvalues ${\beta _{x \pm }}$ for given $k_y$ and $E_{{k_y}}$.  The corresponding eigenvectors are
\begin{equation}\label{cab33}
{\Phi _{x \pm }} = \left( {\begin{split}
{\frac{{E_{{k_y}} + {\lambda _x}{\beta _{x \pm }}}}{{\sqrt {{{( {E_{{k_y}} + {\lambda _x}{\beta _{x \pm }}} )}^2} + {{( {{\gamma _y} + {e^{i{k_y}}}{\lambda _y}} )}^2}} }}}\\
{\frac{{{\gamma _y} + {e^{{\rm{i}}{k_y}}}{\lambda _y}}}{{\sqrt {{{( {E_{{k_y}} + {\lambda _x}{\beta _{x \pm }}} )}^2} + {{( {{\gamma _y} + {e^{i{k_y}}}{\lambda _y}} )}^2}} }}}
\end{split}} \right).
\end{equation}
To fulfill Eq.\ (\ref{cab30}), ${\Phi _{x + }}$ and ${\Phi _{x - }}$ must be linearly dependent. Therefore,  we can reduce the condition for the existence of edge  modes as
\begin{equation}\label{cab34}
{\rm{det}}\left[ {{\Phi _{x + }}\quad{\Phi _{x - }}} \right] = 0.
\end{equation}
Combining Eqs.\ (\ref{cab32}) and (\ref{cab34}) implies
\begin{subequations}
    \begin{align}
{e^{{\rm{i}}{k_y}}} &=  - \frac{{{\gamma _y}}}{{{\lambda _y}}},\label{cab35}\\
{E_{{k_y}}} &=  - {\lambda _x}{\beta _x}.
    \end{align}
\end{subequations}
Equation (\ref{cab35}) implies ${e^{{\rm{i}}{k_y}}}\in \mathbb{Z}$, i.e.,  $k_y = 0$ or $\pi$ with $E_0=E_{\pi}$. 

\subsection{Skin corner modes} \label{sec:3C}

The real-space Hamiltonian for the skin corner modes can be written as
\begin{equation}
{H} = \sum\limits_{m,n} {( {{\rm{\Psi }}_{m,n}^\dag {M}{{\rm{\Psi }}_{m,n}} + {\rm{\Psi }}_{m + 1,n}^\dag {T_{x - }}{{\rm{\Psi }}_{m,n}}+ {\rm{\Psi }}_{m,n}^\dag {T_{x + }}{{\rm{\Psi }}_{m - 1,n}} + {\rm{\Psi }}_{m,n + 1}^\dag {T_{y - }}{{\rm{\Psi }}_{m,n}} + {\rm{\Psi }}_{m,n}^\dag {T_{y + }}{{\rm{\Psi }}_{m,n - 1}}} )}
\end{equation}
with 
\begin{subequations}\label{bb2}
    \begin{align}
M &= {\gamma _y}{\sigma _x},\\
{T_{x \pm }} &= {\lambda _x}\frac{{{\sigma _0} \pm {\sigma _z}}}{2},\\
{T_{y \pm }} &= {\lambda _y}\frac{{{\sigma _x} \pm {\rm{i}}{\sigma _y}}}{2}.
    \end{align}
\end{subequations}
We  construct  Harper’s equation for the 2D wave function as
\begin{equation}
M{\Psi _{m,n}} + {T_{x - }}{\Psi _{m + 1,n}} + {T_{x + }}{\Psi _{m - 1,n}} + {T_{y - }}{\Psi _{m,n + 1}} + {T_{y + }}{\Psi _{m,n - 1}} = E{\Psi _{m,n}}
\end{equation}
with the boundary conditions
\begin{subequations}\label{cab39}
    \begin{align}
M{\Psi _{1,1}} + {T_{x - }}{\Psi _{2,1}} + {T_{y - }}{\Psi _{1,2}} &= E{\Psi _{1,1}}, \label{63a}\\
M{\Psi _{1,{L_y}}} + {T_{x - }}{\Psi _{2,{L_y}}} + {T_{y + }}{\Psi _{1,{L_y} - 1}} &= E{\Psi _{1,{L_y}}},\\
M{\Psi _{{L_x},1}} + {T_{x + }}{\Psi _{{L_x} - 1,1}} + {T_{y - }}{\Psi _{{L_x},2}}&= E{\Psi _{{L_x},1}},\\
M{\Psi _{{L_x},{L_y}}} + {T_{x + }}{\Psi _{{L_x} - 1,{L_y}}} + {T_{y + }}{\Psi _{{L_x},{L_y} - 1}}& = E{\Psi _{{L_x},{L_y}}} \label{63d}.
    \end{align}
\end{subequations}
 Equations (\ref{63a}) to (\ref{63d}) can be simplified as 
\begin{subequations}\label{cab40}
    \begin{align}
{{T_{x + }}{\Psi _{0,n}} = 0,} \label{64a}\\
{{T_{x - }}{\Psi _{{L_x} + 1,n}} = 0,}\\
{{T_{y + }}{\Psi _{m,0}} = 0,}\\
{{T_{y - }}{\Psi _{m,{L_y} + 1}} = 0. \label{64d}}
    \end{align}
\end{subequations}
The generalized Bloch Hamiltonian is
\begin{equation}\label{cab41}
\mathcal{H}( {{\beta _x},{\beta _y}} ) = M + {\beta _x}{T_{x - }} + \beta _x^{ - 1}{T_{x + }} + {\beta _y}{T_{y - }} + \beta _y^{ - 1}{T_{y + }}
\end{equation}
and we obtain the eigenvalue equation
\begin{equation}\label{cab42}
{E^2} + ( {{\beta _x} - \beta _x^{ - 1}} ){\lambda _x}E - ( {{\beta _y} + \beta _y^{ - 1}} ){\lambda _y}{\gamma _y} - \lambda _x^2 - \lambda _y^2 - \gamma _y^2 = 0,
\end{equation}
which results in two eigenvalues ${\beta _{x \pm,y }}$ $({\beta _{x ,y\pm }})$ for given $\beta_y$ ($\beta_x$) and $E$. The corresponding eigenvectors are
\begin{equation}\label{cab43}
{\Phi _{x \pm ,y}} = \left( {\begin{split}
{\frac{{E + {\lambda _x}{\beta _{x \pm }}}}{{\sqrt {{{(E + {\lambda _x}{\beta _{x \pm }})}^2} + {{({\gamma _y} + {\beta _y}{\lambda _y})}^2}} }}}\\
{\frac{{{\gamma _y} + {\beta _y}{\lambda _y}}}{{\sqrt {{{(E + {\lambda _x}{\beta _{x \pm }})}^2} + {{({\gamma _y} + {\beta _y}{\lambda _y})}^2}} }}}
\end{split}} \right),{\Phi _{x,y \pm }} = \left( {\begin{split}
{\frac{{E + {\lambda _x}{\beta _x}}}{{\sqrt {{{(E + {\lambda _x}{\beta _x})}^2} + {{({\gamma _y} + {\beta _{y \pm }}{\lambda _y})}^2}} }}}\\
{\frac{{{\gamma _y} + {\beta _{y \pm }}{\lambda _y}}}{{\sqrt {{{(E + {\lambda _x}{\beta _x})}^2} + {{({\gamma _y} + {\beta _{y \pm }}{\lambda _y})}^2}} }}}
\end{split}} \right).
\end{equation}
To fulfill Eq.\ (\ref{cab40}), ${\Phi _{x + ,y}}$ (${\Phi _{x,y+}}$) and ${\Phi _{x - ,y}}$ (${\Phi _{x,y-}}$) must be linearly dependent. Therefore,  we can reduce the condition for the existence of  corner skin modes as
\begin{subequations}\label{cab44}
\begin{align}
{\rm{det}}\left[ {{\Phi _{x +,y }}\quad{\Phi _{x -,y }}} \right] = 0,\label{cab44a}\\
{\rm{det}}\left[ {{\Phi _{x ,y+ }}\quad{\Phi _{x ,y- }}} \right] = 0.\label{cab44b}
\end{align}
\end{subequations}
Combining Eqs.\ (\ref{cab42}), (\ref{cab44a}), and  (\ref{cab44b}) implies
\begin{subequations}
\begin{align}
{\beta _y} &=  - \frac{{{\gamma _y}}}{{{\lambda _y}}},\\
 E &=  - {\lambda _x}{\beta _x}.\label{c55b}
\end{align}
\end{subequations}
Equations\ (\ref{cab44a}) and (\ref{cab44b}) can be rewritten as 
\begin{subequations}\label{cab56}
\begin{align} 
{\rm{det}}\left[ {{\Phi _{x + ,y + }} \quad {\Phi _{x - ,y + }}} \right] = 0,\\
{\rm{det}}\left[ {{\Phi _{x+ ,y - }} \quad {\Phi _{x - ,y - }}} \right] = 0,\\
{\rm{det}}\left[ {{\Phi _{x+ ,y + }} \quad {\Phi _{x +,y - }}} \right] = 0,\\
{\rm{det}}\left[ {{\Phi _{x- ,y + }} \quad {\Phi _{x - ,y - }}} \right] = 0.
\end{align}
\end{subequations}
Equations\ (\ref{64a}) to (\ref{64d}) imply
\begin{subequations}\label{bb}
    \begin{align}
&{T_{x + }}( {{\Phi _{x +,y  + }} + \beta _x^{{L_x} + 1}{\Phi _{ x-,y  + }}} ) = {T_{x + }}( {{\Phi _{ x+ ,y - }} + \beta _x^{{L_x} + 1}{\Phi _{x -,y  - }}} ) = 0,\\
&{T_{x - }}( {\beta _x^{{L_x} + 1}{\Phi _{x + ,y + }} + {\Phi _{ x-,y  + }}} ) = {T_{x - }}( {\beta _x^{{L_x} + 1}{\Phi _{ x+ ,y - }} + {\Phi _{ x-  ,y- }}} ) = 0,\\
&{T_{y + }}( {{\Phi _{x + ,y + }} + \beta _y^{{L_y} + 1}{\Phi _{x +  ,y- }}} ) = {T_{y + }}( {{\Phi _{x - ,y + }} + \beta _y^{{L_y} + 1}{\Phi _{x -  ,y- }}} ) = 0,\\
&{T_{y - }}( {\beta _y^{{L_y} + 1}{\Phi _{ x+ ,y  + }} + {\Phi _{x +  ,y - }}} ) = {T_{y - }}( {\beta _y^{{L_y} + 1}{\Phi _{x -,y   + }} + {\Phi _{ x- ,y  - }}} ) = 0
    \end{align}
\end{subequations}

and we obtain 
\begin{equation}
\det \left[ {\begin{array}{*{20}{c}}
{E + {\lambda _x}{\beta _x}}&{\beta _x^{2{L_x} + 2}( {E + {\lambda _x}{\beta _x}} )}\\
{{\gamma _y} + {\beta _y}{\lambda _y}}&{\beta _y^{2{L_y} + 2}( {{\gamma _y} + {\beta _y}{\lambda _y}} )}
\end{array}} \right] = 0.
\end{equation}
For sufficiently large $L_x$ and $L_y$, we have
\begin{equation}
\left| {{\beta _x}} \right| = \left| {{\beta _y}} \right|
\end{equation}
and can prove
\begin{equation}
{\rm{det}}\left[ {{\Phi _{x + ,y + }} \quad {\Phi _{x - ,y - }}} \right] = 0 \quad {\rm{and}} \quad {\rm{det}}\left[ {{\Phi _{x + ,y - }}\quad {\Phi _{x - ,y +}}} \right] = 0.
\end{equation}
The wave function can be written as
\begin{equation}
\Psi_{m,n} = \beta _x^m\beta _y^n{\Phi _{x + ,y + }} + \beta _x^m\beta _y^{{L_y} + 1 - n}{\Phi _{x + ,y - }} + \beta _x^{{L_x} + 1 - m}\beta _y^n{\Phi _{x - ,y + }} + \beta _x^{{L_x} + 1 - m}\beta _y^{{L_y} + 1 - n}{\Phi _{x - ,y - }}.
\end{equation}
Since we have $\left| {{\beta _x}} \right| = \left| {{\beta _y}} \right| \neq 1$, the eigenmodes are localized at the corners (NHSE). With the definition  ${\beta _x} := \left| {{\beta _x}} \right|{e^{{\rm{i}}\theta }}, \theta  \in \left[ {0,2\pi } \right]$ and Eq.\ (\ref{c55b}), the spectrum of these corner skin modes is given by
\begin{equation}
E =  - {\gamma _y}\frac{{{\lambda _x}}}{{{\lambda _y}}}{e^{{\rm{i}}\theta }},\quad \theta  \in \left[ {0,2\pi } \right]
\end{equation}
and forms a circle in the complex-energy plane.

\vfill
\subsection{Spectra of the edge modes}\label{sec:3d}

We denote the combination of PBC in the $x$-direction and OBC in the $y$-direction, for example, as PBC$x$-OBC$y$.
 Considering a ribbon geometry and Eq.\ (\ref{eq2}) with the parameters given below Eq.\ (\ref{s10}), the simulated spectrum $j_{\rm{sim.}}$ of the second-order NHSE is plotted in the $j_{\rm{sim.}}$-$k_x$ ($j_{\rm{sim.}}$-$k_y$) space in Fig.\ \ref{sfigure1}\textcolor{blue}{(a)} (\ref{sfigure1}\textcolor{blue}{(b)}) for PBC$x$-OBC$y$ (OBC$x$-PBC$y$). The color represents  the inverse participation ratio  ${\rm{IPR}}=\sum\limits_n {{{| {{\Psi_{n,k}}} |}^4}} /{( {\sum\limits_n {{{| {{\Psi_{n,k}}} |}^2}} } )^2}$, where a larger value corresponds to a more localized wavefunction. For simplicity, all  spectra are given in  normalized units (nu.), i.e., in multiples of $\sqrt {{L_1}/{C_1}} \mathop {\Omega^{-1}}$. In contrast to the first-order topological insulator,  second-order topological insulator, and first-order NHSE, the second-order NHSE has a gapless edge  spectrum in the $y$-direction, see Fig.\ \ref{sfigure1}\textcolor{blue}{(a)}, and has no edge spectrum in the $x$-direction, see Fig.\ \ref{sfigure1}\textcolor{blue}{(b)}. 

 \begin{figure*}[ht!]
\begin{minipage}{0.49\linewidth}
\subfigure{\includegraphics[height=6.5cm]{Supplementary/sfigureedgekx.pdf}\label{s:1a}}
\end{minipage}
\begin{minipage}{0.49\linewidth}
\subfigure{\includegraphics[height=6.5cm]{Supplementary/sfigureedgeky.pdf}\label{s:1b}}
\end{minipage}
\begin{picture}(0,0)
 \put(-453,80){(a)}  \put(-220,80){(b)} 
\end{picture}
\caption{Spectra of the circuit Laplacian (a)  as a function of $k_x$ for PBC$x$-OBC$y$ and (b) as a function of $k_y$ for OBC$x$-PBC$y$.}
\label{sfigure1}
\end{figure*}

\newpage
\section{Physics-graph-informed machine learning}\label{sec:4}

Experimental determination of the $S$-parameters is time-consuming for a large circuit. Physics-graph-informed machine learning (PGIML) enables accurate estimation of the experimental spectra with a reasonable number of measurements. The algorithm is general, as it does not rely on a specific parametrization of the Hamiltonian, and is readily applicable to any system.
\subsection{From graph to circuit}\label{sec:4A0}
A graph (circuit) is  a mathematical structure to model pairwise relations between vertices (nodes) connected by edges (positions). A circuit is a collection of interconnected components which obey Kirchhoff’s laws. Kirchhoff’s voltage law requires the sum of the voltages in any closed loop to be zero and Kirchhoff’s current law requires the sum of the currents meeting at a node to be zero. Thus, circuits are characterised by linear relationships among the voltage variables and  among the current variables. According to graph theory (network topology), a $N$-port electrical circuit can be converted into a  matrix ${\textit{\textbf{G}}}=({\textit{\textbf{P}}},{\textit{\textbf{S}}})$ of complex-weighted directed bipartite graphs $G_{ab}=(P_{ab},S_{ab})$ with the matrix ${\textit{\textbf{P}}}$  of positions $P_{ab}=({a},{b})$ and  the $S$-matrix ${\textit{\textbf{S}}}$ of scattering-parameters ($S$-parameters) $S_{ab}$, where $a,b \in \{1,2,\dots, N\} $ denotes the ports \cite{west2001introduction}. We define the set of graphs as $\mathcal{G}=(\mathcal{P},\mathcal{S})=\{G_{ab}| a,b \in \{1,2,\dots, N\}\}$ with the set of positions $\mathcal{P}$ and the set of $S$-parameters $\mathcal{S}$.


\subsection{$K$-means method}\label{sec:4A}

The $K$-means method is an unsupervised method to classify data points into $K$ mutually exclusive clusters such that the data points in the same cluster are as similar as possible (i.e., high intra-class similarity) and the data points in different clusters are as dissimilar as possible (i.e., low inter-class similarity) \cite{Krishna1999, Likas2003}. Each cluster is classified by its centroid (=  mean of the values assigned to the cluster). We  define the distance between  two $S$-parameters as
\begin{equation}
    d(S_i,S_j) = \|S_i-S_j\|^2_2.
\end{equation}
Given $\mathcal{G = (P, S)}$, representing a circuit with $N$ nodes, the $K$-means method, see Alg.\ \ref{alg:kmeans}, classifies the set ${\mathcal{S}} = \{S_1,S_2,\dots,S_{N^2}\}$ of $S$-parameters (relabel the set of indices $\{(a,b)|a,b\in{1,2,\dots,N}\}\to \{1,2,\dots,N^2\}$) into clusters ${{\mathcal{S}}_1,{\mathcal{S}}_2,\dots,{\mathcal{S}}_K}$  and  the set ${\mathcal{P}} = \{P_1,P_2,\dots,P_{N^2}\}$ of positions into clusters ${{\mathcal{P}}_1,{\mathcal{P}}_2,\dots,{\mathcal{P}}_K}$. Firstly, $K$ centroids are randomly selected
from ${\mathcal{S}}$. Secondly, the distances  to all centroids are calculated for every point in ${\mathcal{S}}$, and the point is assigned to the closest cluster. Thirdly, the centroid is recomputed for every cluster. Steps two and three are repeated until convergence or the maximal number of iterations is reached. Alg.\ \ref{alg:kmeans} partitions $\mathcal{G}$ into subgraphs ${{\mathcal{G}}_1,{\mathcal{G}}_2,\dots,{\mathcal{G}}_K}$ (clusters) based on  the  $S$-parameters.

\begin{algorithm}[H]
	\begin{algorithmic}[1]
		\Input
		\State ${\mathcal{S}} = \{S_1,S_2,\dots,S_{N^2}\}$: Set of $S$-parameters
		\State $K$: Number of clusters
		\EndInput
		\Output
		\State $\{{{\mathcal{S}}_1,{\mathcal{S}}_2,\dots,{\mathcal{S}}_K}\}$:  Set of clusters of $S$-parameters
		\State $\{{{\mathcal{P}}_1,{\mathcal{P}}_2,\dots,{\mathcal{P}}_K}\}$:  Set of clusters of positions 
		\State $\{C_1,C_2,\dots,C_K\}$:  Set of centroids
		\EndOutput
		\State Define centroids $C_1,C_2,\dots,C_K$  by randomly selecting $K$ points from ${\mathcal{S}}$
		\While{Convergence and maximal number of iterations are not reached}
		\For{$n=1$ to ${N^2}$}
		\State $\kappa = \argmin_{n} \|S_n-C_{\kappa}\|^2_2$
		\State $(\tilde{{\mathcal{P}}}_{\kappa},\tilde{{\mathcal{S}}}_{\kappa}).{\rm{append}}(P_n,S_n)$
		\EndFor
		\For{$\kappa=1$ to $K$}
    	\State $C_{\kappa}= {\rm{mean}}(\tilde{{\mathcal{S}}}_{\kappa})$
		\State $({{\mathcal{P}}}_{\kappa},{{\mathcal{S}}}_{\kappa}) = (\tilde{{\mathcal{P}}}_{\kappa},\tilde{{\mathcal{S}}}_{\kappa})$
		\State $(\tilde{{\mathcal{P}}}_{\kappa},\tilde{{\mathcal{S}}}_{\kappa})= (\emptyset,\emptyset)$
		\EndFor
		\EndWhile
		\State\Return $\{{{\mathcal{S}}_1,{\mathcal{S}}_2,\dots,{\mathcal{S}}_K}\}$, $\{{{\mathcal{P}}_1,{\mathcal{P}}_2,\dots,{\mathcal{P}}_K}\}$, and $\{C_1,C_2,\dots,C_K\}$
	\end{algorithmic}
		\caption{$K$-means}
	\label{alg:kmeans}
\end{algorithm}

\subsection{Reconstruction}\label{sec:4b}

We  classify the set of simulated $S$-parameters $\mathcal{S}_{{\rm{sim.}}}=\{S_{{\rm{sim.}},1},S_{{\rm{sim.}},2},\dots,S_{{\rm{sim.}},N^2}\}$ into clusters ${\mathcal{S}}_{{\rm{sim.}},1},$ ${\mathcal{S}}_{{\rm{sim.}},2},$ $\dots,{\mathcal{S}}_{{\rm{sim.}},K}$ and
 obtain for every cluster ${\mathcal{S}}_{{\rm{sim.}},\kappa}$  the centroid $C_{{\rm{sim.}},\kappa}$ which may not belong to ${\mathcal{S}}_{{\rm{sim.}}}$. 
The set of simulated graphs $\mathcal{G}_{\rm{sim.}}$  is an approximation of the set of experimental graphs $\mathcal{G}_{\rm{exp.}}$. As $\mathcal{G}_{\rm{sim.}}$, $\mathcal{G}_{\rm{exp.}}$ are isomorphic and  $\mathcal{G}_{\rm{sim.},\kappa}$ and $ \mathcal{G}_{\rm{exp.},\kappa}$ are isomorphic. Therefore, $\mathcal{G}_{\rm{sim.}}$ provides informative priors  to  $\mathcal{G}_{\rm{exp.}}$ by the mapping ${{\hat G}_{{\rm{sim}}.,\kappa }} \to {{\hat G}_{\exp .,\kappa }}$, where $\hat{G}_{\rm{sim.}}=(\hat{P}_{\rm{sim.}},\hat{S}_{\rm{sim.}})$ and $\hat{G}_{\rm{exp.}}=(\hat{P}_{\rm{exp.}},\hat{S}_{\rm{exp.}})$ are  representative graphs of $\mathcal{G}_{\rm{sim.}}$ and $\mathcal{G}_{\rm{exp.}}$, respectively.  By encoding the $S$-matrix with the representative experimental $S$-parameters ${{\hat S}_{\rm{exp.},1 }},{{\hat S}_{\rm{exp.},2}},\dots,{{\hat S}_{\rm{exp.},\kappa }}$, we obtain the reconstructed experimental $S$-matrix ${\hat{{\textit{\textbf{S}}}}_{\rm{exp.}}}$, see Alg.\ \ref{alg:framework}.


\begin{algorithm}[H]
	\caption{Reconstruction}
	\label{alg:framework}
	\begin{algorithmic}[1]
		\Input
		\State ${\mathcal{S}}_{{\rm{sim.}}}=\{S_{{\rm{sim.}},1},S_{{\rm{sim.}},2},\dots,S_{{\rm{sim.}},N^2}\}$: Set of simulated $S$-parameters 
		\State $K$: Number of clusters
		\EndInput
		\Output
		\State  $\hat{{\textit{\textbf{S}}}}_{\rm{exp.}}$: Reconstructed experimental $S$-matrix
		\EndOutput
		\State Classify ${\mathcal{S}}_{{\rm{sim.}}}$ into clusters ${\mathcal{S}}_{{\rm{sim.}},1},{\mathcal{S}}_{{\rm{sim.}},2},\dots,{\mathcal{S}}_{{\rm{sim.}},K}$ and ${\mathcal{P}}_{{\rm{sim.}}}$ into clusters ${\mathcal{P}}_{{\rm{sim.}},1},{\mathcal{P}}_{{\rm{sim.}},2},\dots,{\mathcal{P}}_{{\rm{sim.}},K}$  with centroids $C_{{\rm{sim.}},1},C_{{\rm{sim.}},2},\dots,C_{{\rm{sim.}},K}$ by Alg.\  \ref{alg:kmeans}
		\For{$\kappa=1$ to $K$}
		\State$\hat{S}_{{\rm{sim.}},\kappa}=\argmin_{S\in {\mathcal{S}}_{{\rm{sim.}},\kappa}}\|S-C_{{\rm{sim.}},\kappa}\|_2^2$
		\State  $\hat{S}_{{\rm{sim.}},\kappa} \to \hat{P}_{{\rm{sim.}},\kappa} \to (a,b)$
		\State Measure $\hat{S}_{{\rm{exp.}},\kappa}$ between port $a$ and $b$
		    \For{$n=1$ to $N^2$} 
		    \If{$n\in {\mathcal{P}}_{{\rm{sim.}},\kappa} $}
	    	\State  $n \to (a,b)$
	    	\State
	    	${\hat{\textit{\textbf{S}}}}_{{\rm{exp.}}}={\hat{\textit{\textbf{S}}}}_{{\rm{exp.}}}+\hat{S}_{{\rm{exp.}},\kappa} {\textit{\textbf{E}}}_{ab}$
	    	\EndIf
	    	\EndFor
		\EndFor
		\State\Return ${\hat{\textit{\textbf{S}}}}_{{\rm{exp.}}}$
	\end{algorithmic}
\end{algorithm}

\subsection{Execution of the PGIML method} \label{sec:4d}

We consider a $10\times 10$ lattice and  Eq.\ (\ref{eq2}) with the parameters below Eq.\ (\ref{s10}). Fig.\ \ref{sfigure2}\textcolor{blue}{(a)} shows the simulated $J$-matrix ${\textit{\textbf{J}}}_{\rm{sim.}}$, which can be transformed into the simulated $S$-matrix ${{\textit{\textbf{S}}}}_{\rm{sim.}}$, see Sec. \ref{spa}. To check the validity of the PGIML method,  we compare ${{\textit{\textbf{S}}}}_{\rm{sim.}}$ with the reconstructed simulated $S$-matrix $\hat{{\textit{\textbf{S}}}}_{\rm{sim.}}$. The $K$-means method is used to classify $\mathcal{S}_{\rm{sim.}}$ into $K=40$ clusters, see Fig.\ \ref{sfigure2}\textcolor{blue}{(b)}. For the $\kappa$-th cluster with centroid $C_{{\rm{sim.}},\kappa}$, we select $\hat{P}_{{\rm{sim.}},\kappa}$ such that $\hat{S}_{{\rm{sim.}},\kappa}={\argmin}_{S\in {\mathcal{S}}_{{\rm{sim.}},\kappa}}\|S-C_{{\rm{sim.}},\kappa}\|_2^2$.
By encoding the $S$-matrix with ${{\hat S}_{\rm{sim.},1 }},{{\hat S}_{\rm{sim.},2}},\dots,{{\hat S}_{\rm{sim.},\kappa }}$, see Fig. \ref{sfigure2}\textcolor{blue}{(c)}, we obtain ${\hat{{\textit{\textbf{S}}}}_{\rm{sim.}}}$, which can be transformed back into the reconstructed simulated $J$-matrix  $\hat{{\textit{\textbf{J}}}}_{\rm{sim.}}$, see Fig.\ \ref{sfigure2}\textcolor{blue}{(d)}.


\begin{figure*}[!htpb]
\begin{minipage}{0.49\linewidth}
\subfigure{\includegraphics[height=6.5cm]{Supplementary/s2a}\label{s:2a}}
\end{minipage}
\begin{minipage}{0.49\linewidth}
\subfigure{\includegraphics[height=6.5cm]{Supplementary/s2b}\label{s:2b}}
\end{minipage}
\begin{minipage}{0.49\linewidth}
\subfigure{\includegraphics[height=6.5cm]{Supplementary/s2c}\label{s:2c}}
\end{minipage}
\begin{minipage}{0.49\linewidth}
\subfigure{\includegraphics[height=6.5cm]{Supplementary/s2d}\label{s:2d}}
\end{minipage}
\begin{picture}(0,0)
\put(-453,270){(a)}\put(-220,270){(b)} \put(-453,80){(c)}  \put(-220,80){(d)} 
\end{picture}
\caption{(a) Simulated $J$-matrix, (b) clusters of the simulated $S$-parameters, (c) representative simulated $S$-parameters, and (d) reconstructed simulated $J$-matrix.}
\label{sfigure2}
\end{figure*}
\newpage
To evaluate the accuracy of the algorithm,  we calculate the mean squared error (MSE) as loss function,
\begin{equation} \label{eq80}
    {\rm{MSE}} = \sum_{n=1}^{N^2}\frac{\|{\hat{S}}_{{\rm{sim.},n}}-S_{{\rm{sim.},n}}\|^2_2}{{N^2}},
\end{equation}
which measures the compactness of the clustering and should be as small as possible. Figure\ \ref{sfigure3} shows the MSE versus the number of clusters for different boundary conditions. When the MSE is smaller than $10^{-9}$, we assume $\hat{\textit{\textbf{J}}}_{\rm{sim.}}$ to be sufficiently similar to ${{\textit{\textbf{J}}}}_{\rm{sim.}}$, i.e., $K$ is large enough to determine the $S$-matrix (underfitting and wellfitting regimes in Fig.\ \ref{sfigure3}). We find a larger MSE for OBC$x$-OBC$y$, i.e., we have to choose a larger $K$ than for the other boundary conditions.
The spectra $j_{\rm{sim.}}$ of  ${\textit{\textbf{J}}}_{\rm{sim.}}$ and $\hat{j}_{\rm{sim.}}$ of ${\hat{\textit{\textbf{J}}}}_{\rm{sim.}}$ for the cases of underfitting and wellfitting  are compared in Fig.\ \ref{sfigure4} for different boundary conditions. In the case of  underfitting, we observe a poor predictive performance, while in the case of wellfitting the reconstructed results resemble the simulated results. Our estimator also provides remarkable accuracy for a large system, such as a $30\times30$ lattice, but a larger number of clusters is required,  see Fig.\ \ref{sfigure5}. 
 \begin{figure}[htpb]
\centering
\includegraphics[width=1\linewidth]{Supplementary/sfigure3.pdf}
\caption{MSE versus the number of clusters for different boundary conditions. }
\label{sfigure3}
\end{figure}

\newpage

 \begin{figure*}[ht!]
\begin{minipage}{0.32\linewidth}
\subfigure{\includegraphics[height=4cm]{Supplementary/sfigure4a.pdf}\label{s:4a}}
\end{minipage}
\begin{minipage}{0.32\linewidth}
\subfigure{\includegraphics[height=4cm]{Supplementary/sfigure4b.pdf}\label{s:4b}}
\end{minipage}
\begin{minipage}{0.32\linewidth}
\subfigure{\includegraphics[height=4cm]{Supplementary/sfigure4c.pdf}\label{s:4c}}
\end{minipage}
\begin{minipage}{0.32\linewidth}
\subfigure{\includegraphics[height=4cm]{Supplementary/sfigure4d.pdf}\label{s:4d}}
\end{minipage}
\begin{minipage}{0.32\linewidth}
\subfigure{\includegraphics[height=4cm]{Supplementary/sfigure4e.pdf}\label{s:4e}}
\end{minipage}
\begin{minipage}{0.32\linewidth}
\subfigure{\includegraphics[height=4cm]{Supplementary/sfigure4f.pdf}\label{s:4f}}
\end{minipage}
\begin{minipage}{0.32\linewidth}
\subfigure{\includegraphics[height=4cm]{Supplementary/sfigure4g.pdf}\label{s:4g}}
\end{minipage}
\begin{minipage}{0.32\linewidth}
\subfigure{\includegraphics[height=4cm]{Supplementary/sfigure4h.pdf}\label{s:4h}}
\end{minipage}
\begin{minipage}{0.32\linewidth}
\subfigure{\includegraphics[height=4cm]{Supplementary/sfigure4i.pdf}\label{s:4i}}
\end{minipage}
\begin{minipage}{0.32\linewidth}
\subfigure{\includegraphics[height=4cm]{Supplementary/sfigure4j.pdf}\label{s:4j}}
\end{minipage}
\begin{minipage}{0.32\linewidth}
\subfigure{\includegraphics[height=4cm]{Supplementary/sfigure4k.pdf}\label{s:4k}}
\end{minipage}
\begin{minipage}{0.32\linewidth}
\subfigure{\includegraphics[height=4cm]{Supplementary/sfigure4l.pdf}\label{s:4l}}
\end{minipage}
\begin{picture}(0,0)
\put(-450,410){(a)}\put(-298,410){(b)} \put(-144,410){(c)}  \put(-450,290){(d)}\put(-298,290){(e)} \put(-144,290){(f)}
\put(-450,170){(g)}\put(-298,170){(h)} \put(-144,170){(i)} 
\put(-450,50){(j)}\put(-298,50){(k)} \put(-144,50){(l)} 
 \put(-460,345){\rotatebox{90}{\colorbox{lightgray}{\tiny{PBC$x$-PBC$y$}}}}
 \put(-460,225){\rotatebox{90}{\colorbox{lightgray}{\tiny{PBC$x$-OBC$y$}}}}
  \put(-460,105){\rotatebox{90}{\colorbox{lightgray}{\tiny{OBC$x$-PBC$y$}}}}
 \put(-460,-15){\rotatebox{90}{\colorbox{lightgray}{\tiny{OBC$x$-OBC$y$}}}}
\end{picture}
\caption{Spectra of the  $J$-matrix on a $ 10\times  10$ lattice for different boundary conditions: (a,\ d,\ g,\ j) simulated, (b,\ e,\ h,\ k) reconstructed with  $K=5$ (underfitting), and  (c,\ f,\ i,\ l) reconstructed with  $K=40$ (wellfitting).}
\label{sfigure4}
\end{figure*}

\newpage

 \begin{figure*}[ht!]
\begin{minipage}{0.32\linewidth}
\subfigure{\includegraphics[height=4cm]{Supplementary/sfigurelaa.pdf}\label{s:5a}}
\end{minipage}
\begin{minipage}{0.32\linewidth}
\subfigure{\includegraphics[height=4cm]{Supplementary/sfigurelab.pdf}\label{s:5b}}
\end{minipage}
\begin{minipage}{0.32\linewidth}
\subfigure{\includegraphics[height=4cm]{Supplementary/sfigurelac.pdf}\label{s:5c}}
\end{minipage}
\begin{picture}(0,0)
\put(-450,50){(a)}\put(-298,50){(b)} \put(-144,50){(c)} 
 \put(-460,-15){\rotatebox{90}{\colorbox{lightgray}{\tiny{OBC$x$-OBC$y$}}}}
\end{picture}
\caption{ Spectra of the $J$-matrix on a $ 30 \times  30$ lattice  for  OBC$x$-OBC$y$: (a) simulated, (b)  reconstructed with $K=80$ (underfitting ), and (c) reconstructed with $K=250$ (wellfitting).}
\label{sfigure5}
\end{figure*}

\section{Experimental implementation}\label{sec:5}

We introduce the experimental setup, fitting model,  and selection of measurements.

\subsection{Experimental setup}\label{sec:5A}
To map the lattice Hamiltonian to a circuit Laplacian, we eliminate the diagonal elements of the circuit Laplacian at the resonance frequency by choosing appropriate inductivities and capacitances and canceling the contributions of the inductivities (capacitances) at the nodes by grounding components with matched capacitances (inductivities). As shown in Fig.\ \ref{scheme}, we divide the circuit into corner (NW, SW, NE, SE), edge ($\rm{ N_A}$, $\rm{S_B}$, $\rm{W_A}$, $\rm{W_B}$, $\rm{E_A}$, $\rm{E_B}$), and bulk ($\rm{M_A}$, $\rm{M_B}$) sites ($\rm{N=north}$, $\rm{S=south}$, $\rm{W=west}$, $\rm{E=east}$, $\rm{M=middle}$).  The  unit cell of the circuit is shown in Fig.\ \ref{sfigure6}\textcolor{blue}{(a)}. The boundary conditions are realized by adjusting the switches to the ground and connections between the boundaries. For instance, to create PBC$x$, we turn on the switches ${T_{\rm W \to S}}$ and turn off the switch ${T_{\rm{W_A}}}$.  We present the grounding components for the different boundary conditions in Table\ \ref{table1}.  The circuit sites are connected by pairs of inductors $(L_1, L_2)$,  pairs of capacitors $(C_1, C_2)$, and voltage feedback operational amplifiers (nonreciprocal couplings; Texas Instruments, LM6171), as shown in Fig.\ \ref{sfigure6}\textcolor{blue}{(b)}, which block the input current while maintaining the output current. The inductors and capacitors are Hermitian components while the resistors and voltage feedback operational amplifiers are non-Hermitian components, see Fig.\ \ref{sfigure6}\textcolor{blue}{(c)}.

\afterpage{\clearpage}
\begin{sidewaysfigure}
\centering
\includegraphics[width=1.0\linewidth]{sfigure7.pdf}
\caption{ Circuit implementation of the second order NHSE.}
\label{scheme}
\end{sidewaysfigure}

\begin{figure}[htpb]
\centering
\includegraphics[width=0.8\linewidth]{Supplementary/sifigureFOLLOWER.pdf}
\caption{ (a) Unit cell at the boundary of the circuit and switches to control the boundary condition, (b) voltage follower, and (c)  functions of the components. }
\label{sfigure6}
\end{figure}

\begin{table}[!h]
\centering
\begin{tabular}{|p{3.6cm}||p{2.9cm}|p{2.9cm}|p{2.9cm}| p{2.9cm}|}
\hline
 & OBC$x$-OBC$y$ & OBC$x$-PBC$y$ &PBC$x$-OBC$y$ & PBC$x$-PBC$y$ \\
\hline
$\rm{NW}$ & $2{C_1} \parallel  {L_1}/2 \parallel  {L_2}$ & ${C_1} \parallel {L_1}/2 \parallel{L_2}$ & ${C_1} \parallel {L_1}/2 \parallel {L_2}$ & ${L_1}/2 \parallel {L_2}$\\
\hline
$\rm{SW}$ & ${C_1} \parallel {L_2}$  & ${L_2}$ & ${C_1} \parallel {L_2}$ & ${L_2}$  \\
\hline
$\rm{NE} $& ${C_1} \parallel{L_1}/2 \parallel{L_2}$  & $ {L_1}/2 \parallel{L_2}$ & ${C_1} \parallel{L_1}/2 \parallel {L_2}$ & ${L_1}/2 \parallel {L_2}$ \\
\hline
$\rm{SE}$  & $ {C_1} \parallel {L_1} \parallel {L_2}$  & ${L_1} \parallel{L_2}$ & ${C_1} \parallel{L_2}$ & ${L_2}$  \\
\hline
$\rm{N_A}$ & ${C_1} \parallel {L_1}/2 \parallel {L_2}$  & $ {L_1}/2 \parallel {L_2}$ & ${C_1} \parallel{L_1}/2 \parallel {L_2}$ & $ {L_1}/2 \parallel {L_2}$  \\
\hline
$\rm{S_B}$ & ${C_1} \parallel {L_2}$  & $ {L_2}$ & ${C_1} \parallel{L_2}$ & ${L_2}$  \\
\hline
$\rm{W_A}$ & ${C_1}\parallel  {L_1}/2 \parallel {L_2}$ &${C_1} \parallel{L_1}/2 \parallel {L_2}$ &$ {L_1}/2 \parallel{L_2}$ &${L_1}/2 \parallel {L_2}$\\
\hline
$\rm{W_B}$ & ${L_2}$ & ${L_2}$ & ${L_2}$ & ${L_2}$\\
\hline
$\rm{E_A}$ & ${L_1}/2 \parallel {L_2}$ & ${L_1}/2 \parallel {L_2}$ & ${L_1}/2 \parallel {L_2}$ & ${L_1}/2 \parallel {L_2}$  \\
\hline
$\rm{E_B}$ & ${L_1} \parallel {L_2}$  & ${L_1} \parallel{L_2}$ & $ {L_2}$ & ${L_2}$  \\
\hline
$\rm{M_A}$ & ${L_1}/2 \parallel{L_2}$  & ${L_1}/2 \parallel{L_2}$ & ${L_1}/2 \parallel {L_2}$ & ${L_1}/2 \parallel {L_2}$  \\
\hline
$\rm{M_B}$ & ${L_2}$  & $  {L_2}$ & ${L_2}$ & $ {L_2}$  \\
\hline
\end{tabular}
\caption{ Grounding components for different boundary conditions, where $\parallel$ indicates parallel connection. \label{table1}}
\end{table}

\subsection{Fitting model}\label{sec:5B}

To take into account the parasitic resistances $R_{L_{j}}$ of the inductors as the most prominent experimental imperfections, we adjust the inductances as $L^{'}_{j} = L_{j}+R_{L_{j}} /({\rm{i}} \omega), j=1,2$, i.e., Eq.\ (\ref{eq2}) is rewritten as 
\begin{equation} \label{eqnn}
{\textit{\textbf{J}}}_{\rm{fit.}}({\textit{\textbf{k}}},\omega ) = {\rm{i}}\omega \left[ {\begin{aligned}
{\frac{{{L^{'}_1}{L^{'}_2}}}{{( {{L^{'}_1} + 2{L^{'}_2}} ){\omega ^2}}} - 2{C_1} - {C_2} + {C_1}{e^{ - {\rm{i}}{k_x}}}}&{\qquad \qquad \qquad {{C_2}+{C_1}{e^{ - {\rm{i}}{k_y}}}}}\\
{{{C_2}+{C_1}{e^{{\rm{i}}{k_y}}}}\qquad \qquad \qquad}&{\frac{{{L^{'}_1}{L^{'}_2}}}{{( {{L^{'}_1} + {L^{'}_2}} ){\omega ^2}}} - {C_1} - {C_2} - \frac{{{L^{'}_1}}}{{{\omega ^2}}}{e^{{\rm{i}}{k_x}}}}
\end{aligned}} \right].
 \end{equation}
For $R_{L_{j}}=7 \mathop \Omega$, we obtain the spectra $j_{\rm{fit.}}$ of the fitted $J$-matrix in Fig.\ \ref{sfigure8} for different boundary conditions. Compared with Figs.\ \ref{sfigure4}\textcolor{blue}{(a,\ d,\ g,\ j)}, the spectra in Figs.\  \ref{sfigure8}\textcolor{blue}{(a,\ b)}  show  deviations in terms of non-degenerate modes due to the parasitic resistances, as observed in the experiments, while the spectra in Figs.\ \ref{sfigure8}\textcolor{blue}{(c,\ d)} show no such deviations due to the fact that there are no degenerate modes.

\begin{figure*}[htp!]
\begin{minipage}{0.49\linewidth}
\subfigure{\includegraphics[height=4cm]{Supplementary/sfigurefita.pdf}\label{s:8a}}
\end{minipage}
\begin{minipage}{0.49\linewidth}
\subfigure{\includegraphics[height=4cm]{Supplementary/sfigurefitb.pdf}\label{s:8b}}
\end{minipage}
\begin{minipage}{0.49\linewidth}
\subfigure{\includegraphics[height=4cm]{Supplementary/sfigurefitc.pdf}\label{s:8c}}
\end{minipage}
\begin{minipage}{0.49\linewidth}
\subfigure{\includegraphics[height=4cm]{Supplementary/sfigurefitd.pdf}\label{s:8d}}
\end{minipage}
\begin{picture}(0,0)
\put(-453,170){(a)}\put(-220,170){(b)} \put(-453,55){(c)}  \put(-220,55){(d)} 
\end{picture}
\caption{Spectra of the fitted $J$-matrix for (a) PBC$x$-PBC$y$, (b) PBC$x$-OBC$y$, (c) OBC$x$-PBC$y$, and (d) OBC$x$-OBC$y$.}
\label{sfigure8}
\end{figure*}

\subsection{Selection of measurements}\label{sec:5C}
We choose $K= 100$, i.e., more than the $K= 40$ clusters considered in Sec. \ref{sec:4d} due to the imperfections of the components and instability of the voltage feedback operational amplifiers. More clusters provide not only more accuracy of the reconstruction according to Fig. \ref{sfigure3} but also offer better correction of the components. The selected sites for the measurements are presented in Table\ \ref{stable2}. For example,  the measurement for cluster $\kappa=1$ for PBC$x$-PBC$y$ (${\hat{S}}_{\rm{exp.},1}$) is executed between sites $(0,0)$ and $(2,2)$, denoted as $(0,0) \to (2,2)$.  As shown in Fig.\ \ref{sfigure9}, every site is measured at least once. If the measured result agrees with the simulated result, all grounding and connecting components operate correctly; if not, the welding must be checked.

\begin{figure*}[ht!]
\begin{minipage}{0.49\linewidth}
\subfigure{\includegraphics[height=5cm]{Supplementary/sfigureport1.pdf}\label{s:9a}}
\end{minipage}
\begin{minipage}{0.49\linewidth}
\subfigure{\includegraphics[height=5cm]{Supplementary/sfigureport2.pdf}\label{s:9b}}
\end{minipage}
\begin{picture}(0,0)
 \put(-453,70){(a)}  \put(-220,70){(b)} 
\end{picture}
\caption{Statistics of the measurements for $S_{ab}$ between (a) site $a$  and  (b) site $b$. The sites are labeled $0,1,\dots,9$ along the $x$ and $y$-directions. }
\label{sfigure9}
\end{figure*}


\newpage
\begin{longtable}[ht!]
{|p{1cm}||p{2.5cm}|p{2.5cm}|p{2.5cm}| p{2.5cm}|}
\hline
$\kappa$ & PBC$x$-PBC$y$ & PBC$x$-OBC$y$ &OBC$x$-PBC$y$ & OBC$x$-OBC$y$ \\
\hline
1   & $(0,0) \to (2,2)$ & $(0,0) \to (0,0)$ & $(0,0) \to (0,0)$ & $(0,0) \to (0,0)$ \\
2   & $(0,3) \to (0,3)$ & $(0,1) \to (3,2)$ & $(0,1) \to (0,1)$ & $(0,0) \to (1,0)$ \\
3   & $(0,3) \to (3,2)$ & $(0,2) \to (0,2)$ & $(0,1) \to (0,2)$ & $(0,1) \to (0,0)$ \\
4   & $(0,4) \to (1,3)$ & $(0,3) \to (1,0)$ & $(0,1) \to (1,2)$ & $(0,1) \to (0,2)$ \\
5   & $(0,5) \to (7,5)$ & $(0,3) \to (9,1)$ & $(0,2) \to (0,2)$ & $(0,1) \to (0,3)$ \\
6   & $(0,6) \to (9,7)$ & $(0,4) \to (0,4)$ & $(0,2) \to (0,3)$ & $(0,2) \to (0,4)$ \\
7   & $(0,7) \to (1,7)$ & $(0,5) \to (0,5)$ & $(0,2) \to (1,1)$ & $(0,3) \to (0,2)$ \\
8   & $(0,7) \to (9,9)$ & $(0,6) \to (5,2)$ & $(0,2) \to (1,2)$ & $(0,4) \to (0,3)$ \\
9   & $(0,9) \to (0,9)$ & $(0,8) \to (1,5)$ & $(0,3) \to (0,1)$ & $(0,6) \to (0,6)$ \\
10  & $(1,1) \to (2,8)$ & $(0,9) \to (0,9)$ & $(0,3) \to (0,2)$ & $(0,6) \to (0,7)$ \\
11  & $(1,2) \to (0,5)$ & $(1,1) \to (1,1)$ & $(0,4) \to (1,5)$ & $(0,6) \to (2,6)$ \\
12  & $(1,4) \to (0,6)$ & $(1,1) \to (9,2)$ & $(0,4) \to (2,2)$ & $(0,7) \to (0,7)$ \\
13  & $(1,4) \to (1,6)$ & $(1,3) \to (1,1)$ & $(0,5) \to (1,4)$ & $(0,8) \to (0,9)$ \\
14  & $(1,4) \to (2,4)$ & $(1,4) \to (0,1)$ & $(0,6) \to (2,6)$ & $(0,8) \to (1,8)$ \\
15  & $(1,5) \to (0,6)$ & $(1,5) \to (0,5)$ & $(0,8) \to (0,6)$ & $(0,9) \to (0,9)$ \\
16  & $(1,5) \to (9,9)$ & $(1,5) \to (2,4)$ & $(0,8) \to (0,7)$ & $(1,0) \to (1,0)$ \\
17  & $(1,8) \to (0,8)$ & $(1,7) \to (0,8)$ & $(1,0) \to (0,9)$ & $(1,0) \to (2,0)$ \\
18  & $(2,0) \to (2,0)$ & $(1,7) \to (8,7)$ & $(1,2) \to (2,2)$ & $(1,1) \to (0,1)$ \\
19  & $(2,2) \to (0,1)$ & $(1,8) \to (0,8)$ & $(1,3) \to (0,4)$ & $(1,2) \to (3,3)$ \\
20  & $(2,5) \to (2,5)$ & $(1,8) \to (1,8)$ & $(1,4) \to (1,4)$ & $(1,4) \to (1,4)$ \\
21  & $(2,9) \to (9,9)$ & $(1,8) \to (2,8)$ & $(1,5) \to (0,5)$ & $(1,5) \to (0,6)$ \\
22  & $(3,1) \to (2,1)$ & $(1,9) \to (1,9)$ & $(1,6) \to (0,7)$ & $(1,5) \to (2,5)$ \\
23  & $(3,3) \to (4,1)$ & $(2,2) \to (3,2)$ & $(1,9) \to (0,8)$ & $(1,6) \to (1,8)$ \\
24  & $(3,5) \to (1,5)$ & $(2,3) \to (3,1)$ & $(2,0) \to (1,3)$ & $(1,8) \to (0,7)$ \\
25  & $(3,6) \to (2,4)$ & $(2,5) \to (5,1)$ & $(2,2) \to (2,0)$ & $(1,8) \to (2,8)$ \\
26  & $(3,7) \to (5,8)$ & $(2,7) \to (0,9)$ & $(2,2) \to (2,2)$ & $(1,8) \to (4,8)$ \\
27  & $(3,8) \to (0,6)$ & $(2,7) \to (3,6)$ & $(2,5) \to (3,5)$ & $(1,9) \to (0,9)$ \\
28  & $(3,8) \to (3,8)$ & $(2,9) \to (1,9)$ & $(2,6) \to (3,6)$ & $(1,9) \to (2,9)$ \\
29  & $(3,9) \to (2,1)$ & $(2,9) \to (2,5)$ & $(2,7) \to (0,7)$ & $(2,0) \to (3,0)$ \\
30  & $(3,9) \to (3,9)$ & $(3,1) \to (4,2)$ & $(2,7) \to (4,9)$ & $(2,1) \to (0,3)$ \\
31  & $(4,1) \to (4,0)$ & $(3,2) \to (3,0)$ & $(2,8) \to (1,0)$ & $(2,1) \to (1,3)$ \\
32  & $(4,1) \to (5,3)$ & $(3,5) \to (1,5)$ & $(2,8) \to (2,8)$ & $(2,1) \to (2,3)$ \\
33  & $(4,1) \to (7,2)$ & $(3,8) \to (4,7)$ & $(3,0) \to (3,0)$ & $(2,2) \to (1,4)$ \\
34  & $(4,4) \to (4,4)$ & $(3,9) \to (1,8)$ & $(3,2) \to (1,2)$ & $(2,4) \to (1,5)$ \\
35  & $(4,5) \to (3,8)$ & $(4,0) \to (2,0)$ & $(3,5) \to (3,5)$ & $(2,4) \to (3,4)$ \\
36  & $(4,5) \to (6,5)$ & $(4,1) \to (3,1)$ & $(3,6) \to (3,6)$ & $(2,6) \to (2,6)$ \\
37  & $(4,5) \to (8,8)$ & $(4,1) \to (6,1)$ & $(3,6) \to (4,6)$ & $(2,9) \to (0,9)$ \\
38  & $(4,6) \to (1,5)$ & $(4,2) \to (4,4)$ & $(3,6) \to (6,6)$ & $(3,0) \to (3,0)$ \\
39  & $(4,9) \to (1,8)$ & $(4,2) \to (6,2)$ & $(3,8) \to (4,0)$ & $(3,0) \to (4,0)$ \\
40  & $(4,9) \to (4,2)$ & $(4,3) \to (3,3)$ & $(3,9) \to (2,9)$ & $(3,1) \to (1,1)$ \\
41  & $(5,1) \to (3,1)$ & $(4,5) \to (2,6)$ & $(4,0) \to (4,0)$ & $(3,4) \to (3,4)$ \\
42  & $(5,1) \to (4,1)$ & $(4,6) \to (6,5)$ & $(4,0) \to (4,4)$ & $(3,4) \to (4,4)$ \\
43  & $(5,1) \to (4,2)$ & $(4,7) \to (3,7)$ & $(4,0) \to (5,0)$ & $(3,5) \to (2,6)$ \\
44  & $(5,3) \to (6,5)$ & $(4,8) \to (4,8)$ & $(4,0) \to (5,9)$ & $(3,9) \to (2,9)$ \\
45  & $(5,4) \to (6,3)$ & $(4,9) \to (3,9)$ & $(4,1) \to (2,1)$ & $(4,0) \to (4,0)$ \\
46  & $(5,5) \to (3,7)$ & $(4,9) \to (6,9)$ & $(4,2) \to (4,2)$ & $(4,0) \to (5,0)$ \\
47  & $(5,5) \to (4,5)$ & $(5,0) \to (7,0)$ & $(4,2) \to (6,2)$ & $(4,1) \to (4,1)$ \\
48  & $(5,6) \to (7,3)$ & $(5,2) \to (5,2)$ & $(4,2) \to (7,1)$ & $(4,2) \to (4,2)$ \\
49  & $(5,7) \to (3,7)$ & $(5,2) \to (6,2)$ & $(4,3) \to (4,4)$ & $(4,8) \to (5,8)$ \\
50  & $(5,7) \to (5,9)$ & $(5,6) \to (4,7)$ & $(4,7) \to (5,5)$ & $(5,0) \to (5,0)$ \\
51  & $(5,9) \to (3,8)$ & $(5,6) \to (7,5)$ & $(4,9) \to (2,8)$ & $(5,0) \to (6,0)$ \\
52  & $(6,1) \to (6,1)$ & $(5,7) \to (7,5)$ & $(5,0) \to (5,0)$ & $(5,2) \to (6,1)$ \\
53  & $(6,2) \to (2,3)$ & $(6,0) \to (5,0)$ & $(5,0) \to (6,1)$ & $(5,8) \to (5,8)$ \\
54  & $(6,4) \to (7,4)$ & $(6,0) \to (5,1)$ & $(5,1) \to (2,2)$ & $(5,8) \to (6,8)$ \\
55  & $(6,5) \to (5,3)$ & $(6,0) \to (7,0)$ & $(5,4) \to (6,4)$ & $(5,9) \to (5,9)$ \\
56  & $(6,5) \to (5,5)$ & $(6,2) \to (8,0)$ & $(5,6) \to (5,6)$ & $(6,0) \to (6,0)$ \\
57  & $(6,6) \to (6,0)$ & $(6,3) \to (4,2)$ & $(5,7) \to (4,6)$ & $(6,0) \to (7,0)$ \\
58  & $(6,6) \to (8,6)$ & $(6,3) \to (5,3)$ & $(6,0) \to (7,0)$ & $(6,1) \to (5,1)$ \\
59  & $(6,7) \to (1,3)$ & $(6,3) \to (8,4)$ & $(6,1) \to (3,1)$ & $(6,1) \to (6,2)$ \\
60  & $(6,7) \to (7,7)$ & $(6,5) \to (5,5)$ & $(6,1) \to (6,0)$ & $(6,4) \to (7,4)$ \\
61  & $(6,7) \to (9,8)$ & $(6,6) \to (0,6)$ & $(6,2) \to (5,9)$ & $(6,5) \to (6,4)$ \\
62  & $(6,8) \to (5,5)$ & $(6,6) \to (5,4)$ & $(6,2) \to (8,3)$ & $(6,6) \to (6,6)$ \\
63  & $(6,8) \to (6,7)$ & $(6,7) \to (3,8)$ & $(6,5) \to (6,2)$ & $(6,8) \to (8,8)$ \\
64  & $(6,8) \to (7,8)$ & $(6,7) \to (6,8)$ & $(6,7) \to (6,0)$ & $(6,9) \to (6,8)$ \\
65  & $(6,9) \to (4,7)$ & $(6,7) \to (6,9)$ & $(6,8) \to (6,8)$ & $(7,0) \to (6,0)$ \\
66  & $(7,1) \to (6,1)$ & $(6,8) \to (6,5)$ & $(7,0) \to (8,9)$ & $(7,0) \to (7,0)$ \\
67  & $(7,3) \to (9,2)$ & $(6,8) \to (6,9)$ & $(7,2) \to (9,2)$ & $(7,0) \to (8,0)$ \\
68  & $(7,3) \to (9,3)$ & $(6,8) \to (8,9)$ & $(7,6) \to (5,4)$ & $(7,0) \to (8,2)$ \\
69  & $(7,4) \to (7,4)$ & $(6,9) \to (2,9)$ & $(7,6) \to (8,3)$ & $(7,2) \to (7,2)$ \\
70  & $(7,4) \to (8,2)$ & $(6,9) \to (5,9)$ & $(7,6) \to (8,6)$ & $(7,3) \to (7,0)$ \\
71  & $(7,6) \to (1,6)$ & $(7,3) \to (6,3)$ & $(7,7) \to (5,8)$ & $(7,4) \to (7,1)$ \\
72  & $(7,6) \to (1,7)$ & $(7,4) \to (5,8)$ & $(7,8) \to (6,8)$ & $(7,4) \to (8,4)$ \\
73  & $(7,6) \to (8,7)$ & $(7,5) \to (6,5)$ & $(7,8) \to (7,8)$ & $(7,5) \to (8,3)$ \\
74  & $(7,7) \to (3,7)$ & $(7,7) \to (6,7)$ & $(7,9) \to (9,0)$ & $(7,5) \to (9,6)$ \\
75  & $(7,7) \to (7,4)$ & $(7,7) \to (7,6)$ & $(8,0) \to (8,0)$ & $(7,6) \to (6,6)$ \\
76  & $(7,9) \to (6,9)$ & $(7,8) \to (1,8)$ & $(8,0) \to (9,0)$ & $(7,6) \to (8,7)$ \\
77  & $(8,2) \to (0,1)$ & $(7,8) \to (4,8)$ & $(8,1) \to (8,9)$ & $(7,8) \to (5,9)$ \\
78  & $(8,2) \to (6,2)$ & $(8,0) \to (1,1)$ & $(8,1) \to (9,0)$ & $(8,0) \to (8,0)$ \\
79  & $(8,2) \to (7,4)$ & $(8,0) \to (2,8)$ & $(8,2) \to (9,3)$ & $(8,0) \to (9,0)$ \\
80  & $(8,2) \to (9,9)$ & $(8,0) \to (8,0)$ & $(8,3) \to (9,4)$ & $(8,2) \to (9,3)$ \\
81  & $(8,3) \to (6,5)$ & $(8,1) \to (9,1)$ & $(8,4) \to (9,3)$ & $(8,4) \to (6,3)$ \\
82  & $(8,4) \to (4,2)$ & $(8,7) \to (6,7)$ & $(8,5) \to (7,7)$ & $(8,4) \to (9,3)$ \\
83  & $(8,4) \to (8,1)$ & $(8,8) \to (5,9)$ & $(8,6) \to (9,9)$ & $(8,4) \to (9,4)$ \\
84  & $(8,4) \to (8,5)$ & $(8,8) \to (9,9)$ & $(8,8) \to (7,7)$ & $(8,6) \to (8,6)$ \\
85  & $(8,5) \to (1,4)$ & $(8,9) \to (6,9)$ & $(8,9) \to (0,1)$ & $(8,9) \to (1,8)$ \\
86  & $(8,5) \to (9,3)$ & $(8,9) \to (7,9)$ & $(8,9) \to (8,0)$ & $(8,9) \to (5,9)$ \\
87  & $(8,6) \to (0,7)$ & $(9,0) \to (0,0)$ & $(9,1) \to (9,0)$ & $(9,0) \to (9,0)$ \\
88  & $(8,6) \to (1,6)$ & $(9,2) \to (7,2)$ & $(9,1) \to (9,1)$ & $(9,0) \to (9,1)$ \\
89  & $(8,7) \to (6,9)$ & $(9,2) \to (9,1)$ & $(9,1) \to (9,9)$ & $(9,1) \to (9,1)$ \\
90  & $(8,9) \to (6,9)$ & $(9,4) \to (0,2)$ & $(9,2) \to (8,1)$ & $(9,1) \to (9,2)$ \\
91  & $(8,9) \to (7,9)$ & $(9,5) \to (1,5)$ & $(9,2) \to (9,0)$ & $(9,5) \to (8,5)$ \\
92  & $(8,9) \to (9,6)$ & $(9,6) \to (2,6)$ & $(9,2) \to (9,1)$ & $(9,5) \to (9,4)$ \\
93  & $(9,1) \to (1,3)$ & $(9,6) \to (7,4)$ & $(9,3) \to (8,3)$ & $(9,6) \to (9,6)$ \\
94  & $(9,1) \to (5,2)$ & $(9,6) \to (8,9)$ & $(9,3) \to (8,4)$ & $(9,6) \to (9,7)$ \\
95  & $(9,2) \to (7,0)$ & $(9,6) \to (9,6)$ & $(9,4) \to (9,4)$ & $(9,7) \to (7,7)$ \\
96  & $(9,3) \to (7,1)$ & $(9,7) \to (6,8)$ & $(9,4) \to (9,5)$ & $(9,7) \to (8,6)$ \\
97  & $(9,7) \to (0,0)$ & $(9,7) \to (9,4)$ & $(9,6) \to (8,7)$ & $(9,8) \to (9,7)$ \\
98  & $(9,8) \to (9,8)$ & $(9,8) \to (8,7)$ & $(9,7) \to (7,7)$ & $(9,9) \to (8,9)$ \\
99  & $(9,9) \to (0,9)$ & $(9,9) \to (0,9)$ & $(9,9) \to (8,8)$ & $(9,9) \to (9,8)$ \\
100 & $(9,9) \to (9,9)$ & $(9,9) \to (6,9)$ & $(9,9) \to (9,0)$ & $(9,9) \to (9,9)$\\
\hline
\caption{Sites selected for the measurements.}
\label{stable2}
\end{longtable}

\newpage
\vfill
\newpage
\vfill
\bibliography{subn}